\documentclass[aps,prd,final,groupedaddress,showpacs,showkeys,floatfix,preprintnumbers,nofootinbib,letterpaper]{revtex4}
\usepackage{graphicx}
\usepackage{float}

\usepackage[latin1]{inputenc}                    
\usepackage{latexsym}                            
\usepackage{amsfonts}                            
\usepackage{amssymb}                             
\usepackage{amsmath}                             
\usepackage[mathscr]{eucal}                      
\usepackage{dcolumn}                             
\usepackage{subfigure}
\usepackage[english]{babel}
\usepackage{theorem}                            
\usepackage{hyperref}
\usepackage{slashed}
\usepackage{cancel}
\usepackage{xcolor}

\definecolor{red}{rgb}{1.0, 0, 0}

\newcommand{\alfive}{{\alpha_5}}
\newcommand{\alfivep}{{\alpha_5^\prime}}
\newcommand{\alfivehat}{{\hat\alpha_5}}

\newcommand{\CP}{{CP}}
\newcommand{\CPbar}{{\overline{CP}}}
\newcommand{\CPviol}{{\cancel{CP}}}

\newcommand{\EucConv}{{\mathcal Euc}}

\newcommand{\MTwoConv}{{{\mathcal M}2}}

\newcommand{\llangle}{\langle\langle}
\newcommand{\rrangle}{\rangle\rangle}
\newcommand{\Tr}{\mathrm{Tr}}
\newcommand{\Dslash}{\slashed{D}}

\newcommand{\mcF}{{\mathcal F}}
\newcommand{\mcA}{{\mathcal A}}
\newcommand{\mcB}{{\mathcal B}}
\newcommand{\mcC}{{\mathcal C}}
\newcommand{\mcD}{{\mathcal D}}
\newcommand{\mcE}{{\mathcal E}}
\newcommand{\mcG}{{\mathcal G}}
\newcommand{\mcR}{{\mathcal R}}
\newcommand{\mcS}{{\mathcal S}}

\newcommand{\mcO}{{\mathcal O}}
\newcommand{\mcM}{{\mathcal M}}
\newcommand{\mcH}{{\mathcal H}}
\newcommand{\mcL}{{\mathcal L}}

\newcommand{\mcK}{{\mathcal K}}
\newcommand{\invplus}{{\phantom{+}}}

\begin{document}

\title{On Lattice Calculation of Electric Dipole Moments and Form Factors of the Nucleon}

\author{M.~Abramczyk}
\affiliation{Physics Department, University of Connecticut, Storrs, CT 06269, USA}

\author{S.~Aoki}
\affiliation{Center for Gravitational Physics, Yukawa Institute for Theoretical Physics,
  Kyoto University, Kyoto 606-8502, Japan}

\author{T.~Blum}
\affiliation{Physics Department, University of Connecticut, Storrs, CT 06269, USA}
\affiliation{RIKEN/BNL Research Center, Brookhaven National Laboratory, Upton, NY 11973, USA}

\author{T.~Izubuchi}
\affiliation{RIKEN/BNL Research Center, Brookhaven National Laboratory, Upton, NY 11973, USA}

\author{H.~Ohki}
\affiliation{RIKEN/BNL Research Center, Brookhaven National Laboratory, Upton, NY 11973, USA}

\author{S.~Syritsyn}
\affiliation{Jefferson Lab, 12000 Jefferson Ave, Newport News, VA 23606, USA}
\affiliation{Kavli Institute for Theoretical Physics, UC Santa Barbara, CA 93106, USA}
\affiliation{Department of Physics and Astronomy, Stony Brook University, Stony Brook, NY 11794, USA}

\preprint{RBRC-1226}

\begin{abstract}
We analyze commonly used expressions for computing the nucleon electric 
dipole form factors (EDFF) $F_3$ and moments (EDM) on a lattice
and find that they lead to spurious contributions from the Pauli form factor $F_2$ 
due to inadequate definition of these form factors when parity mixing of lattice nucleon 
fields is involved.
Using chirally symmetric domain wall fermions, we calculate the proton and the neutron EDFF
induced by the CP-violating quark chromo-EDM interaction using the corrected expression.
In addition, we calculate the electric dipole moment of the neutron using background
electric field that respects time translation invariance and boundary conditions,
and find that it decidedly agrees with the new formula but not the old formula for $F_3$.
Finally, we analyze some selected lattice results for the nucleon EDM and observe that after 
the correction is applied, they either agree with zero or 
are substantially reduced in magnitude, thus reconciling their difference from 
phenomenological estimates of the nucleon EDM.
\end{abstract}

\maketitle

\section{Introduction}

The origin of nuclear matter can be traced back to the excess of nucleons 
over antinucleons in the early Universe and it is one of the greatest puzzles
in Physics known as the baryonic asymmetry of the Universe (BAU).
One of the required conditions for the BAU is violation of the $\CP$ symmetry (\CPviol).
In the Standard Model~(SM), the CKM matrix phases lead to $\CP$ violations in weak interactions, 
but their magnitudes are not sufficient to explain the BAU, 
and signs of additional $\CPviol$ are actively sought in experiments.
The most promising ways to look for $\CPviol$ are measurements of
electric dipole moments (EDM) of atoms, nucleons and nuclei.
In particular, the Standard Model prediction for the neutron EDM is five orders 
below the current experimental bound, and represents a negligible background.
Near-future EDM experiments plan to improve this bound by 2 orders of magnitude,
and are capable of constraining various Beyond-the-Standard-Model (BSM) extensions 
of particle physics, purely from low-energy nuclear and atomic high-precision experiments.
%
Knowledge of nucleon structure and interactions is necessary to interpret these 
experiments in terms of quark and gluon effective operators and put constraints 
on proposed extensions of the Standard Model, in particular SUSY and GUT models 
as sources of additional $\CPviol$. 
Connecting the quark- and gluon- to hadron-level effective $\CPviol$ 
interactions is an urgent task for lattice QCD (an extensive
review of EDM phenomenology can be found in Ref.~\cite{Engel:2013lsa}).

The proton and the neutron can have electric dipole moments only if the symmetry of 
the Standard Model Lagrangian is broken by additional $P$-,$T$-odd interactions.
The only such dimension-4 operator is the QCD $\bar\theta$-angle 
($\bar\theta$ stands for the physically-relevant combination of the QCD $\theta$ 
angle and quark mass phases).
The $\bar\theta$-induced nucleon EDMs (nEDMs) have been calculated on a lattice 
from energy shifts in uniform background electric 
field~\cite{Aoki:1989rx,Shintani:2006xr,Shintani:2008nt}
or extracting the $P$-odd electric dipole form factor (EDFF) $F_3(Q^2)$
from nucleon matrix elements of the vector current in $\CPviol$ vacuum~\cite{
Shintani:2005xg,Berruto:2005hg,Aoki:2008gv,Guo:2015tla,Shindler:2015aqa,
Alexandrou:2015spa,Shintani:2015vsx}.
Nucleon EDMs 
have been studied using QCD 
sum rules, quark models, and chiral perturbation theory (see Refs.~\cite{
Crewther:1979pi,Pich:1991fq,Pospelov:1999ha,Pospelov:2000bw,Hisano:2012sc,Borasoy:2000pq,
Mereghetti:2010kp} 
to name a few).
On a lattice, quark EDM-induced nucleon EDMs have been recently
computed on a lattice in partially-quenched framework~\cite{Bhattacharya:2015esa}.
Another important dimension-5(6)\footnote{
  These operators are sometimes referred to as ``dimension-6'' because in the SM they
  contain a factor of the Higgs field.} 
operator is the $\CP$-odd quark-gluon interaction,
also known as the chromo-electric dipole moment (cEDM)
\begin{equation}
\label{eqn:qcedm}
\mcL_{cEDM} = i \sum_{\psi=u,d} \frac{\tilde \delta_\psi}2
    \bar\psi (T^a G^q_{\mu\nu}) \sigma^{\mu\nu}\gamma_5 \psi\,,\\
\end{equation}
and calculations of cEDM-induced nEDMs have recently started using Wilson
fermions~\cite{Bhattacharya:2016oqm,Bhattacharya:2016rrc}.

In this paper, we report several important achievements in studying nucleon EDMs 
on a lattice. 
First, we argue that the commonly accepted methodology for computing electric 
dipole form factors of spin-1/2 particles on a lattice has a problem to identify the electric
dipole moment form factor.
In particular, in the standard analysis of the nucleon-current correlators~\cite{
Shintani:2005xg,Berruto:2005hg,Aoki:2008gv,Guo:2015tla,Shindler:2015aqa,Alexandrou:2015spa,
Shintani:2015vsx}, the electric dipole form factor 
$F_3$ receives large and likely dominant contribution from spurious mixing with 
the Pauli form factor $F_2$.
The energy shift methods~\cite{Aoki:1989rx,Shintani:2006xr,Shintani:2008nt} are not affected by
such mixing, but their precision has not been sufficient to detect the discrepancy.
This problem affects all the previous lattice calculations of the nucleon EDFFs and EDMs from
nucleon-current correlators, including those studying the $\bar\theta$-angle~\cite{
Shintani:2005xg,Berruto:2005hg,Aoki:2008gv,Guo:2015tla,Shindler:2015aqa,Alexandrou:2015spa} 
as well as the more recent studying chromo-EDM~\cite{Bhattacharya:2016oqm,Bhattacharya:2016rrc}.
We demonstrate the problem formally in Sec~\ref{sec:edm_ff_alg}
and also derive the correction for the results of Refs.~\cite{
Shintani:2005xg,Berruto:2005hg,Aoki:2008gv,Guo:2015tla,Shindler:2015aqa,Alexandrou:2015spa,
Shintani:2015vsx} to subtract the spurious mixing with $F_2$.
In addition, in Sec.~\ref{sec:edm_dirac_eshift_euc} we study the energy shift of a neutral
particle on a Euclidean lattice in uniform background electric field.
We introduce the uniform electric field preserving translational invariance and periodic 
boundary conditions on a lattice~\cite{Detmold:2009dx}.
In order to satisfy these conditions, the electric field value has to be analytically continued to 
the imaginary axis upon Wick rotation from Minkowski to Euclid, and we demonstrate that 
the eigenstates of a fermion having an EDM are shifted by a purely imaginary value.
In Sec.~\ref{sec:bgem}, we apply this formalism to the analysis of neutron correlators 
computed in the presence of the quark chromo-EDM interaction~(\ref{eqn:qcedm}).

Calculation of the neutron EDM in the background field is independent from parity mixing
ambiguities, and it allows us to validate our new formula for the EDFF $F_3$ numerically.
The difference is evident only if the nucleon ``parity-mixing'' angle $\alfive$ is large, 
$\alfive\gtrsim1$.
The calculations with quark chromo-EDM generate very strong parity mixing compared to the 
$\bar\theta$-angle, which is beneficial for our numerical check.
In Section~\ref{sec:lat_ff} we calculate the proton and neutron EDFFs $F_{3p,n}(Q^2)$ 
induced by the quark chromo-EDM interaction~(\ref{eqn:qcedm}), as well as the regular 
$\CP$-even Dirac and Pauli form factors $F_{1,2}$.
In Sec.~\ref{sec:ff_eshift_cmp} we compare the EDM results from the form factor and 
the energy-shift calculations, 
providing a numerical confirmation of the validity of our new EDFF analysis.
Finally, in Section~\ref{sec:prev_works} we analyze some select results for nucleon EDM 
induced by $\bar\theta$-angle availiable in the literature~\cite{
Shintani:2005xg,Berruto:2005hg,
Guo:2015tla,
Alexandrou:2015spa,Shintani:2015vsx} and attempt to correct them according to our findings.

\section{CP-odd form factors of spin-1/2 particle}\label{sec:edm_ff_alg}

\subsection{Form factors and parity mixing}\label{sec:edm_ff_mink}

In this section we argue that the ubiquitously used expression for computing $\CP$-odd electric
dipole form factor $F_3$ \emph{on a lattice} does not correspond to the electric dipole moment
measured in experiments and leads to a finite and perhaps dominant contribution from the Pauli 
form factor $F_2$ to the reported values of EDFF $F_3$ and EDM of the proton and the neutron. 
First, we recall the lattice framework for calculation of the $\CP$-violating form factor $F_3$
first introduced in Ref.~\cite{Shintani:2005xg}, and later used without substantial
changes in the subsequent papers~\cite{Berruto:2005hg,Aoki:2008gv,Guo:2015tla,Shindler:2015aqa,
Alexandrou:2015spa,Shintani:2015vsx} studying the QCD $\theta$-term, as well as more recent
developments~\cite{Bhattacharya:2016oqm,Bhattacharya:2016rrc} to study the quark chromo-EDM.

\begin{figure}[ht!]
\centering
\includegraphics[width=0.2\textwidth]{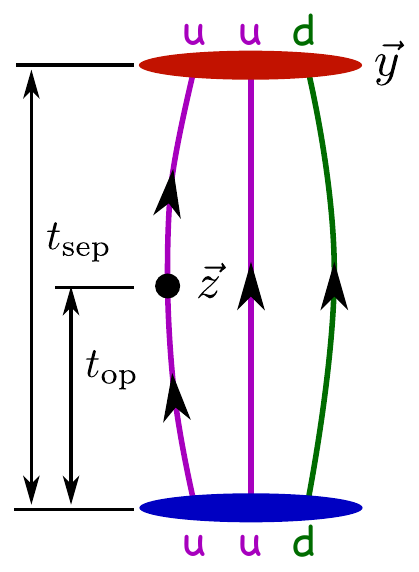}
\caption{Dependence of the nucleon three-point function (\ref{eqn:threept_cpviol}) 
  on the source-sink separation $t_\text{sep}$ and operator insertion $t_\text{op}$.
\label{fig:nucleon-3pt-coord}}
\end{figure}

To compute nucleon form factors on a lattice, one evaluates nucleon two- and three-point functions
(see Fig.~\ref{fig:nucleon-3pt-coord})
in presence of $\CP$-violating ($\CPviol$) interactions
\begin{align}
\label{eqn:twopt_cpviol}
C_{N\bar N}^\CPviol(\vec p,t) 
  &= \sum_{\vec x} e^{-i\vec p\cdot \vec x}\langle N(\vec x,t) \bar{N}(0)\rangle_\CPviol\,,\\
\label{eqn:threept_cpviol}
C_{NJ\bar N}^\CPviol(\vec p^\prime,t_{sep};\vec q,t_{op})
  &= \sum_{\vec y,\vec z} e^{-i\vec p^\prime\cdot \vec y+i\vec q\cdot\vec z}
              \langle N(\vec y,t_{sep}) J^\mu(\vec z,t_{op}) \bar{N}(0)\rangle_\CPviol\,.
\end{align}
The subscript $\CPviol$ indicates that these correlation functions are evaluated in $\CPviol$
vacuum, either with a finite value of the relevant $\CPviol$ coupling or an infinitesimal value,
i.e. performing first-order Taylor expansion of the correlation functions.
As argued in Ref.~\cite{Shintani:2005xg}, as well as earlier model
calculations~\cite{Pospelov:1999ha,Pospelov:2000bw}, the $\CPviol$ background leads to a 
$\CPviol$ phase in the
nucleon mass in the Dirac equation that governs the on-shell nucleon fields $N,\,\bar{N}$ 
\begin{equation}
(i\slashed{\partial} - m_{N^\prime}e^{-2i\alfive\gamma_5})N(x) = 0\,.
\end{equation}
where the real-valued $m_{N^\prime}>0$ is the nucleon ground state mass in the new vacuum.
The spinor wave functions $\tilde{u}_p,\,\bar{\tilde{u}}_p$ for the new ground states
\begin{equation}
\label{eqn:ferm_pmixed}
\langle\Omega|N|p,\sigma\rangle_\CPviol
  = Z_{N^\prime} \tilde{u}_{p,\sigma}  
  = Z_{N^\prime} e^{i\alfive\gamma_5} u_{p,\sigma}\,,
\end{equation}
also satisfy the same Dirac equation
\begin{equation}
\label{eqn:spinor_dirac_euc_tw}
(\slashed{p} - m_{N^\prime} e^{-2i\alfive\gamma_5}) \tilde{u}_p 
  = (\slashed{p} - m_{N^\prime} e^{-2i\alfive\gamma_5}) e^{i\alfive\gamma_5} u_p = 0 \,,
\end{equation}
where the chirally-rotated spinors $\tilde{u}_p,\,\bar{\tilde{u}}_p$ have a 
Lorentz-invariant $\CPviol$ phase similarly to the mass term,
\begin{equation}
\label{eqn:spinor_g5phase}
\tilde{u}=e^{i\alfive\gamma_5} \, u\,, 
\quad
\bar{\tilde{u}} = \bar{u} \, e^{i\alfive\gamma_5}\,,
\end{equation}
where the regular spinors $u_p$, $\bar{u}_p$ satisfy the regular Dirac equation with a real-valued
nucleon mass,
\begin{equation}
\label{eqn:spinor_dirac_reg}
(\slashed{p} - m_{N^\prime}) u_p = 0\,,
\quad
\bar{u}_p (\slashed{p} - m_{N^\prime}) = 0\,
\end{equation}
From the above equation (\ref{eqn:spinor_dirac_reg}) it also follows that the spinors $u_p$,
$\bar{u}_p$ transform under spatial reflection (parity $P$) as the regular spinors,
\begin{equation}
\label{eqn:spinor_parity_reg}
\gamma_4 u_{p=(\vec p,E)} = u_{\tilde{p}=(-\vec p,E)}\,.
\end{equation}

Below we will discuss correlation functions on a Euclidean lattice, which depend on the
Wick-rotated 4-momentum and are more conveniently expressed using the Euclidean matrices
$[\gamma^\mu]_\EucConv$~(\ref{eqn:gamma_euc}). 
Whenever a confusion may arise, we will explicitly specify the type 
$\MTwoConv$ (Minknowski) or $\EucConv$ (Euclidean) of $\gamma$-matrices and 4-vectors
(see App.~\ref{sec:app_mink_euc} for details). 
The Euclidean versions of the Dirac equations~(\ref{eqn:spinor_dirac_reg}) 
for the nucleon spinors are
\begin{equation}
\label{eqn:spinor_dirac_euc_reg}
(i\slashed{p}_\EucConv + m_{N^\prime}) u_p = 0\,,
\quad
\bar{u}_p (i\slashed{p}_\EucConv + m_{N^\prime}) = 0\,,
\end{equation}
where $(-i\slashed{p}_{\EucConv}) = (-i)\,p^\mu_\EucConv \, [\gamma_\mu]_\EucConv
  = E\,[\gamma^4]_\EucConv - i\vec p\cdot [\gamma]_\EucConv$,
in which the Euclidean on-shell 4-momentum $p^\mu_\EucConv=(\vec p, iE)$ 
is contracted with Euclidean $\gamma$-matrices and
$E=\sqrt{m_{N^\prime}^2 + \vec p^{\prime2}}$ is the real-valued on-shell energy of the nucleon.
Due to the chiral phase~(\ref{eqn:spinor_g5phase}), the nucleon propagator on a
lattice~(\ref{eqn:twopt_cpviol}) contains chiral phases $e^{i\alfive\gamma_5}$.
Keeping only the ground state and omitting the exponential time dependence for simplicity,
we get
\begin{equation}
\label{eqn:nucfield_prop_mixed}
C_{N\bar N}(\vec p,t)\Big|_{g.s.}
  \sim \frac{\sum_{\sigma}\tilde{u}_{p,\sigma} \bar\tilde{u}_{p,\sigma}}{2E_{N^\prime}}
  = \frac{-i\slashed{p}_{\EucConv} + m_{N^\prime}e^{2i\alfive\gamma_5}}{2E_{N^\prime}}
  = e^{i\alfive\gamma_5} 
    \Big[\frac{-i\slashed{p}_{\EucConv} + m_{N^\prime}}{2E_{N^\prime}}\Big]
    e^{i\alfive\gamma_5}\,.
\end{equation}
Analogously, the expression for the nucleon-current correlator~(\ref{eqn:threept_cpviol}) 
contains the phases $e^{i\alfive\gamma_5}$:
\begin{equation}
C_{NJ\bar N}(\vec p^\prime,t_{sep};\vec q, t_{op})\Big|_{g.s.}
  \sim \sum_{\sigma^\prime,\sigma} 
    \tilde{u}_{p^\prime,\sigma^\prime}
    \langle p^\prime,\sigma^\prime |J^\mu|p,\sigma\rangle_{\CPviol}
    \bar{\tilde{u}}_{p,\sigma}
  = e^{i\alfive\gamma_5} 
    \big[\sum_{\sigma^\prime,\sigma} u_{p^\prime,\sigma^\prime}
    \langle p^\prime,\sigma^\prime |J^\mu|p,\sigma\rangle_{\CPviol}
    \bar{u}_{p,\sigma} \big] e^{i\alfive\gamma_5}\,.
\end{equation}

The problem with the commonly used expression for the three-point function comes from the fact
that the physical interpretation of a parity-mixed fermion state~(\ref{eqn:ferm_pmixed})
on the lattice is not clear.
In Refs.~\cite{Shintani:2005xg,Berruto:2005hg,Aoki:2008gv,Guo:2015tla,Shindler:2015aqa,
Alexandrou:2015spa,Shintani:2015vsx}, it is assumed that the 
nucleon matrix elements of the vector current in $\CPviol$ vacuum have the form 
(in Minkowski space, up to sign conventions for $F_3$ and $F_A$)
\begin{equation}
\label{eqn:ff_cpviol_mink_wrong}
\langle p^\prime,\sigma^\prime |J^\mu|p,\sigma\rangle_{\CPviol}
  \overset?= \bar{\tilde{u}}_{p^\prime,\sigma^\prime} \big[
    F_1(Q^2) \gamma^\mu 
    + \tilde{F}_2(Q^2) \frac{i\sigma^{\mu\nu}q_\nu}{2m_{N^\prime}}
    - \tilde{F}_3(Q^2) \frac{\gamma_5\sigma^{\mu\nu}q_\nu}{2m_{N^\prime}}
    + F_A(Q^2) \frac{(\slashed{q}q^\mu - \gamma^\mu q^2)\gamma_5}{m_{N^\prime}^2}
  \big] \tilde{u}_{p,\sigma}\,,
\end{equation}
where $q=p^\prime-p$, $Q^2$ = $-[q^2]_{\mcM2}=-(q^4)^2+\vec q^2$, 
$F_1$ and $\tilde{F}_2$ are the Dirac and Pauli form factors, 
$\tilde{F}_3$ is the electric dipole form factor (EDFF),
and $F_A$ is the anapole form factor
(notations $\tilde{F}_{2,3}$ are introduced to avoid confusion with the true $F_{2,3}$ below).
The matrix element expression~(\ref{eqn:ff_cpviol_mink_wrong}), however, disagrees with the 
literature~\cite{ItzyksonZuber-QFT-ffs}, 
\begin{equation}
\label{eqn:ff_cpviol_mink}
\langle p^\prime,\sigma^\prime |J^\mu|p,\sigma\rangle_{\CPviol}
  = \bar{u}_{p^\prime,\sigma^\prime} \big[
    F_1(Q^2) \gamma^\mu 
    + F_2(Q^2) \frac{i\sigma^{\mu\nu}q_\nu}{2m_{N^\prime}}
    - F_3(Q^2) \frac{\gamma_5\sigma^{\mu\nu}q_\nu}{2m_{N^\prime}}
    + F_A(Q^2) \frac{(\slashed{q}q^\mu - \gamma^\mu q^2)\gamma_5}{m_{N^\prime}^2}
  \big] u_{p,\sigma}\,,
\end{equation}
in which the vertex spin matrix 
\begin{equation}
\label{eqn:nuc_cur_vertex}
\Gamma^\mu(p^\prime,p)
  = F_1 \gamma^\mu + (F_2 + i F_3\gamma_5) \frac{i\sigma^{\mu\nu}q_\nu}{2m_{N^\prime}} 
  + F_A\frac{(\slashed{q}q^\mu - \gamma^\mu q^2)\gamma_5}{m_{N^\prime}^2}
\end{equation}
is contracted with the spinors 
satisfying the regular parity transformations of spinors~(\ref{eqn:spinor_parity_reg}).
Only in this case the contribution of the form factor $F_3$ to the matrix element 
$\langle p^\prime,\sigma^\prime |J^\mu|p,\sigma\rangle$ transforms as an axial 4-vector
so that $F_3$ is indeed the $\CP$-odd coupling of the nucleon to the electromagnetic
potential~\cite{ItzyksonZuber-QFT-ffs}.
Let us show that this is not the case if the matrix element of the current has the
form~(\ref{eqn:ff_cpviol_mink_wrong}).
Upon spatial reflection, the true 4-vectors of momenta and current have to transform as
\begin{equation}
(\vec{p}^{(\prime)},\,p^{4(\prime)}) \rightarrow (-\vec{p}^{(\prime)},\,p^{4(\prime)})\,,
\quad (\vec{J},\,J^{4}) \rightarrow (-\vec{J},\,J^{4})\,.
\end{equation}
while the axial vector current $A^\mu$ transforms with the sign opposite to $J^\mu$:
\begin{equation}
(\vec{A},\,A^{4}) \rightarrow (\vec{A},\,-A^{4})\,.
\end{equation}
The chirally-rotated spinors transform as
\begin{align}
\tilde{u}_{\vec p} &\rightarrow 
  \tilde{u}_{-\vec p} = e^{2i\alfive\gamma_5}\gamma_4 \, \tilde{u}_{\vec p}\,, \\
\bar{\tilde{u}}_{\vec p} & \rightarrow 
  \bar{\tilde{u}}_{-\vec p} = \bar{\tilde{u}}_{\vec p} \, \gamma_4 e^{2i\alfive\gamma_5}\,,
\end{align}
up to an irrelevant spinor-diagonal phase factor.
Finally, remembering that the spatial momentum $\vec q$ is also reflected and using the identities
\begin{gather}
\gamma_4\sigma^{i\nu}\,(-\vec q,\,q^4)_\nu\gamma_4 = -\sigma^{i\nu}(\vec q,\,q^4)_\nu\,,
\\
\gamma_4\sigma^{4\nu}\,(-\vec q,\,q^4)_\nu\gamma_4 = \sigma^{4\nu}(\vec q,\,q^4)_\nu\,,
\end{gather}
we observe that a combination of the $\tilde{F}_{2,3}$ form factors transforms as
\begin{equation}
e^{2i\alfive}(\tilde{F}_2 + i\tilde{F}_3) \rightarrow
e^{-2i\alfive}(\tilde{F}_2 - i\tilde{F}_3)\,.
\end{equation}
Therefore, we conclude that the axial-vector contribution of the matrix
element~(\ref{eqn:ff_cpviol_mink_wrong}) appears because of the parity-odd form factor combination
\begin{equation}
\label{eqn:twF23_cpodd}
\mathrm{Im}[e^{2i\alfive}(\tilde{F}_2 + i\tilde{F}_3)]
  = \sin(2\alfive) \tilde{F}_2 + \cos(2\alfive) \tilde{F}_3\,.
\end{equation}
which is different from $F_3$ if $\alfive\ne\pi n$.

Since the expression~(\ref{eqn:ff_cpviol_mink_wrong}) is used in lattice calculations so
ubiquitously, we present extensive arguments that it is not correct.
The form factor $F_A$ is irrelevant for this discussion, and will be omitted\footnote{
  It is also worth noting that $F_A$ is not affected by the parity mixing, unlike $F_{2,3}$.}.
In Appendix~\ref{sec:edm_ff_rest} we directly show that it is the 
expression~(\ref{eqn:ff_cpviol_mink}) that leads to the correct $\CP$-odd EDM coupling 
$\sim\vec E\cdot \vec S$, 
and the forward limit $Q^2\to0$ of form factors $F_2(Q^2)$ and $F_3(Q^2)$ yields the anomalous
magnetic $\kappa$ and electric $\zeta$ dipole moments (in units $e/(2m_N)$), respectively.
In Section~\ref{sec:edm_dirac_eshift} we calculate the mass shift of a particle governed by a
Dirac equation with chirally-rotated mass in background electric field.

In this section we offer several heuristic arguments why
expression~(\ref{eqn:ff_cpviol_mink_wrong}) is not correct.
First, revisiting the form factor expression~(\ref{eqn:ff_cpviol_mink}), we note that 
the only effect of the chiral phase is to mix form factors $F_2$ and $F_3$ into each other,
\begin{equation}
\label{eqn:vertex_mixing}
  \bar{\tilde{u}}_{p^\prime} \big[
    F_1 \gamma^\mu 
    + (\tilde{F}_2 + i\tilde{F}_3\gamma_5) \frac{i\sigma^{\mu\nu}q_\nu}{2m_{N^\prime}}
  \big] \tilde{u}_{p}
= \bar{u}_{p^\prime} \big[
    F_1 \gamma^\mu 
    + e^{2i\alfive\gamma_5} (\tilde{F}_2 + i \tilde{F}_3 \gamma_5)
    \frac{i\sigma^{\mu\nu}q_\nu}{2m_{N^\prime}}
  \big] u_{p}
\end{equation}
while the form factor $F_1$, as well as the omitted $F_A$, are independent of $\alfive$.
Thus, the form factors $\tilde{F}_{2,3}$ computed in Refs.~\cite{Shintani:2005xg,
Berruto:2005hg,Aoki:2008gv,Guo:2015tla,Shindler:2015aqa,Alexandrou:2015spa,Shintani:2015vsx}
are linear combinations of the true form factors $F_{2,3}$ 
\begin{equation}
\label{eqn:F23_mixing}
(F_2 + i F_3 \gamma_5) = e^{2i\alfive\gamma_5} (\tilde{F}_2 + i \tilde{F}_3 \gamma_5)\,,
\quad\text{ or }
\left\{\begin{array}{rl}
\tilde{F}_2 &= \invplus \cos(2\alfive)\,F_2 + \sin(2\alfive)\,F_3 \,,\\
\tilde{F}_3 &= -\sin(2\alfive)\,F_2 + \cos(2\alfive)\,F_3\,,\\
\end{array}\right.
\end{equation}
which is also consistent with Eq.~(\ref{eqn:twF23_cpodd}).

It is easy to see that the effect of the phase $e^{i\alfive\gamma_5}$ can be completely
removed by a field redefinition $N^\prime= e^{-i\alfive\gamma_5} N$.
After this transformation, the on-shell nucleon field $N^\prime$ will satisfy a Dirac equation 
with the real-valued mass $m_{N^\prime}$,
\begin{equation}
(i\slashed{\partial} - m_{N^\prime})N^\prime(x) = 0\,.
\end{equation}
A similar transformation for the nucleon correlators 
$C_{N[J]\bar N}\rightarrow C^\prime_{N[J]\bar N}
  = e^{-i\alfive\gamma_5} C_{N[J]\bar N} e^{-i\alfive\gamma_5}$ 
will remove any dependence on $\alfive$ altogether.
Note, however, that this is the case only if Eq.~(\ref{eqn:ff_cpviol_mink}) is used for the 
nucleon matrix elements of the current.
Thus, this phase is purely conventional and similar to the operator normalization $Z_{N^\prime}$,
in that physical quantities cannot depend on it.
In a lattice calculation, however, this phase is not known in advance and must 
be determined numerically to be removed from the two- and three-point
correlators~(\ref{eqn:twopt_cpviol},\ref{eqn:threept_cpviol}).

To make this point evident, suppose one calculated nucleon form factors in $\CP$-even QCD but
using unconventional nucleon interpolating fields $e^{i\alfivep\gamma_5} N$ with some arbitrary 
$\alfivep$.
If Eq.~(\ref{eqn:ff_cpviol_mink_wrong}) was used, the definition of $\tilde{F}_{2,3}$  would
depend on this arbitrarily chosen $\alfivep$.
Consequently, because of the spurious mixing~(\ref{eqn:F23_mixing}), 
the electric dipole form factor would obtain the non-zero value $\tilde{F}_3 = -F_2\sin(2\alfivep)$
in absence of any $\CPviol$ interactions.
Analogously, the apparent nucleon magnetic moment 
$\tilde\mu_{N^\prime}=\tilde{G}_M(0)=F_1(0)+\tilde{F}_2$ 
would have contributions from both $F_2$ and $F_3$.
In $\CP$-even QCD vacuum, $F_3=0$ and the mixing~(\ref{eqn:F23_mixing}) would just reduce
the contribution of $F_2$ to $\tilde\mu_{N^\prime}$ by a factor of $\cos(2\alfivep)$.
This would happen because the spin operator $\Sigma^k=\frac12\epsilon^{ijk}\sigma^{ij}$
was ``sandwiched'' between chirally-rotated 4-spinors and
\begin{equation}
(2\vec S)^k = \Sigma^k 
  = \bar{\tilde{u}}_{p^\prime} \frac{\Sigma^k}{2m_{N^\prime}} \tilde{u}_p
  = \xi^{\prime\dag}\sigma^k\xi\,\cos(2\alfive)
\end{equation}
where the initial and final momenta $\vec p,\vec p^\prime\approx0$
and $\xi,\xi^\prime$ are the corresponding 2-spinors.

The resolution to this apparent paradox is hinted by the modified form of the Gordon identity
for the spinors $\tilde{u}_p,\,\bar{\tilde{u}}_p$.
Since these spinors satisfy the Dirac equation with the chirally rotated 
mass~(\ref{eqn:spinor_dirac_euc_tw}), the Gordon identity takes the form
\begin{equation}
\label{eqn:gordon_id_tw}
\bar{\tilde{u}}_{p^\prime} \gamma^\mu \tilde{u}_p 
  = \bar{\tilde{u}}_{p^\prime} \big[
      \frac{(p^\prime + p)^\mu + i\sigma^{\mu\nu}(p^\prime - p)_\nu}
           {2 m_{N^\prime} e^{2i\alfive\gamma_5}} 
    \big]\tilde{u}_p\,,
\end{equation}
which is obtained from the standard Gordon identity by replacing 
$m_{N^\prime}\rightarrow m_{N^\prime} e^{2i\alfive\gamma_5}$.
The form of the nucleon-current vertex must be compatible with the form of the Gordon identity,
which, among other things, relates form factors $F_{1,2}$ to $G_M$.
Therefore, to make the nucleon-current vertex compatible with the spinors
$\tilde{u}_p,\,\bar{\tilde{u}}_p$, 
the nucleon mass in the $\tilde{F}_{2,3}$ terms in Eq.~(\ref{eqn:ff_cpviol_mink_wrong}) 
has to be adjusted similarly to Eq.~(\ref{eqn:gordon_id_tw}),
which leads back to the correct expression~(\ref{eqn:ff_cpviol_mink}).

Finally, we emphasize that Eqs.~(\ref{eqn:ff_cpviol_mink_wrong}) and
(\ref{eqn:ff_cpviol_mink}) result in different prescriptions for analyzing the three-point
nucleon-current correlators:
\begin{equation}
C_{NJ\bar N}(\vec p,t_{sep};\vec q,t_{op})\Big|_{g.s.}
  \overset{?}{=} 
    e^{-E_{N^\prime}^\prime(t_{sep}-t_{op})-E_{N^\prime}t_{op}}
    e^{i\alfive\gamma_5}
    \frac{-i\slashed{p}^\prime_{\EucConv} + m_{N^\prime}}{2E^\prime_{N^\prime}}
    \big\{e^{i\alfive\gamma_5}\big\}^?
    \Gamma^\mu_\EucConv
    \big\{e^{i\alfive\gamma_5}\big\}^?
    \frac{-i\slashed{p}_{\EucConv} + m_{N^\prime}}{2E_{N^\prime}}
    e^{i\alfive\gamma_5}
\end{equation}
where phase factors in curly braces $\big\{e^{i\alfive\gamma_5}\big\}^?$ are present only if 
one uses the (incorrect) Eq.(\ref{eqn:ff_cpviol_mink_wrong}).
In the above equation, we have introduced the Euclidean nucleon-current vertex
\begin{equation}
\label{eqn:nuc_cur_vertex_euc}
\Gamma^\mu_\EucConv(p^\prime, p) =[F_1\gamma^\mu + (F_2 + i\gamma_5 F_3)
    \frac{\sigma^{\mu\nu}q_\nu}{2m_{N^\prime}}]_\EucConv
\end{equation}


\subsection{EDM energy shift from Dirac equation}\label{sec:edm_dirac_eshift}

We argued in the previous section that one has to use regular ''even'' spinors satisfying
Eq.~(\ref{eqn:spinor_parity_reg}) to evaluate the nucleon matrix elements even if the QCD vacuum
is $\CP$-broken, contrary to the previous works~\cite{Shintani:2005xg,Berruto:2005hg,Aoki:2008gv,
Guo:2015tla,Shindler:2015aqa,Alexandrou:2015spa,Shintani:2015vsx}.
Most of the ambiguity must have resulted from the notion that in a $\CP$-broken vacuum, particles
are no longer parity eigenstates, hence, the argument goes, the nucleon must be described by a
parity-mixed spinor.
This argument is rather confusing because parity transformations of fermion fields are fixed only
up to a phase factor, and only a fermion-antifermion pair may have definite parity.
To clarify this question, in this section we calculate the energy spectrum of a particle described
by the Dirac operator $\tilde\Dslash_N$ with the complex mass $me^{-2i\alfive\gamma_5}$ and with
magnetic and electric dipole interactions in the form~(\ref{eqn:ff_cpviol_mink_wrong}) in the
background of uniform magnetic and electric fields.
Such an operator is exactly the nucleon effective operator on a lattice.
The zero modes of this operator (i.e., the poles of its Green's function) must correspond to 
particle eigenstates, and their calculation avoids the spinor phase ambiguity completely.
The energy shifts linear in these fields are then identified with the magnetic $\kappa$ and 
electric $\zeta$ dipole moments, respectively.

The effective action for the Euclidean lattice nucleon field in the $\CPviol$ vacuum and
point-like electromagnetic interaction introduced via ``long derivative'' is
\begin{equation}
\mcL_\text{int} = \bar N \, \big[i\slashed{\partial} 
      - Q \gamma^\mu A_\mu - m e^{-2i\alfive\gamma_5} \big]\,N\,,
\end{equation}
where we neglect the momentum transfer dependence of the nucleon form factors
for simplicity, setting $F_1$ to a ``point-like'' value $Q=F_1(0)=\mathrm{const}$.
In the absence of electromagnetic potential $A_\mu$, the nucleon
propagator $\langle N\bar N \rangle$ takes the form~(\ref{eqn:nucfield_prop_mixed}).
We add effective point-like anomalous magnetic $\tilde\kappa=\tilde{F}_2(0)$
and electric $\tilde{\zeta}=\tilde{F}_3(0)$ dipole interactions 
to the interaction vertex
\begin{equation}
Q \gamma^\mu \rightarrow 
  Q \gamma^\mu + \tilde\kappa\frac{i\sigma^{\mu\nu}q_\nu}{2m} 
  - \tilde{\zeta}\frac{\gamma_5\sigma^{\mu\nu}q_\nu}{2m}
\end{equation}
Using conventions~(\ref{eqn:field_momenta}-\ref{eqn:Fmunu_mom}) as well as
(\ref{eqn:Fmunu},\ref{eqn:Fmunu_dual}), the Dirac equation for $N$ becomes
\begin{equation}
\label{eqn:dslash_eff_twisted}
\big[\slashed{p} - Q\gamma^\mu A_\mu 
  - (\tilde\kappa + i\tilde{\zeta}\gamma_5)\,\frac12 F_{\mu\nu}\frac{\sigma^{\mu\nu}}{2m}
      - m e^{-2i\alfive\gamma_5} \big] N
  = 0
\end{equation}
We are going to find the energy levels of the particle in presence of constant field strength 
$F_{\mu\nu}$. 
To avoid irrelevant complications, we consider only a neutral particle with $Q=0$. 
Using the identity~(\ref{eqn:gamma5sigma}) to trade $\gamma_5$ for
$F_{\mu\nu}\to\tilde{F}_{\mu\nu}$, we obtain the Dirac operator in the block-diagonal form 
in the chiral basis~(\ref{eqn:gamma_chiral}):
\begin{equation}
\label{eqn:dirac_op_block_tw}
\slashed{p} - \frac12(\tilde\kappa F_{\mu\nu} 
      - \tilde{\zeta}\tilde F_{\mu\nu})\frac{\sigma^{\mu\nu}}{2m} 
  - m e^{-2i\alfive\gamma_5}
  =\left(\begin{array}{cc} 
    -M & E-\vec p\cdot\vec\sigma \\ E+\vec p\cdot\vec\sigma & -M^\dag \end{array}\right)\,, 
\end{equation}
where 
$M = me^{2i\alfive} - \frac{1}{2m}(\tilde\kappa - i \tilde\zeta)(\vec H + i\vec E)\cdot\vec\sigma$.
In the rest frame, $\vec p=(\vec0,E_0)$, and the operator~(\ref{eqn:dirac_op_block_tw}) has 
solutions if
\begin{equation}
\mathrm{det}\left(\begin{array}{cc} -M & E_0 \\ E_0 & -M^\dag \end{array}\right) = 0
\quad\Leftrightarrow\quad
    \mathrm{det}(E_0^2 - M^\dag M) = \mathrm{det}(E_0^2 - M M^\dag) = 0
\end{equation}
Up to terms linear in $\tilde\kappa$, $\tilde\zeta$, the normal operator $M^\dag M$ is
\begin{equation}
\begin{aligned}
M^\dag M &= m^2 - \frac12\big[ e^{ 2i\alfive}(\tilde\kappa + i\tilde \zeta)(\vec H - i\vec E) 
              + e^{-2i\alfive}(\tilde\kappa - i\tilde \zeta)(\vec H + i\vec E)\big] \cdot\vec\sigma
      + O(\tilde\kappa^2,\tilde \zeta^2)
\\
  &= m^2 - \big[\kappa\vec H + \zeta\vec E\big]\cdot\vec\sigma + O(\kappa^2,\zeta^2)
\end{aligned}
\end{equation}
where in the last line we redefined 
$e^{2i\alfive}(\tilde\kappa + i\tilde \zeta)=(\kappa + i \zeta)$, 
which is the same transformation as Eq.~(\ref{eqn:F23_mixing}).
Finally, the energy of the particle's interaction with the E\&M background is
\begin{equation}
\label{eqn:eshift_mink}
E_0 - m = -\frac{\kappa}{2m} \vec H\cdot\hat\Sigma 
  - \frac{\zeta}{2m} \vec E \cdot \hat\Sigma + O(\kappa^2,\zeta^2)
\end{equation}
where $\hat\Sigma$ is the unit vector of the particle's spin.
From the interaction energy, we conclude that indeed 
\begin{equation}
\kappa=F_2(0)\,, \quad \zeta=F_3(0)
\end{equation} are the
particle's magnetic and electric dipole moments. 
For a neutral particle such as the neutron, the form factor $F_2(0)$ is indeed the full magnetic 
moment.

Thus, we have shown that if particle's field is governed by the Dirac equation with a complex
mass~(\ref{eqn:dslash_eff_twisted}), electric and magnetic dipole moments have to be properly
adjusted $(\tilde{\kappa},\,\tilde{\zeta})\rightarrow(\kappa,\,\zeta)$.
This adjustment follows from redefining the field and the operator
\begin{equation}
\label{eqn:chiral_redef}
N \rightarrow N^\prime = e^{-i\alfive\gamma_5} N\,,
\quad
\tilde{\slashed{D}}_N\rightarrow \slashed{D}_N
  = e^{i\alfive\gamma_5} \tilde{\slashed{D}}_N e^{i\alfive\gamma_5}
\end{equation}
to remove the complex (chiral) phase from the mass,
where $\tilde{\slashed{D}}_N(\slashed{D}_N)$ contains $\tilde\kappa(\kappa)$ and $\tilde\zeta(\zeta)$.

\subsection{EDM energy shift in Euclidean space}\label{sec:edm_dirac_eshift_euc}
In order to verify our findings, in this paper we calculate the EDM of the neutron on a lattice
using two methods: from the energy shift in the background electric field method and using the
new formula for the $\CP$-odd form factor $F_3$. 
The electric field is introduced following Ref.~\cite{Detmold:2009dx} and preserving the 
(anti)periodic boundary condition in time.
Such electric field~\cite{Detmold:2009dx} is analytically continued to an imaginary value.
If the particle's electric dipole moment is finite and real-valued, the energy shift will be
imaginary, which might present a problem in the analysis of corresponding lattice correlators.
However, in our methodology, the $\CP$-odd interaction is always infinitesimal, and so are 
the electric dipole moments and the corresponding energy shifts, which are extracted from the 
first-order Taylor expansion of the nucleon correlation functions in the $\CP$-odd interaction.
In this paper, we study only neutral particles, because analysis of charged particle propagators
is more complicated~\cite{Detmold:2010ts}.

\begin{figure}[ht!]
\centering
\includegraphics[width=.3\textwidth]{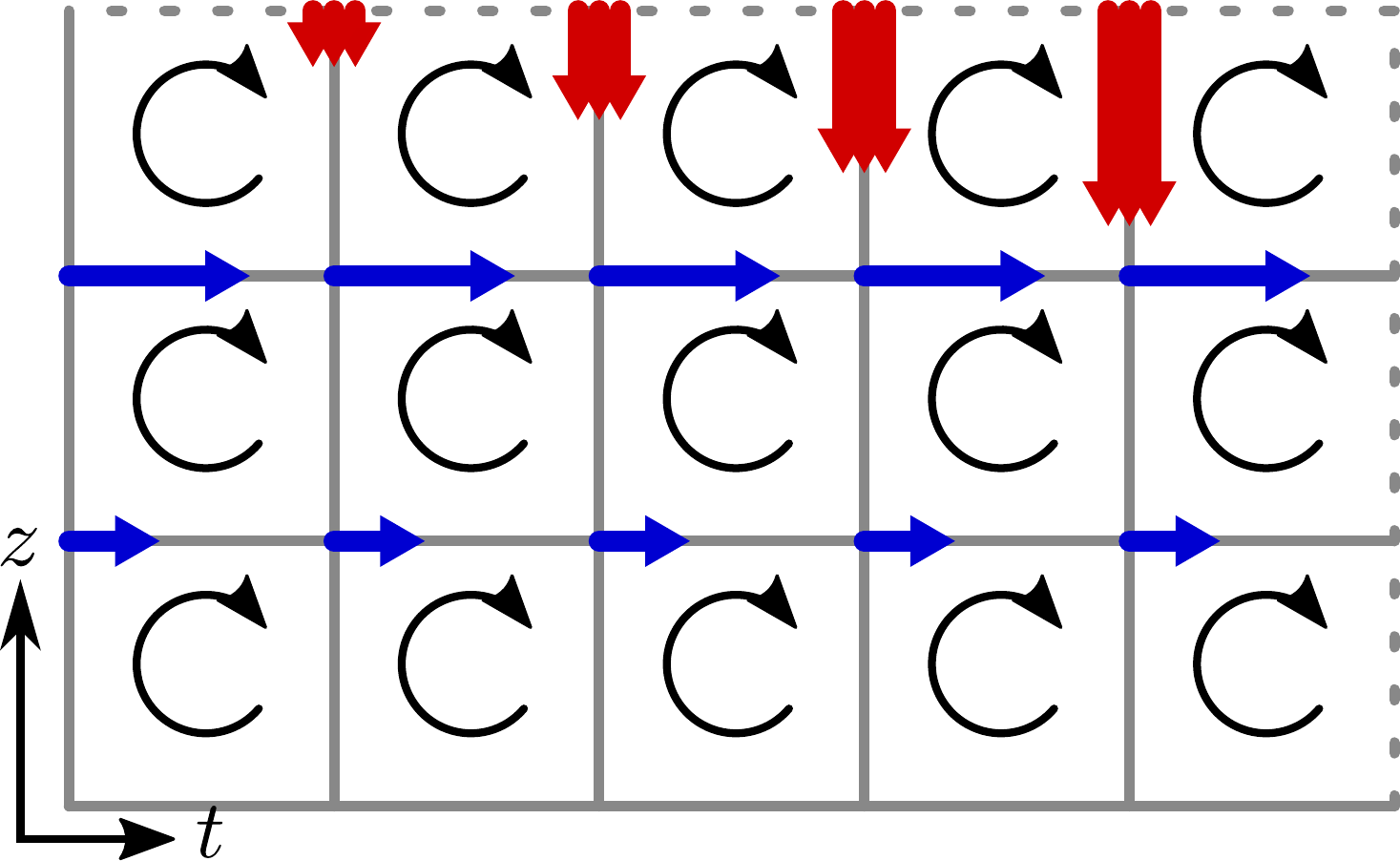}
\caption{Constant background electric field on a periodic lattice, following
  Ref.~\cite{Detmold:2009dx}.\label{fig:bgelectric}}
\end{figure}

In this section, we repeat the calculation of Sec.~\ref{sec:edm_dirac_eshift} for a neutral
particle on a Euclidean lattice, which has on-shell Euclidean 4-momentum 
$p_\EucConv=(\vec p,iE)$, with energy $E=\sqrt{E_0^2 + \vec p^2}$ up to discretization errors.
The energy at rest $E_0$ is modified from the mass $m$ due to electric and magnetic dipole
interactions.
To avoid any confusion, we imply no relation between the Minkowski $\vec E,\vec H$ and Euclid
$\vec\mcE,\vec\mcH$ electric and magnetic fields. 
Instead, we introduce \emph{ad hoc} uniform Abelian fields on a lattice 
(see Fig.~\ref{fig:bgelectric}) preserving boundary conditions in both space and
time~\cite{Detmold:2009dx} that probe the MDM and EDM: 
magnetic $\epsilon^{ijk}\mcH^k = (\partial_i \mcA_j - \partial_j \mcA_i) = n_{ij} \Phi_{ij}$
(no summation over $i,j$)
\begin{equation}
\label{eqn:euc_const_bgmag}
\left\{\begin{array}{ll}
  \mcA_{x,j} &= n_{ij}\,\Phi_{ij} \, x_i \\
  \mcA_{x,i}|_{x_i=L_i-1} &= -n_{ij}\,\Phi_{ij}\, L_i x_j 
\end{array}\right.
\end{equation}
and electric $\mcE^k = (\partial_k \mcA_4 - \partial_4 \mcA_k) = n_{k4} \Phi_{k4}$ 
\begin{equation}
\label{eqn:euc_const_bgel}
\left\{\begin{array}{ll}
  \mcA_{x,4} &= n_{k4}\,\Phi_{k4} \, x_k \\
  \mcA_{x,k}|_{x_k=L_k-1} &= -n_{k4}\,\Phi_{k4}\, L_k x_4
\end{array}\right.\,,
\end{equation}
where $\Phi_{\mu\nu} = \frac{6\pi}{L_\mu L_\nu}$ is the quantum of flux through a plaquette
$(\mu\nu)$ and $n_{\mu\nu}$ is the corresponding number of quanta.
The fractional quark charges $Q_u=2/3$, $Q_d=-1/3$ and periodic boundary conditions require 
that the flux through the edge of the lattice is $L_\mu L_\nu \cdot \Phi_{\mu\nu} = 3\cdot2\pi$.
From potentials~(\ref{eqn:euc_const_bgmag},\ref{eqn:euc_const_bgel}), the Euclidean field strength 
tensor $\mcF_{\mu\nu} = \partial_\mu \mcA_\nu - \partial_\nu \mcA_\mu$
\begin{equation}
\label{eqn:fmunu_euc_lat}
\mcF_{\mu\nu} 
= \left(\begin{array}{r|rrrr}
  & 1 & 2 & 3 & 4\\
\hline
    1 &        0  &   \mcH^3  &  -\mcH^2  & \mcE^1 \\
    2 &  -\mcH^3  &        0  &   \mcH^1  & \mcE^2 \\
    3 &   \mcH^2  &  -\mcH^1  &        0  & \mcE^3 \\
    4 &  -\mcE^1  &  -\mcE^2  &  -\mcE^3  &      0 
\end{array}\right)
\end{equation}
with
$\vec\mcH=(n_{23}\Phi_{23}, n_{31}\Phi_{31}, n_{12}\Phi_{12})$ and
$\vec\mcE=(n_{14}\Phi_{14}, n_{24}\Phi_{24}, n_{34}\Phi_{34})$.

We start from the effective EDM and MDM coupling in the nucleon-current vertex.
The Dirac operator for the nucleon on a lattice is 
$\slashed{D}+m=\gamma^\mu(\partial_\mu + i Q \mcA_\mu) + m$, which we extend to include the
point-like effective interactions from Eq.~(\ref{eqn:nuc_cur_vertex_euc})
\begin{equation}
\label{eqn:dirac_op_euc}
\big[i\slashed{p} +m +iQ\slashed{\mcA}_q 
  -\big(\frac12 \sigma^{\mu\nu}\mcF_{\mu\nu}\big) \, \frac{\kappa +i\zeta\gamma_5}{2m}
  \big]_\EucConv
\end{equation}
with $\kappa=F_2(0)$ and $\zeta=F_3(0)$.
We use Euclidean matrices $\gamma^\mu$~(\ref{eqn:gamma_euc}) 
and $[\gamma_{5}]_\EucConv=(\gamma^1\gamma^2\gamma^3\gamma^4)_\EucConv$~(\ref{eqn:gamma5_euc})
\footnote{
  The results are manifestly independent from the basis of $\gamma$-matrices used, 
  if the relation between $\gamma_5$ and $\gamma^\mu$ is kept unchanged.}
and the plain-wave fields $\psi_p(x)$ and $\mcA_{q,\mu}(x)$ depending on 
the Euclidean 4-momenta $p$, $q$ as
\begin{equation}
\begin{gathered}
\label{eqn:momenta_euc}
\psi_p(x)\sim e^{i p x}\,,
\,\, \partial_\mu\psi_p(x) \leftrightarrow i p_\mu u_p\,;
\quad
\mcA_{q,\mu}(x) \sim e^{i(p^\prime-p) x} = e^{i q x}\,, 
\,\, \partial_\nu \mcA_{q,\mu}(x) \leftrightarrow i q_\nu \mcA_\mu\,.
\end{gathered}
\end{equation}
The mass $m$ in the above equation (\ref{eqn:dirac_op_euc}) is chosen real and positive, 
since any chiral phase factor may be removed with a field redefinition~(\ref{eqn:chiral_redef}), 
which at the same time rotates the dipole couplings $(\kappa,\,\zeta)$ into the physical magnetic 
and electric dipole moments, as has been shown in Sec.~\ref{sec:edm_dirac_eshift}.
After setting the charge $Q=0$ and the momentum $\vec p=0$, we use
\begin{equation}
\sigma_\EucConv^{ij} 
  = \epsilon^{ijk}\left(\begin{array}{cc}-\sigma^k & \\ & -\sigma^k\end{array}\right)\,,
\quad
\sigma_\EucConv^{i4} 
  = \left(\begin{array}{cc} \sigma^i & \\ & -\sigma^i \end{array}\right)\,,
\quad
\end{equation}
and transform the operator~(\ref{eqn:dirac_op_euc}) into the block-diagonal form,
and find the condition for on-shell fermion energies
\begin{equation}
\label{eqn:dirac_op_block_euc}
\det\left(\begin{array}{cc} 
  \mcM_- & -E_0  \\  -E_0 &  \mcM_+ 
\end{array}\right)
=0 
\quad\Leftrightarrow\quad 
    \det(E_0^2 - \mcM_+ \mcM_-) = \det(E_0^2 - \mcM_- \mcM_+) = 0
\end{equation}
where
\begin{equation}
\mcM_\pm = m + \frac1{2m}(\kappa \mp i \zeta)\, (\vec\mcH \pm
\vec\mcE)\cdot\vec\sigma\,.
\end{equation}
The on-shell energies are then determined by the eigenvalues of the spin-dependent operator
\begin{gather}
\label{eqn:dirac_euc_eval_matr}
\mcM_{\mp}\mcM_{\pm}
  =m^2 +\kappa \vec\sigma \cdot \vec\mcH  - \zeta \vec\sigma \cdot i \vec \mcE 
  +O(k^2, \zeta^2)\,,
\\
\label{eqn:eshift_euc}
E_0-m=\frac{\kappa}{2m}\,\vec\mcH\cdot\hat\Sigma - \frac{\zeta}{2m}\,i\vec\mcE\cdot\hat\Sigma
  +O(k^2, \zeta^2)\,,
\end{gather}
where $\hat\Sigma$ is the direction of the particle's spin.
Note that the electric field enters Eq.~(\ref{eqn:eshift_euc}) as $i\vec\mcE$,
with an imaginary factor emphasizing that its value has been analytically continued 
to the imaginary axis, and the corresponding energy shift is purely imaginary.
Equation~(\ref{eqn:eshift_euc}) provides a prescription for extracting the EDM and MDM from
energy shifts of a neutral particle on a lattice in uniform background fields.

\section{cEDM-induced EDM and EDFF on a lattice}

In our initial calculation of cEDM-induced nucleon EDMs,
we use two lattice ensembles with Iwasaki gauge action and $N_f=2+1$ dynamical domain wall
fermions: $16^3\times32$ with $m_\pi\approx420\text{ MeV}$~\cite{Blum:2011pu} 
and $24^3\times64$ with $m_\pi\approx340\text{ MeV}$~\cite{Aoki:2010dy,Blum:2014tka}.
The ensemble parameters are summarized in Tab.~\ref{tab:ens}.
We use identical ensembles, statistics, and spatial sampling per gauge configuration 
in both calculation methods discussed in further sections.

\begin{table}[ht!]
\caption{
  Lattice ensembles on which the simulations were performed. 
  Both ensembles use Iwasaki gauge action and $N_f=2+1$ domain wall fermions.
  The statistics are shown for ``sloppy'' (low-precision) samples.
  The nucleon masses were extracted using 2-state fits.
  For the background electric field method, we quote the quantum of the electric field 
  $\mcE_0 = \frac{6\pi}{a^2L_t L_x}$.
  \label{tab:ens}}
\begin{tabular}{l|cccccc|rrrrr}
\hline\hline
$L_x^3\times L_t\times L_5$ 
  & $a\text{ [fm]}$ & $am_l$ & $am_s$ &  $m_\pi\,[\mathrm{MeV}]$ &  
  $m_N\,[\mathrm{GeV}]$ &   $\mcE_0\,[\mathrm{GeV}^2]$ &
  conf & stat & $N_{ev}$ & $N_{ev}^{\mcE=1,2}$ & $N_{CG}$ \\
\hline
$16^3\times32\times16$ &  
  0.114(2) & 0.01 & 0.032 & 422(7) & 1.250(28) & 0.110 & 
  500 & 16000 & 200 & 150 & 100 \\
$24^3\times64\times16$ & 
  0.1105(6) & 0.005 & 0.04 & 340(2) & 1.178(10) & 0.0388 &
  100 & 3200 & 200 & 200 & 200 \\
\hline\hline
\end{tabular}
\end{table}

We use \emph{all-mode-averaging}~\cite{Shintani:2014vja} framework to optimize sampling, 
in which we  approximate quark propagators with truncated-CG solutions 
to a M\"obius operator~\cite{Brower:2005qw}.
We use the M\"obius operator with short 5th dimension $L_{5s}$ and complex $s$-dependent
coefficients $b_s + c_s = \omega_s^{-1}$ (later referred to as ``zMobius'') that approximates 
the same 4d effective operator as the Shamir operator with the full $L_{5f}=32$ (DSDR) 
or $L_{5f}=16$ (Iwasaki).
The approximation is based on the domain wall-overlap equivalence
\begin{gather}
[\Dslash^\text{DWF}]_{4d} = \frac{1+m_q}2 - \frac{1-m_q}2 \gamma_5
\epsilon_{L_5}(H_T)\,,
\quad
H_T = \gamma_5\frac{\Dslash_W}{2+\Dslash_W}\,,\\
\epsilon^\text{M\"obius}_{L_{5s}}(x) 
  = \frac{\prod_s^{L_{5s}}(1+\omega_s^{-1}x) - \prod_s^{L_{5s}}(1-\omega_s^{-1}x)}
         {\prod_s^{L_{5s}}(1+\omega_s^{-1}x) + \prod_s^{L_{5s}}(1-\omega_s^{-1}x)}
  \approx \epsilon^\text{Shamir}_{L_{5f}}(x) 
  \,.
\end{gather}
where the coefficients $\omega_s$ are chosen so that the function 
$\epsilon^\text{M\"obius}_{L_{5s}}(x)$ is the \emph{minmax} approximation to the 
$\epsilon^\text{Shamir}_{L_{5f}}(x)$.
We find that $L_{5s}=10$ is enough for efficient 4d operator approximation.
Shortened 5th dimension reduces the CPU and memory requirements:
for example, $L_{5f}=16$ is reduced to $L_{5s}=10$ saving $38\%$ of the cost.
We deflate the low-lying eigenmodes of the internal even-odd preconditioned operator,
to make the truncated-CG AMA efficient. 
The numbers of deflation eigenvectors $N_{ev}$ and truncated CG iterations $N_{CG}$ are given in
Tab.~\ref{tab:ens}.
We compute 32 sloppy samples per configuration. 
To correct any potential bias due to the approximate $\Dslash$ operator and the truncated CG
inversion, in addition we compute one exact sample per configuration using the Shamir operator.
The latter is computed iteratively by refining the solution of the ``zMobius'' 
to approach the solution of the Shamir operator, again taking advantage of the short $L_{5s}$
and deflation.

\subsection{Parity-even and -odd nucleon correlators}\label{sec:evenodd_corr}

\begin{figure}[ht!]
\centering
\includegraphics[width=.25\textwidth]{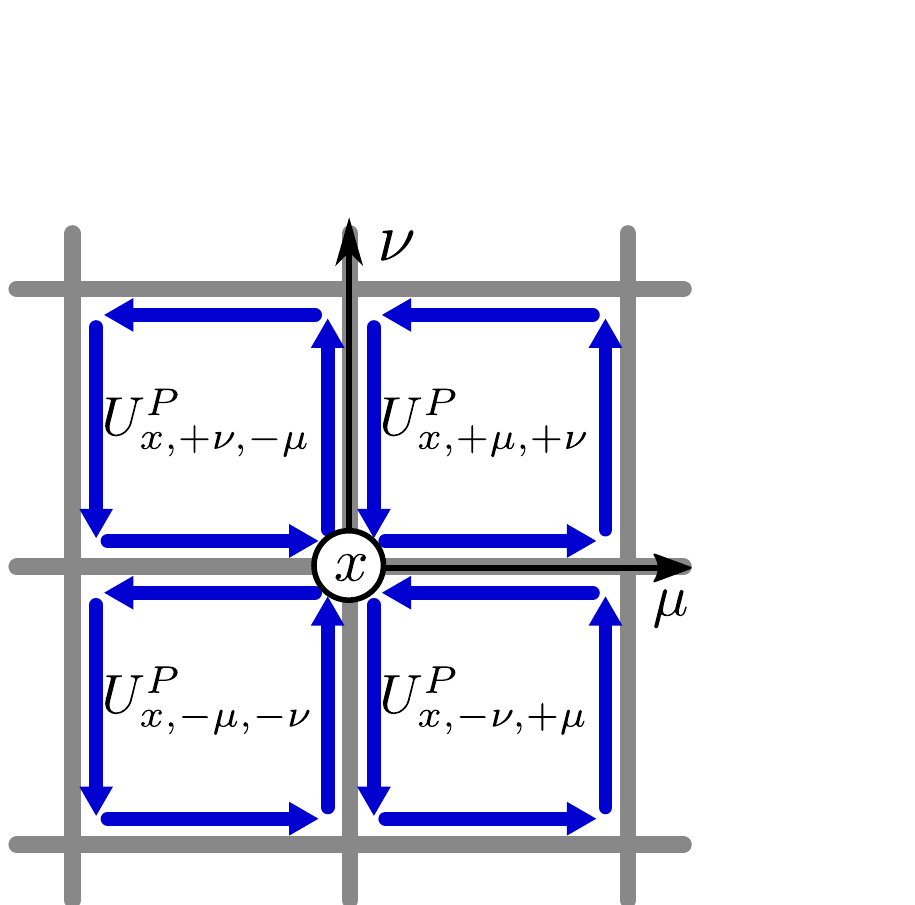}
\caption{``Clover'' definition of the gauge field strength tensor on a lattice.
  \label{fig:link-clover}}
\end{figure}
\begin{figure}[ht!]
\centering
\includegraphics[width=.7\textwidth]{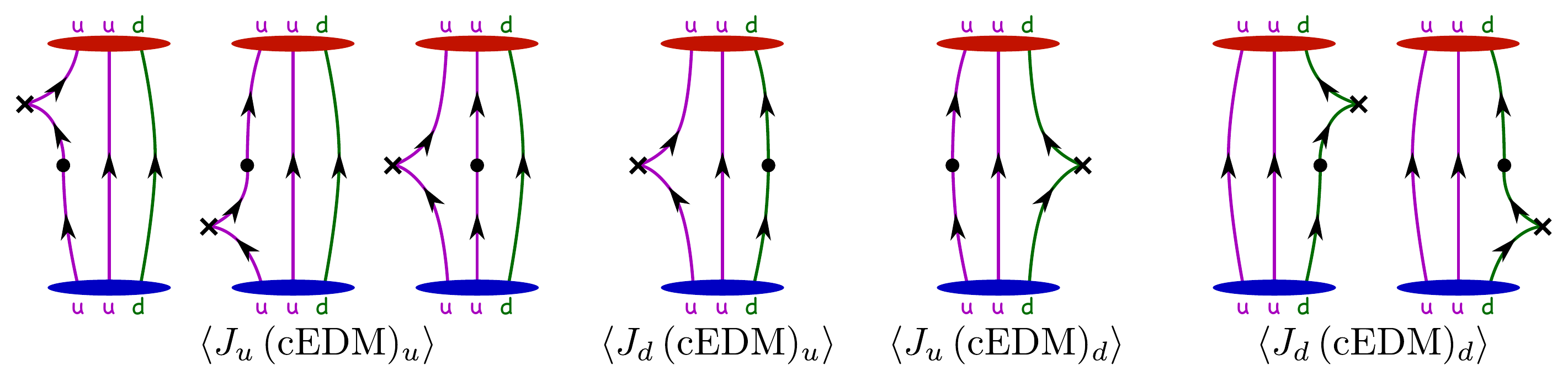}
\caption{Quark-connected contractions of nucleon, quark current, and cEDM operators.
  \label{fig:cedm_contract_conn}}
\end{figure}
\begin{figure}[ht!]
\centering
\includegraphics[width=.5\textwidth]{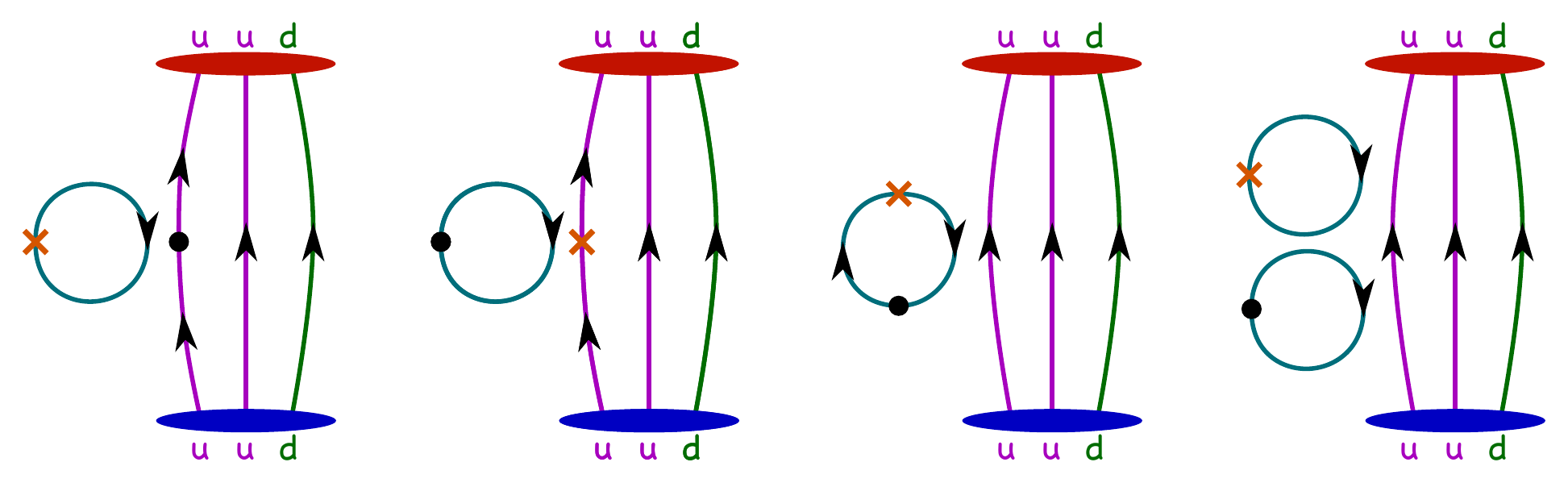}
\caption{Quark-disconnected contractions of nucleon, quark current, and cEDM operators.
  \label{fig:cedm_contract_disc}}
\end{figure}

The EDFF $F_3$ is a parity-odd quantity induced by $\CPviol$ interactions.
To compute the effect of $\CP$-odd interactions, we modify the lattice action 
\begin{equation}
S\to S + i \delta^\CPbar S = S + i\sum_{i,x} c_i [\mcO^{\CPbar}_i]_x
\end{equation}
where $c_i$ are the $\CP$-odd couplings such as the QCD $\theta$-angle, quark (chromo-)EDMs, etc.
We Taylor-expand QCD$+\CPviol$ vacuum averages in the couplings $c_i$.
For example, for the three-point function, we get\footnote{
  In this section, all conventions for correlators, form factors, and momenta are Euclidean.}
\begin{gather}
\label{eqn:corr_cpviol}
\langle N\,[\bar q \gamma^\mu q]\, \bar N \rangle_{\CPviol}
  = \frac1Z\int\,\mcD U\,\mcD\bar\psi\mcD\psi e^{-S - i\delta^\CPbar S}
      \, N\,[\bar q \gamma^\mu q]\, \bar N 
  = C_{NJ\bar N} - i\sum_i c_i \,\delta^\CPbar_i C_{NJ\bar N}
    + O(c_\psi^2)\,,
\end{gather}
where $C_{\ldots}$ and $\delta^\CPbar C_{\ldots}$ stand for the $\CP$-even and $\CP$-odd correlators.
Similarly, we also analyze the effect of $\CPviol$ interaction on the nucleon-only correlators.
In total, we calculate the following two- and three-point $\CP$-even correlators as well as 
three- and four-point $\CP$-odd correlators,
\begin{align*}
C_{N\bar N} &= \langle N \bar N\rangle\,,
\quad 
&\delta^\CPbar_i C_{N\bar N} &= \langle N \bar N\,\cdot\sum_x [\mcO^\CPbar_i]_x \rangle\,,
\\
C_{NJ\bar N} &= \langle N\,[\bar q \gamma^\mu q]\, \bar N\rangle\,,
\quad
&\delta_i^\CPbar C_{NJ\bar N} 
 &= \langle N\,[\bar q \gamma^\mu q]\, \bar N \cdot \sum_x [\mcO^\CPbar_i]_x \rangle\,,
\end{align*}
where $\langle\cdot\rangle$ stand for vacuum averages computed with $\CP$-even QCD action $S$.
In Sec.~\ref{sec:bgem}, we also modify the action $S$ to include the uniform background electric
field as the probe of the electric dipole moment.
In this work, we study only the quark chromo-EDM as the source of $\CP$ violation,
\begin{equation}
\label{eqn:qcedm_lat}
\mcO_{\psi G}^{\CPbar}  
  = \frac12\bar\psi [G_{\mu\nu}]^\text{clov}\sigma^{\mu\nu}\gamma_5\psi
  = \frac12\bar\psi (g_S G^{a,\text{cont}}_{\mu\nu} T^a)\sigma^{\mu\nu}\gamma_5\psi\,,
\end{equation}
where $G^{a,\text{cont}}$ is the continuum color field strength tensor and 
the ``clover'' $[G_{\mu\nu}]^\text{clov}$ gauge field strength tensor 
on a lattice is (see Fig.~\ref{fig:link-clover})
\begin{equation}
\label{eqn:Gmunu_clover}
\begin{aligned}
\big[G_{\mu\nu}\big]^\text{clov} &= \frac1{8i} \big[
  ( U^P_{x,+\hat\mu,+\hat\nu} + U^P_{x,+\hat\nu,-\hat\mu} 
  + U^P_{x,-\hat\mu,-\hat\nu} + U^P_{x,-\hat\nu,+\hat\mu}) 
  - \mathrm{h.c.}\big]
\end{aligned}
\end{equation}
Insertions of the quark-bilinear cEDM density~(\ref{eqn:qcedm_lat}) can generate both connected 
and disconnected contractions, similarly to the quark current.
In this work, we calculate only the fully connected contributions to these correlation functions 
shown in Fig.~\ref{fig:cedm_contract_conn}. 
The disconnected contributions (see Fig.~\ref{fig:cedm_contract_disc}) are typically much more
challenging to calculate, and we will address them in future work. 
Neglecting the disconnected diagrams will not affect the comparison of the form factor and the
energy shift methods, because they are omitted in both calculations.

\begin{figure}[ht!]
\centering
\includegraphics[width=.4\textwidth]{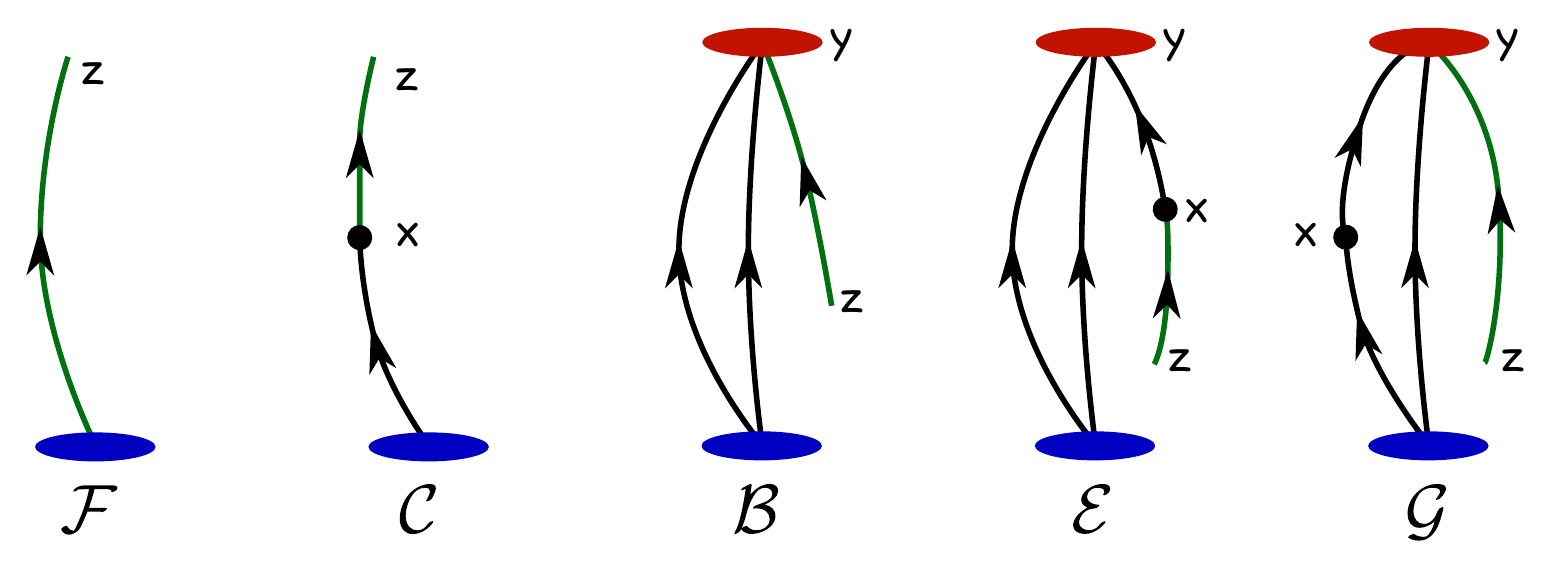}
\caption{Propagators required for computing quark-connected contractions of nucleon, quark current, 
  and cEDM operators.
  \label{fig:cedm_contract_props}}
\end{figure}

To compute the connected diagrams, we insert the quark-bilinear cEDM density~(\ref{eqn:qcedm_lat})
once in every $\psi$-quark line of $C_{N J\bar N}$ diagrams, generating the four-point functions 
shown in Fig.\ref{fig:cedm_contract_conn}.
We evaluate all the connected three- and four-point contractions using the forward 
and the set of sequential propagators shown in Fig.~\ref{fig:cedm_contract_props}.
In addition to the usual one forward $\mcF$ and two backward (sink-sequential) $\mcB$ propagators,
we compute one cEDM-sequential $\mcC$ and four doubly-sequential (\{cEDM, sink\}-sequential)
$(\mcE+\mcG)$ propagators per sample.
For every additional value of the source-sink separation $t_\text{sep}$ and sink momentum $\vec
p^\prime$, additional backward $\mcB$ and doubly-sequential $(\mcE+\mcG)$ propagators must be
computed, i.e.
\begin{equation*}
N_\mcF = N_\mcC = 1\,, 
\quad N_\mcB=N_q N_\text{sep} N_\text{mom}\,,
\quad N_{\mcE+\mcG} = N_q N_\psi N_\text{sep} N_\text{mom}
\end{equation*}
where $N_q$ is the number of separate flavors in the quark current and $N_\psi$ is the number of
separate flavors in the $\CPviol$ operator.
The connected $\CP$-even two- and three-point correlators do not require any additional inversions.
In this scheme, we perform only the minimal number of inversions required for computing 
all the diagrams for the neutron and proton EDM induced by 
a \emph{connected} flavor-dependent quark-bilinear $\CPviol$ interaction 
with the two degenerate flavors $u$ and $d$.
Compared to Ref.~\cite{Bhattacharya:2016oqm}, in which a finite small $O(\epsilon)$
$\CP$-odd perturbation term is added to the quark action that results in modified quark
propagators
\begin{equation}
\Dslash_m^{-1}\rightarrow(\Dslash_m + i\epsilon\sigma^{\mu\nu}\tilde{G}_{\mu\nu})^{-1}\,,
\end{equation}
our four-point contractions correspond to directly computing the first derivative 
$(\partial C_{2,3}^\epsilon / \partial\epsilon)_{\epsilon=0}$, thus avoiding 
any higher-order dependence on $\epsilon$ and obviating the $\epsilon$-extrapolation.
As a cross-check, we have verified our contraction code on a small test lattice by replacing 
propagators $\Dslash_m^{-1}\eta$ with
\begin{equation}
\Dslash_m^{-1}\eta \rightarrow 
  [\Dslash_m^{-1} - \Dslash_m^{-1} (i\epsilon\Gamma) \Dslash_m^{-1}]\eta 
\end{equation}
to approximate $[\Dslash_m + i\epsilon\Gamma]^{-1}\eta$,  
where $\Gamma=\frac12 G_{\mu\nu}\sigma^{\mu\nu}\gamma_5$.
Using these ``$\CP$-perturbed'' propagators, each of which needed two inversions, 
we have computed the nucleon $C^\epsilon_{N\bar N}$ and nucleon-current 
$C^\epsilon_{NJ\bar N}$ correlators, and compared their finite-difference 
$\epsilon$-derivatives to $\delta C_{N\bar N}$ and $\delta C_{NJ\bar N}$.

\begin{figure}[ht!]
\centering
\includegraphics[width=0.49\textwidth]{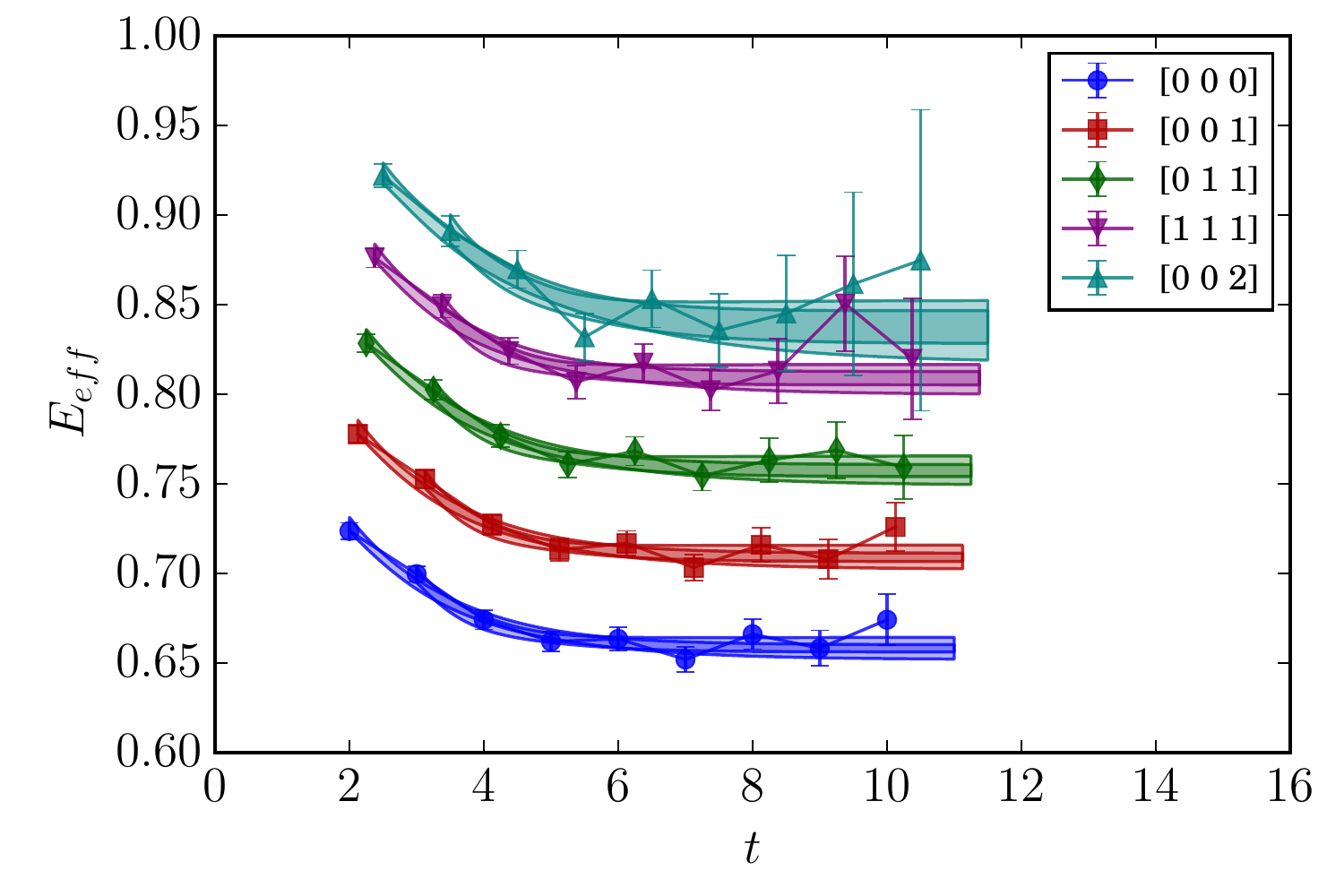}~
\includegraphics[width=0.49\textwidth]{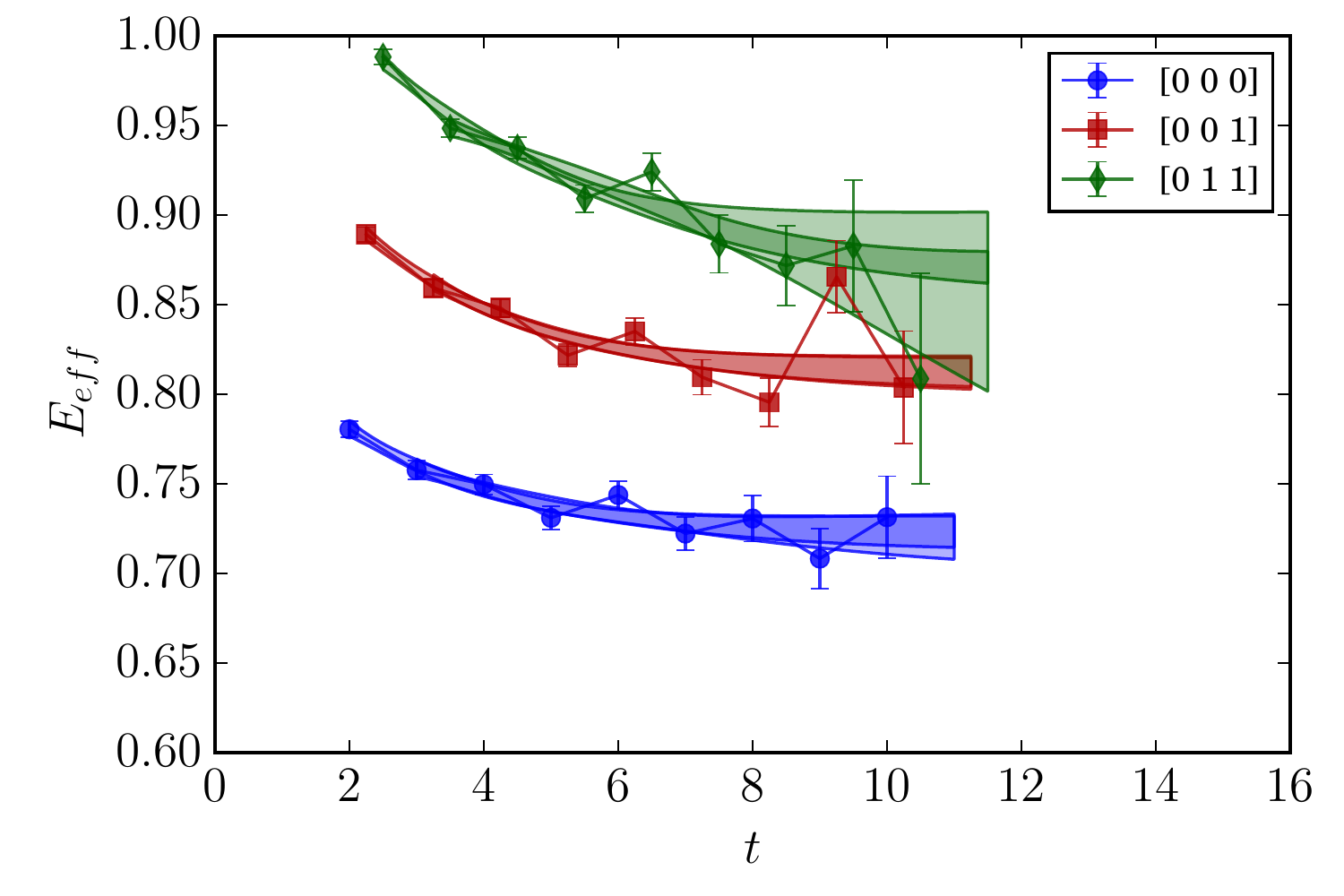}\\
\caption{Effective energy plots from the $24^3\times64$ (left) and $16^3\times32$ (right) lattices, 
  together with two-state fits.
  \label{fig:Eeff}}
\end{figure}

We use only one value of the sink momentum $\vec p^\prime=0$.
We compute nucleon-current three- and four-point correlators with two source-sink separation 
values $t_\text{sep}=\{8,10\}a=\{0.91,1.15\}\text{ fm}$ for the $16^3\times32$ ensemble, 
and three $t_\text{sep}=\{8,10,12\}a=\{0.88,1.11,1.33\}\text{ fm}$ for the $24^3\times64$ ensemble.
For the $24^3\times64$ lattice, we use the Gaussian-smeared sources with APE-smeared gauge links
using parameters optimized for overlap with the ground state~\cite{Syritsyn:2009mx},
while for the $16^3\times32$ ensemble we used the smearing parameters from
Ref.~\cite{Shintani:2015vsx}.
The effective nucleon mass plots for the ensembles are shown in Fig.~\ref{fig:Eeff}.
Correlators $C_{NJ\bar N}$ and $\delta^{\CPbar}C_{NJ\bar N}$ are computed with
the polarization projector
\begin{equation} 
T^+_{S_z+}=\frac{1+\gamma_4}2(1+\Sigma_3)=\frac12(1+\gamma_4)(1-i\gamma_1\gamma_2)\,,
\end{equation},
while correlators $C_{N\bar N}$ and $\delta^\CPbar C_{N\bar N}$ are computed with all 16
polarizations and saved to be used later for disconnected contractions.
We reduce the cost of computing backward propagators with the widely-used ``coherent''
trick combining 2 backward sources from samples separated by $L_t/2$ into one inversion.
Combining 4 samples resulted in a large increase in the statistical uncertainty negating the
cost-saving advantages.

\subsection{Nucleon form factors} \label{sec:lat_ff}

\begin{figure}[ht!]
\centering
\includegraphics[width=0.49\textwidth]{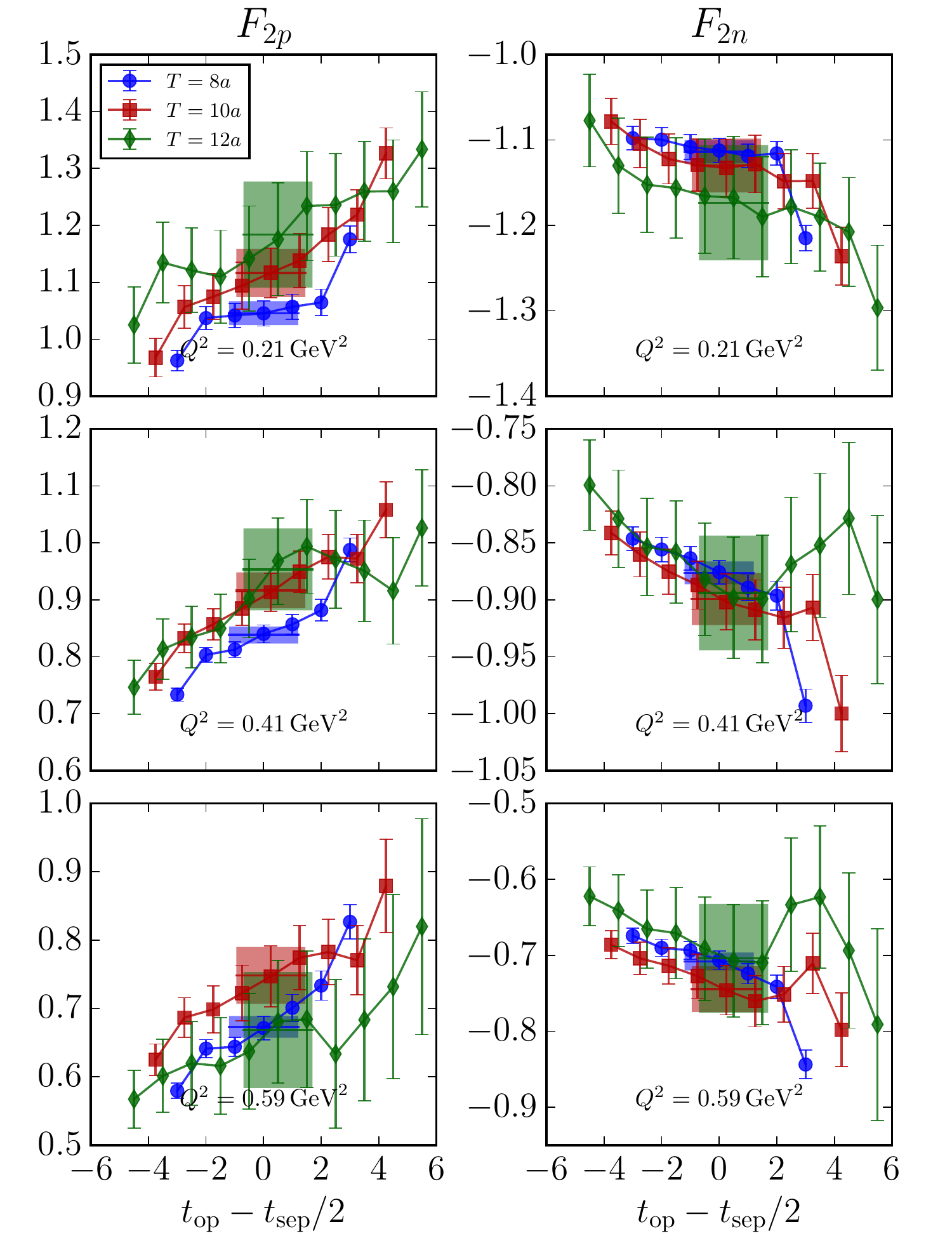}~
\includegraphics[width=0.49\textwidth]{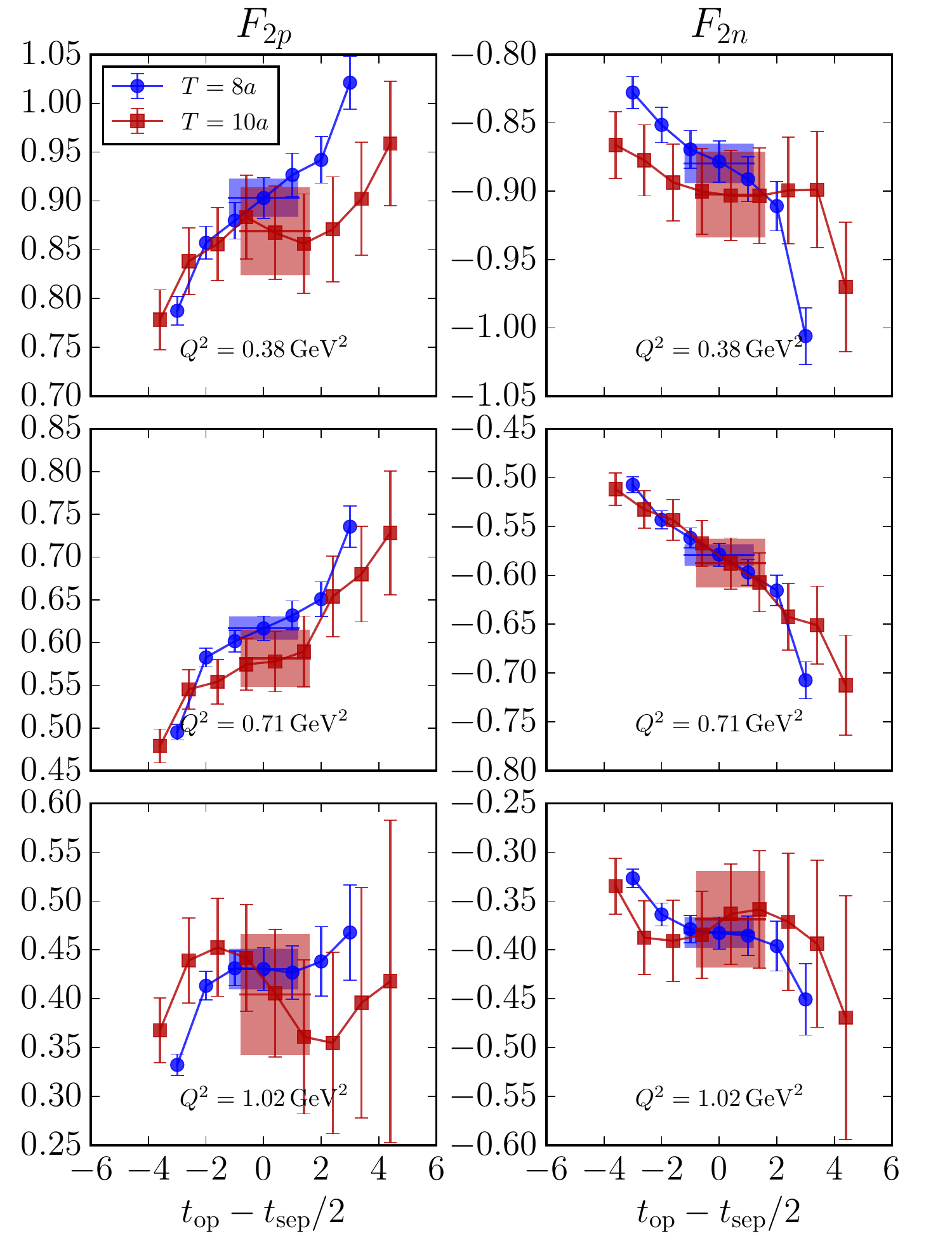}\\
\caption{
  Plateau plots for the neutron and proton Pauli form factors, 
  the three smallest $Q^2>0$ points.
  Results for the $24^3\times64$ (left) and $16^3\times32$ (right) lattices. 
  \label{fig:neut_cpviol_f2_pltx}}
\end{figure}

\begin{figure}[ht!]
\centering
\includegraphics[width=0.49\textwidth]{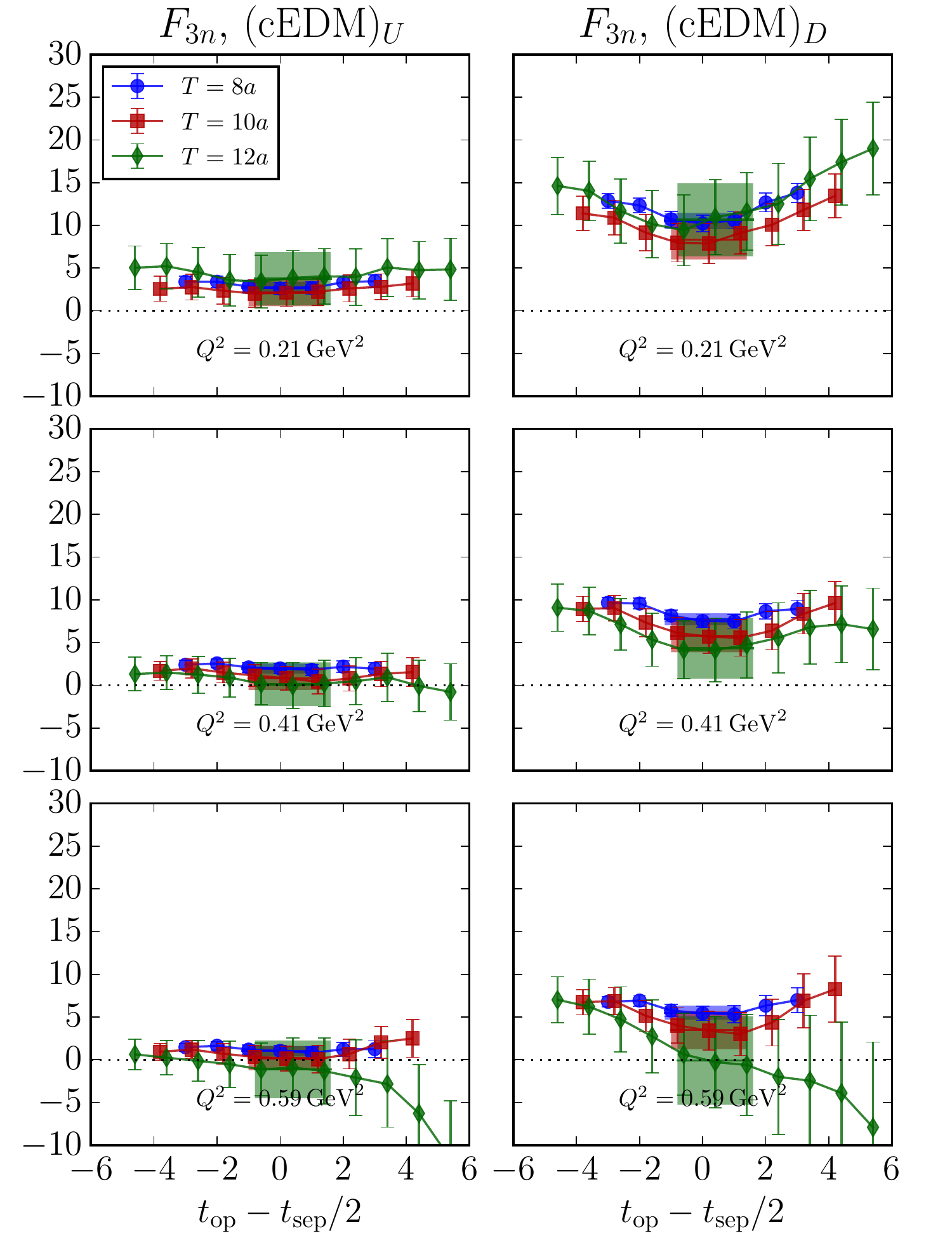}~
\includegraphics[width=0.49\textwidth]{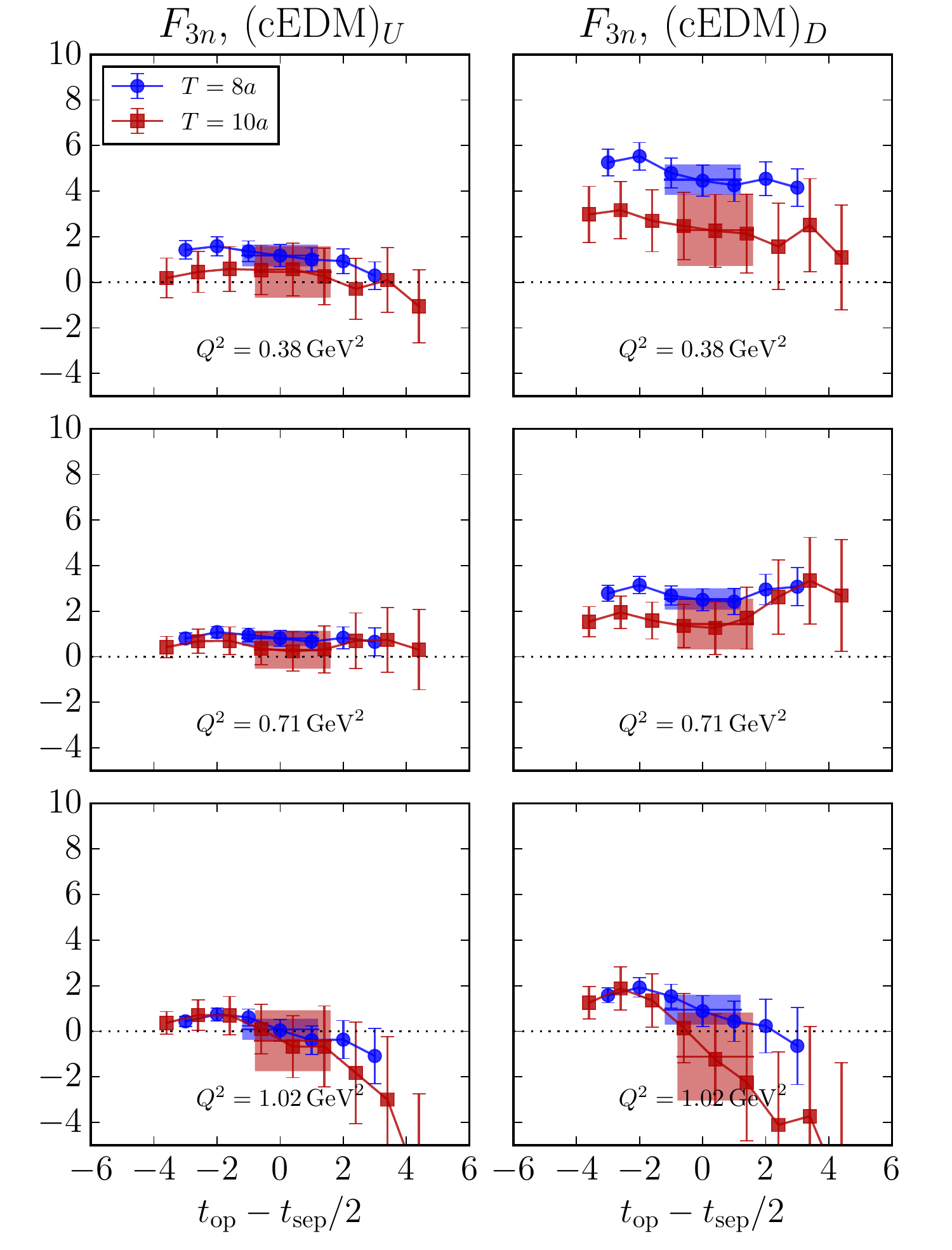}\\
\caption{
  Plateau plots for the neutron EDFF form factor, the 3 lowest $Q^2>0$ points.
  Results for the $24^3\times64$ (left) and $16^3\times32$ (right) lattices. 
  \label{fig:neut_cpviol_f3_pltx}}
\end{figure}

Following the discussion in Sec.\ref{sec:edm_ff_mink}, we use the form factor decomposition that
is different from the previous works~\cite{Shintani:2005xg,Berruto:2005hg,Aoki:2008gv,Guo:2015tla,
Shindler:2015aqa,Alexandrou:2015spa,Shintani:2015vsx},
\begin{equation}
\langle N_{p^\prime} | \bar{q}\gamma^\mu q | N_p\rangle
  = \bar{u}_{p^\prime} \big[ F_1(Q^2) \gamma^\mu
      + F_2(Q^2) \frac{\sigma^{\mu\nu} q_\nu}{2m_N}
      + F_3(Q^2) \frac{i\gamma_5\sigma^{\mu\nu} q_\nu}{2m_N}
      \big] u_p\,.
\end{equation}
where the spinors $u_p$,$\bar u_{p^\prime}$ have positive parity.
Details of evaluating kinematic coefficients for form factors $F_{1,2,3}$
are given in Appendix~\ref{sec:app_kincoeff}.
We use the standard plateau method to evaluate both $\CP$-even and $\CP$-odd matrix elements of
the nucleon 
\begin{equation}
\label{eqn:c3_c2_ratio}
[\delta^\CPbar] \mcR_{NJ\bar N} (t_{sep},t_{op})
  = \frac
         {[\delta^\CPbar] C_{NJ\bar N}(t_{sep},t_{op})}
         {c_2^\prime(t_{sep})}
    \sqrt{\frac{c_2^\prime(t_{sep})}{c_2(t_{sep})} 
          \frac{c_2^\prime(t_{op})}{c_2(t_{op})} 
          \frac{c_2(t_{sep}-t_{op})}{c_2^\prime(t_{sep}-t_{op})} }
\end{equation}
where the two-point functions are projected with the positive-parity polarization matrix
$T^+=\frac12(1+\gamma_4)$,
\begin{equation}
\label{eqn:c2pt_proj_unpol}
c^{(\prime)}_2(t) = \Tr\big[ T^+\cdot C_{N\bar N}(\vec p^{(\prime)}, t) \big] \,.
\end{equation}
The three central points on the ratio plateaus are taken as the estimate of the ground state 
matrix elements.
This is a crude estimate and improved analysis of excited states is necessary for better
control of systematic uncertainties.
However, we find that our results change insignificantly with increasing source-sink separation 
(see Figs.~\ref{fig:neut_cpviol_f2_pltx}, \ref{fig:neut_cpviol_f3_pltx}), therefore we conclude
that excited state effects cannot influence the main conclusions of the paper.

\begin{figure}[ht!]
\centering
\includegraphics[width=0.49\textwidth]{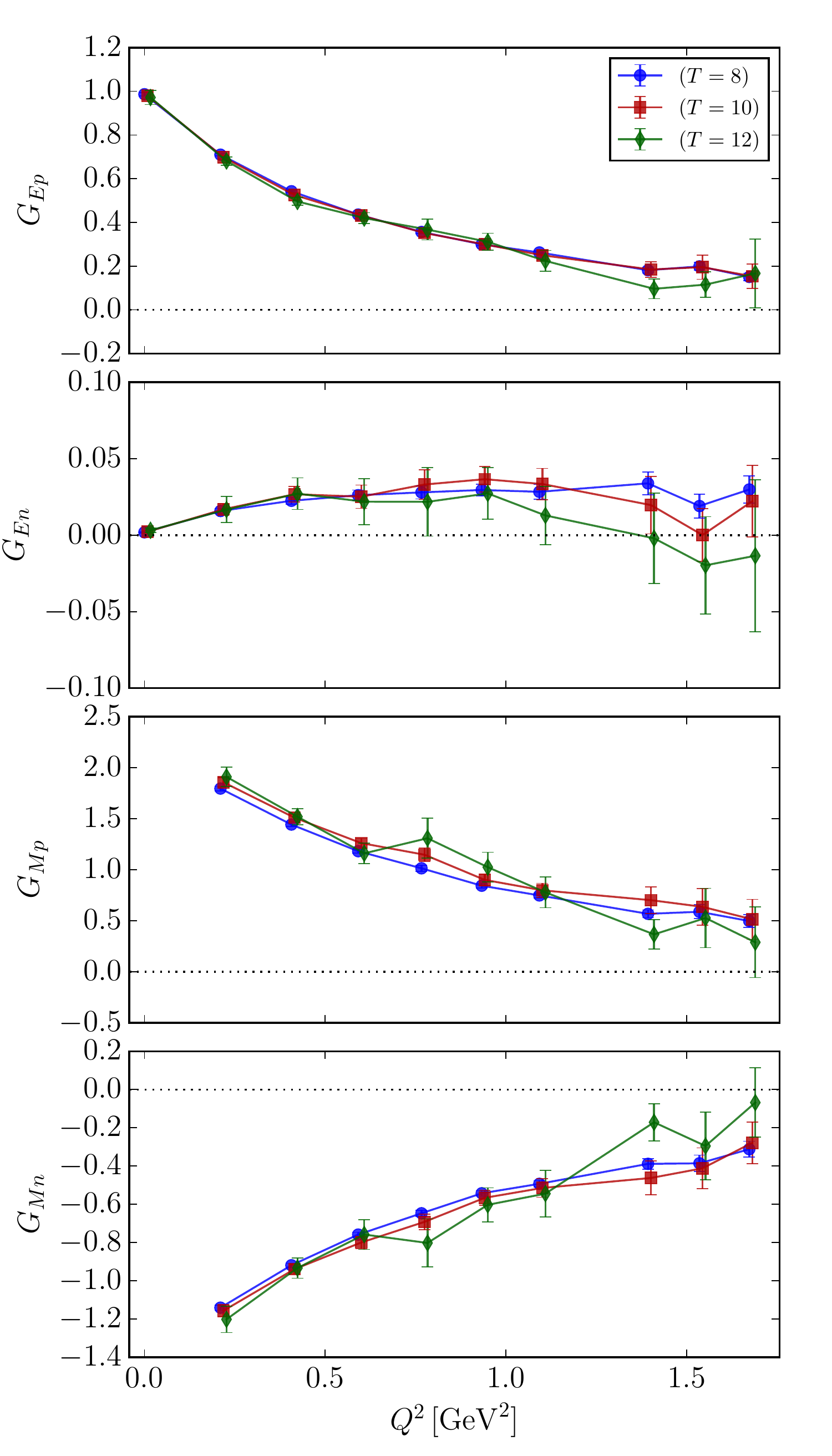}~
\includegraphics[width=0.49\textwidth]{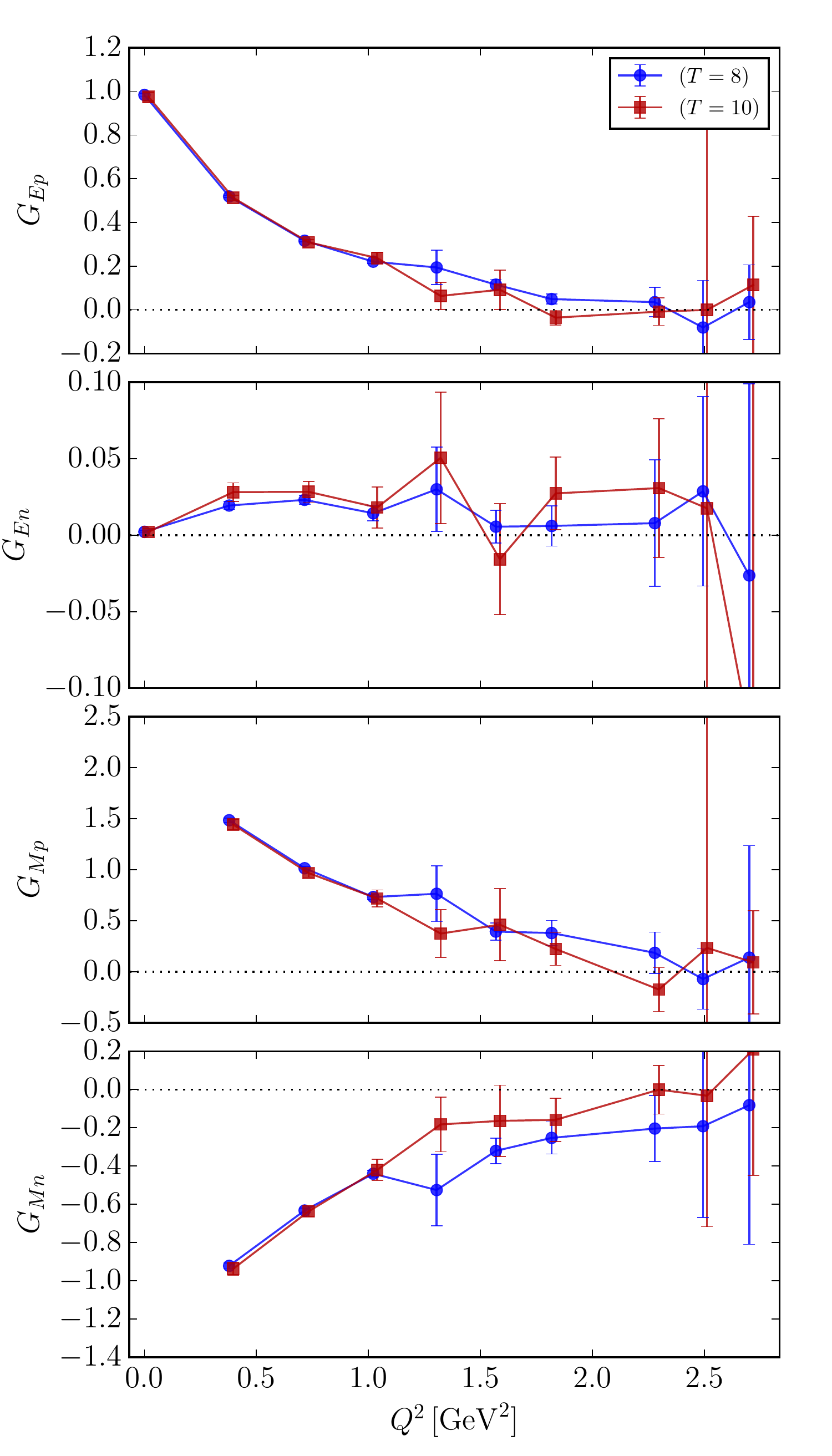}\\
\caption{Nucleon vector current form factors from the $24^3\times64$ (left) and $16^3\times32$
  (right) lattices.
  Disconnected contractions are not included.
  \label{fig:nucl_vec_ff}}
\end{figure}

We calculate the Dirac and Pauli form factors $F_{1,2}$ using a correlated $\chi^2$ fit to the
matrix elements of the quark vector current (``overdetermined analysis''). 
The system of equations for form factors is reduced by combining equivalent equations to
reduce the system dimension and make estimation of the covariance matrix more stable 
(see, e.g., Ref.~\cite{Syritsyn:2009mx} for details).
The quark current operator is renormalized using renormalization constants 
$Z_V=0.71408$ for $24^3\times64$~\cite{Blum:2014tka} and 
the chiral-limit value $Z_V=Z_A=0.7162$ for $16^3\times32$~\cite{Allton:2007hx} ensembles.
We show the momentum dependence of the resulting Sachs electric and magnetic form factors 
\begin{equation}
G_E(Q^2) = F_1(Q^2) - \frac{Q^2}{4m_N^2} F_2(Q^2)\,,
\quad
G_M(Q^2) = F_1(Q^2) + F_2(Q^2)\,,
\end{equation}
for the proton and the neutron (connected-only) for both ensembles in Fig.~\ref{fig:nucl_vec_ff}.
Our data for form factors $G_{E,M}$ show no significant systematic variation with 
increasing the source-sink separation $t_\text{sep}$.

\begin{figure}[ht!]
\centering
\includegraphics[width=0.49\textwidth]{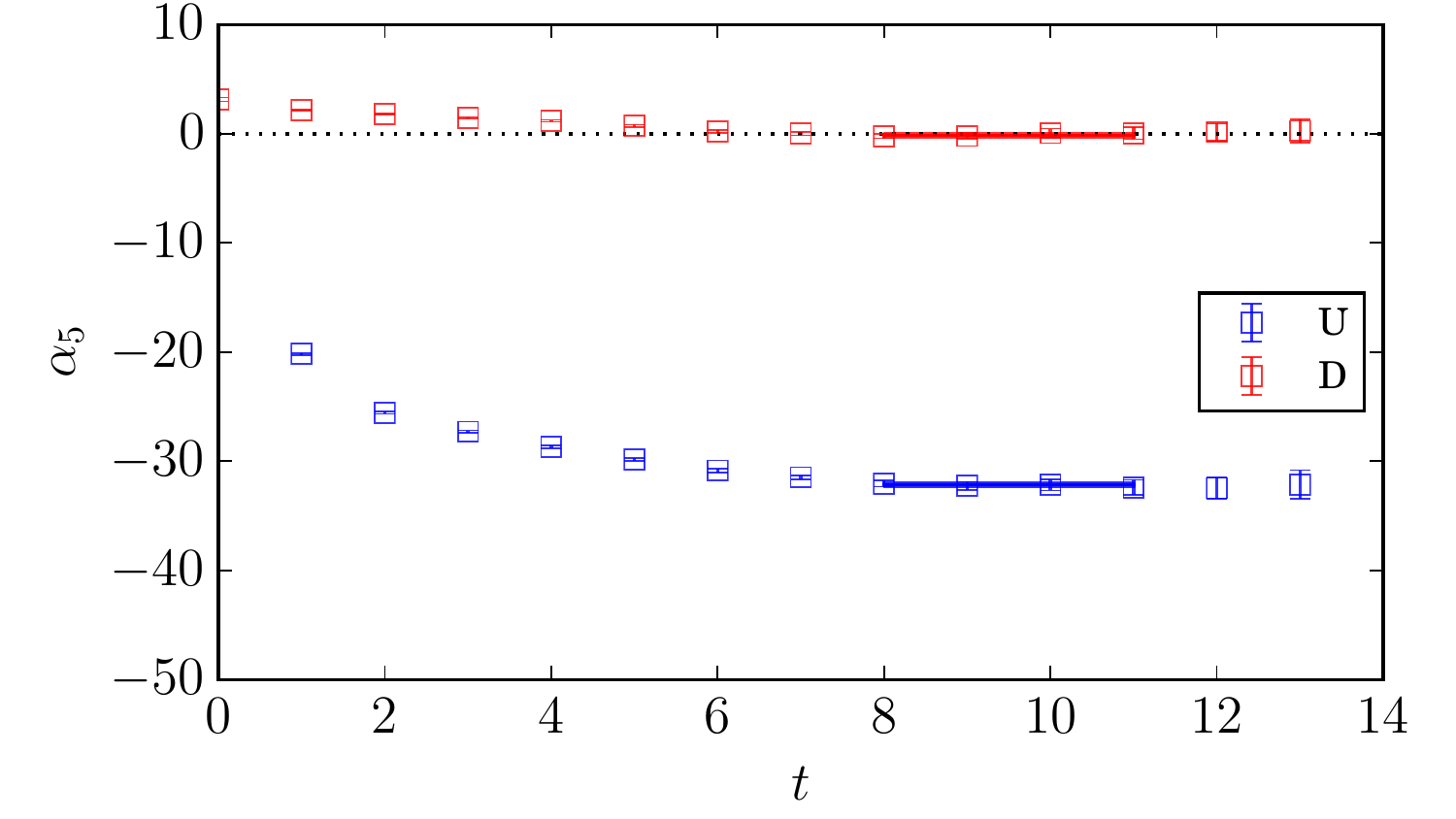}~
\includegraphics[width=0.49\textwidth]{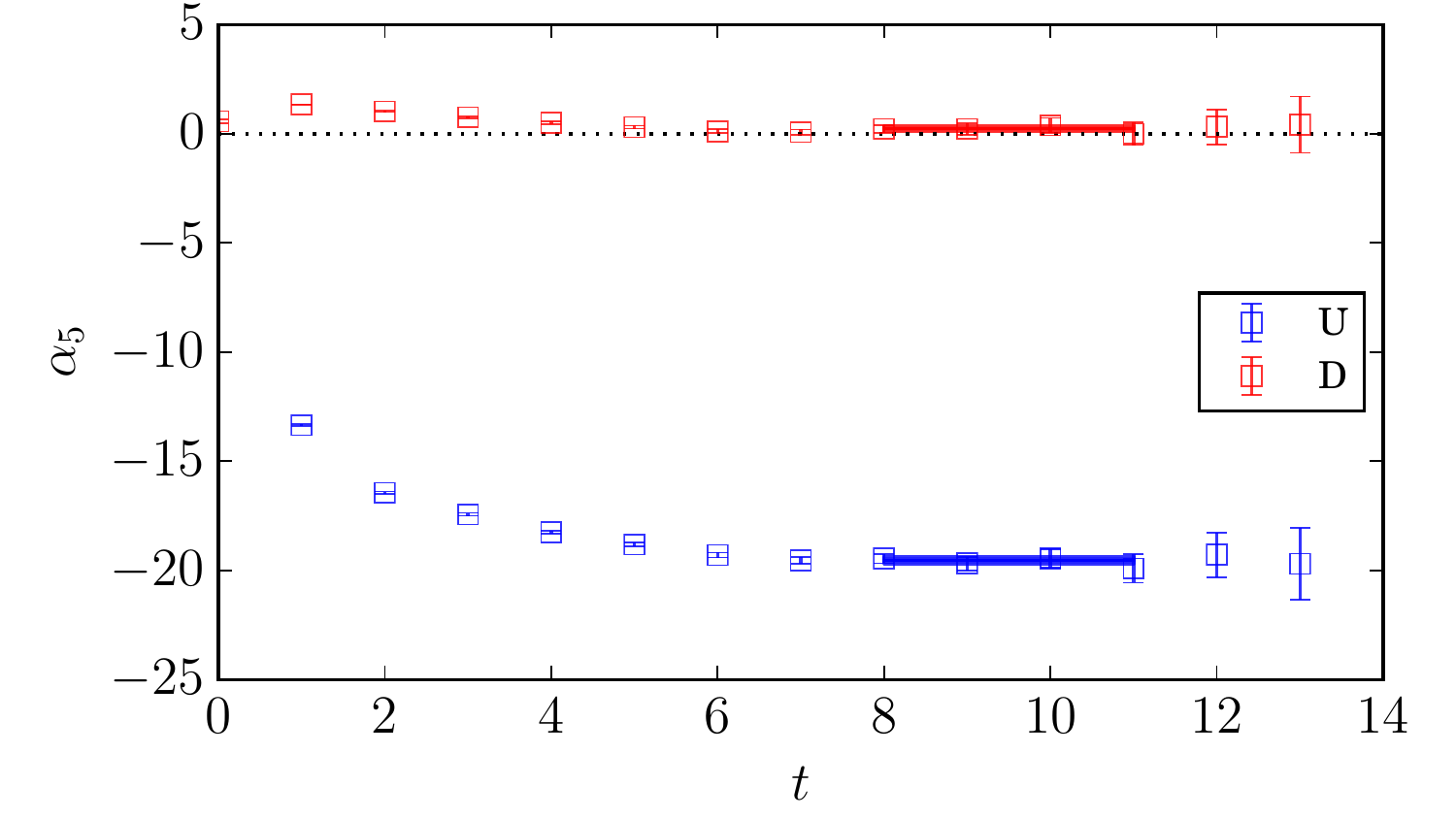}\\
\caption{Chiral rotation angle $\alfive$ of the proton field induced by $u$- and $d$-quark cEDM
  interactions, on the $24^3\times64$ (left) and $16^3\times32$ (right) lattices.
  The angles $\alfive$ for the neutron are related by the $SU(2)_f$ symmetry 
  $u\leftrightarrow d$.
  The chromo-EDM interactions are not renormalized and may include mixing with other operators.
  \label{fig:alfive}}
\end{figure}

In order to compute the form factor $F_3$, we first need to calculate the parity mixing angle
$\alfive$ in order to subtract the $F_{1,2}$ mixing terms.
Expanding the nucleon two-point function $C_{N\bar N}^\CPviol(t)$ to the first order in 
$\alfive\sim c_{\psi G}$ and assuming that the ground state dominates for sufficiently large $t$,
\begin{equation}
\begin{aligned}
C_{N\bar N}(t) - i c_{\psi G} \,\delta^\CPbar C_{N\bar N}(t) +O(c_{\psi G}^2) 
  \stackrel{t\to\infty}= |Z_N|^2\big[\frac{1+\gamma^4}2 + i\alfive\gamma_5 + O(\alfive^2)\big] e^{-m_N t}
\end{aligned}
\end{equation}
we use the projectors $T^+=\frac{1+\gamma_4}2$ and $T^+\gamma_5$ to calculate 
the ``effective'' mixing angle $\alfivehat(t)$ normalized to $c_{\psi G}=1$
\begin{equation}
\label{eqn:alfive_ratio}
\alfivehat^{eff}(t) 
  = -\frac{\Tr\big[ T^+\gamma_5 \cdot \delta^\CPbar C_{N\bar N} (t)\big]}
          {\Tr\big[ T^+ \cdot C_{N\bar N}(t)\big]}
  \stackrel{t\to\infty} = \frac{\alfive}{c_{\psi G}} 
\end{equation}
The time dependence of the ratios~(\ref{eqn:alfive_ratio}) for both ensembles is shown in 
Fig.~\ref{fig:alfive}.
The quark flavors in the cEDM interaction are shown respective to the proton, and 
for the neutron must be switched $u\leftrightarrow d$ due to the isospin symmetry.
The plateau is reached for time $t\geq8$, and we extract the $\alfive$ values from a constant fit 
(weighted average) to points $t=8\ldots11$.
An interesting observation is that the mixing angle depends very strongly on the flavor 
involved in the $\CPviol$ interaction.
Thus, for the proton $P_\delta = u_\delta(u^T C \gamma_5 d)$, in which the $d$-quark enters
together with $u$ as a scalar diquark, the $d$-cEDM does not lead to any parity mixing.

Finally, the electric dipole form factor $F_3$ is calculated from the $\CP$-odd 
four-point correlator $\delta^\CPbar C_{NJ\bar N}$.
Similarly to the extraction of $\alfivehat$ above, we can expand the $\CPviol$ 
three-point function in the $\CP$-odd interaction.
We extract the matrix elements using the ratios~(\ref{eqn:c3_c2_ratio}) of 
polarization-projected three-point functions $\Tr\big[T\cdot\mcR^\CPviol_{NJ\bar N} \big]$
to $\CP$-even two-point functions~(\ref{eqn:c2pt_proj_unpol}).
Expanding the ratio in $\alfive\sim c_{\psi G}$, we get
\begin{equation}
\begin{aligned}
  \Tr\big[T\big(\mcR_{NJ\bar N}
    - i c_{\psi G} \,\delta^\CPbar \mcR_{NJ\bar N} +O(c_{\psi G}^2)\big)\big]
\stackrel{t\to\infty}= 
  \sum_{i=1,2}\big[ \mcK_{\mcR\,i}^{(T)} + i\alfive\mcK_{\mcR\,i}^{(\{T,\gamma_5\})} \big]F_i
    + \mcK_{\mcR\,3}^{(T)} F_3 + O(\alfive^2)
\end{aligned}
\end{equation}
where $\mcK^{(T)}_{\mcR\,1,2,3}$ are the kinematic 
coefficients~(\ref{eqn:V3coeff_pos}-\ref{eqn:V4coeff_posSz}) for form factors $F_{1,2,3}$ 
computed with the polarization matrix $T$ and with $\mcK\to\mcK_\mcR$~(\ref{eqn:kincoeff_ratio}).
Matching the $O(c_{\psi G}^{1})$ terms in the above expansion and neglecting excited states, 
we obtain
\begin{equation}
\label{eqn:F3coeff}
i \mcK_{\mcR\,3}^{(T)} \hat F_3 
  = \Tr\big[T\cdot\delta^\CPbar \mcR_{NJ\bar N}\big]_{g.s.} 
    + \alfivehat \sum_{i=1,2}\mcK_{\mcR\,i}^{(\{T,\gamma_5\})} F_i
\end{equation}
The second term in the RHS of the above equation is the mixing subtraction. 
Its form indicates that the mixing between form factors $F_{1,2}$ and $F_3$ happens only
because of the mixing of the polarization of the nucleon interpolating fields on a lattice.
This is substantially different from expressions used in Refs.~\cite{Shintani:2005xg,
Berruto:2005hg,Aoki:2008gv,Guo:2015tla,Shindler:2015aqa,Alexandrou:2015spa,Shintani:2015vsx},
which also include additional subtraction term $(-2\alfive F_3)$ because of spurious mixing of
$F_2$ and $F_3$ in the vector current vertex~(\ref{eqn:vertex_mixing}).

Although both timelike and spacelike components of the current can be used to calculate $F_3$, 
in practice we find that the time component $J^4$ yields much better precision than 
the spacelike component $J^3$. 
Due to the larger uncertainty of the $J^3$ signal, combining both components did not result in
improved precision of the  $F_3$ form factor.
If only the $J^4$ component is used, the overdetermined fit to matrix elements is not required,
and for $T=T^+_{S_z+} = \frac{1+\gamma_4}{2}(1-i\gamma_1\gamma_2)$ 
from Eqs.~(\ref{eqn:V4coeff_pos},\ref{eqn:V4coeff_posSz})
\begin{equation}
\label{eqn:f3_formula}
(1+\tau) F_3(Q^2)
  = \frac{m_N}{q_3 \mcK_\mcR} \Tr[T^+_{S_z+}\cdot \delta^\CPbar \mcR_{N J^4\bar N}]
    - \alfivehat G_E(Q^2)
\end{equation}
where $\tau$ is the kinematic variable~(\ref{eqn:tau_def}).
It is remarkable that for the neutron the subtraction term $\sim\alfive G_E$ is zero in the
forward limit.
In fact, if one uses the traditional formula for extracting the neutron EDM 
$d_N = F_3(0)/(2m_N)$, a large contribution $(-2\alfive F_2(0))/(2m_N)$ comes from 
the spurious mixing if $\alfive$ is not zero.
In Section~\ref{sec:prev_works} we will discuss the currently available lattice results for the 
neutron and proton EDM induced by the QCD $\theta$-term.

\begin{figure}[ht!]
\centering
\includegraphics[width=0.49\textwidth]{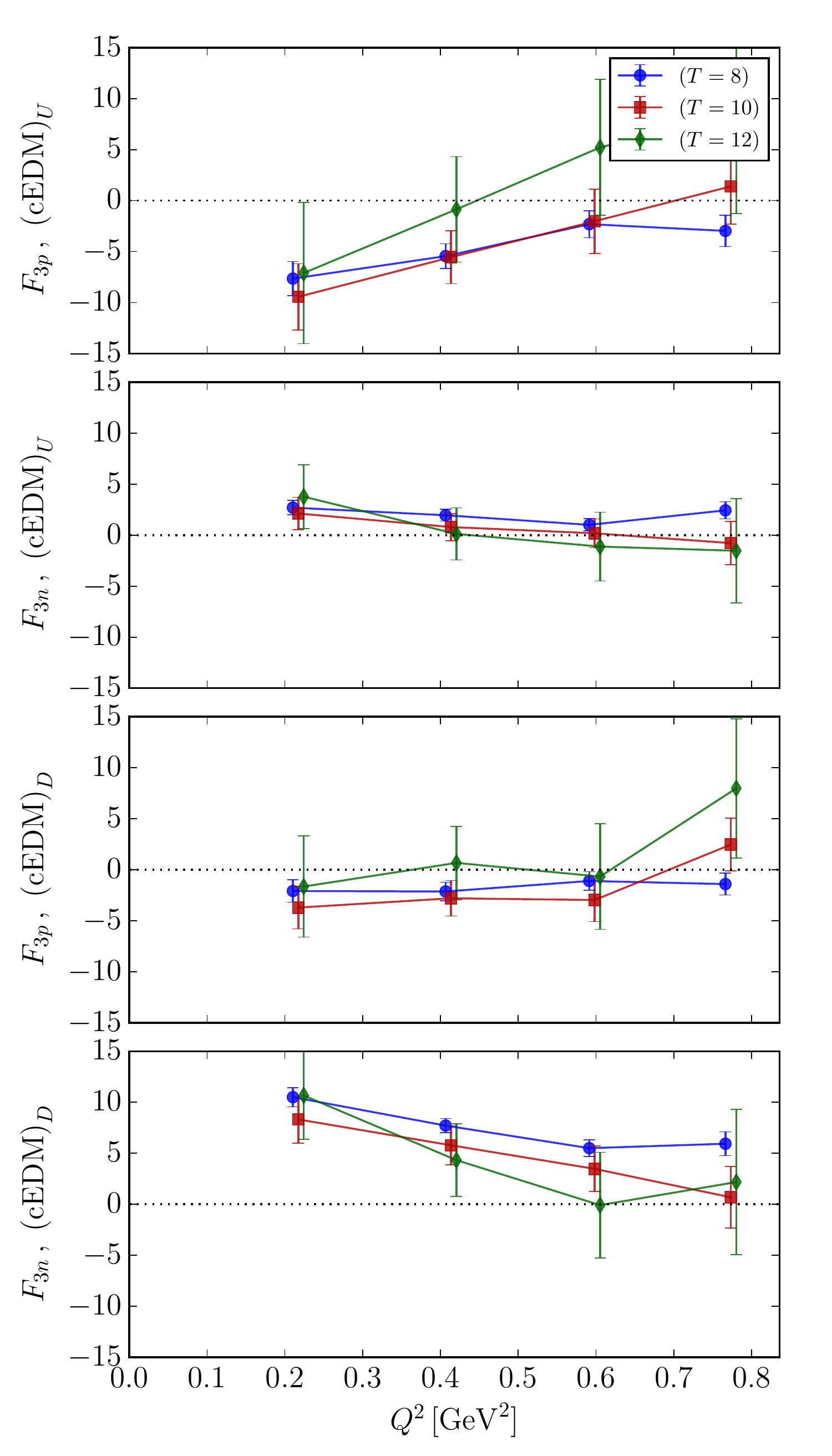}~
\includegraphics[width=0.49\textwidth]{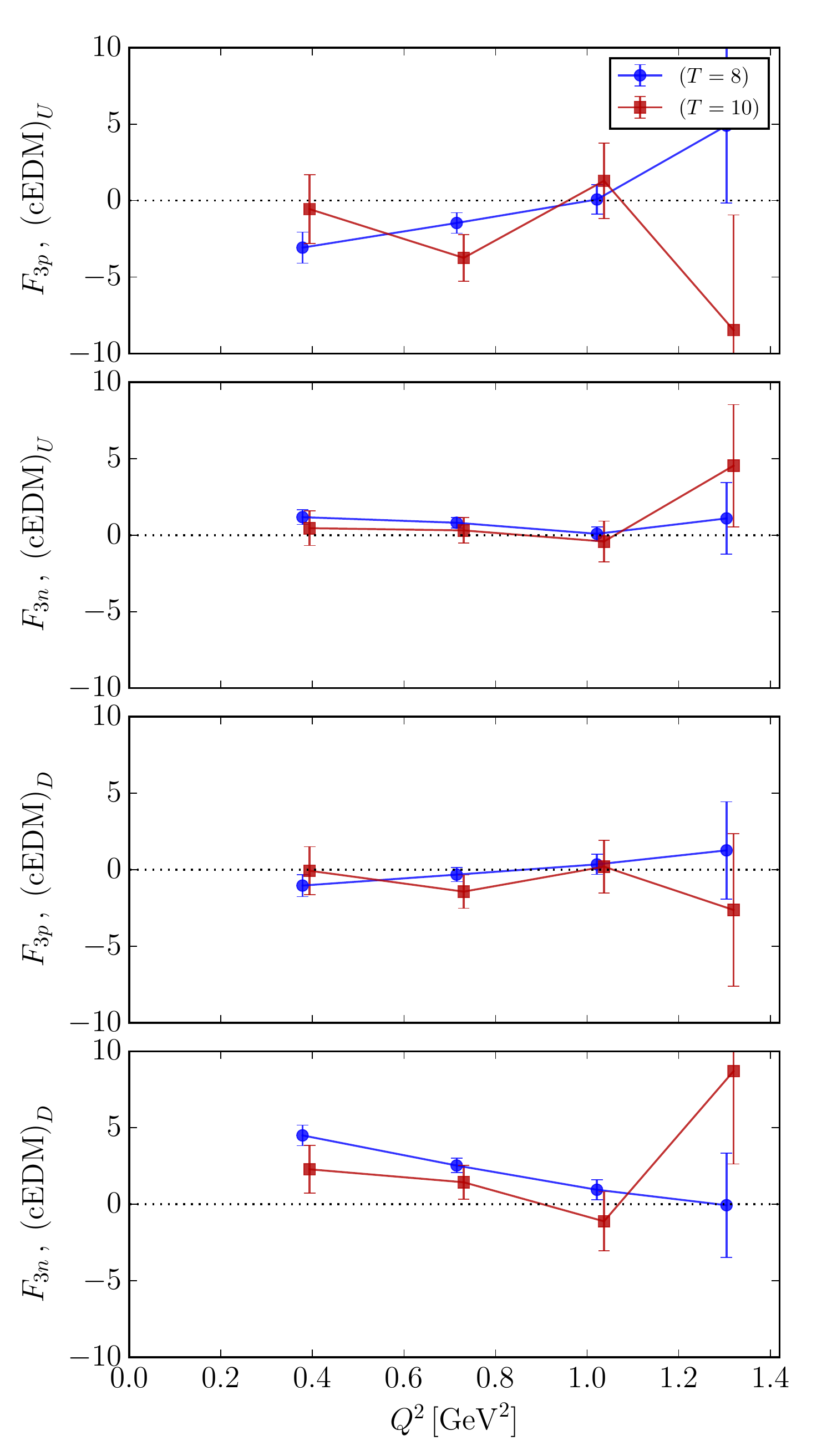}\\
\caption{Nucleon electric dipole form factors $F_3$ induced by 
  $u$- and $d$-quark chromo-EDM interactions, from the $24^3\times64$ (left) and $16^3\times32$
  (right) lattices. 
  The chromo-EDM interactions are not renormalized and may include mixing with other operators.
  Disconnected contractions are not included for either current or cEDM insertion.
  \label{fig:nucl_cpviol_ff}}
\end{figure}

To compute form factors from data with each source-sink separation $t_\text{sep}$, 
we use the value $\alfivehat = \alfivehat^\text{eff}(t_\text{sep})$ in Eq.~(\ref{eqn:f3_formula})
to subtract the mixing.
The results for the EDFF $F_3$ are shown in Fig.~\ref{fig:nucl_cpviol_ff}.
Despite relatively high statistics, the signal for the cEDM-induced form factor is noisy.
There is no significant dependence on the source-sink separation $t_\text{sep}$.
Since the cEDM operator is not renormalized, it can include contributions from other operators of
dimension 5, as well as operators from lower dimension 3~\cite{Bhattacharya:2015rsa}.
One peculiar feature of these results is that, similarly to $\alfive$, the contribution to the
proton EDM comes mostly from the $u$-cEDM, while the contribution to the neutron comes mostly from
the $d$-cEDM.
However, a substantial increase in statistics, as well as more elaborate analysis of excited
states, are required to confirm these observations.

\begin{figure}[ht!]
\centering
\includegraphics[width=0.49\textwidth]{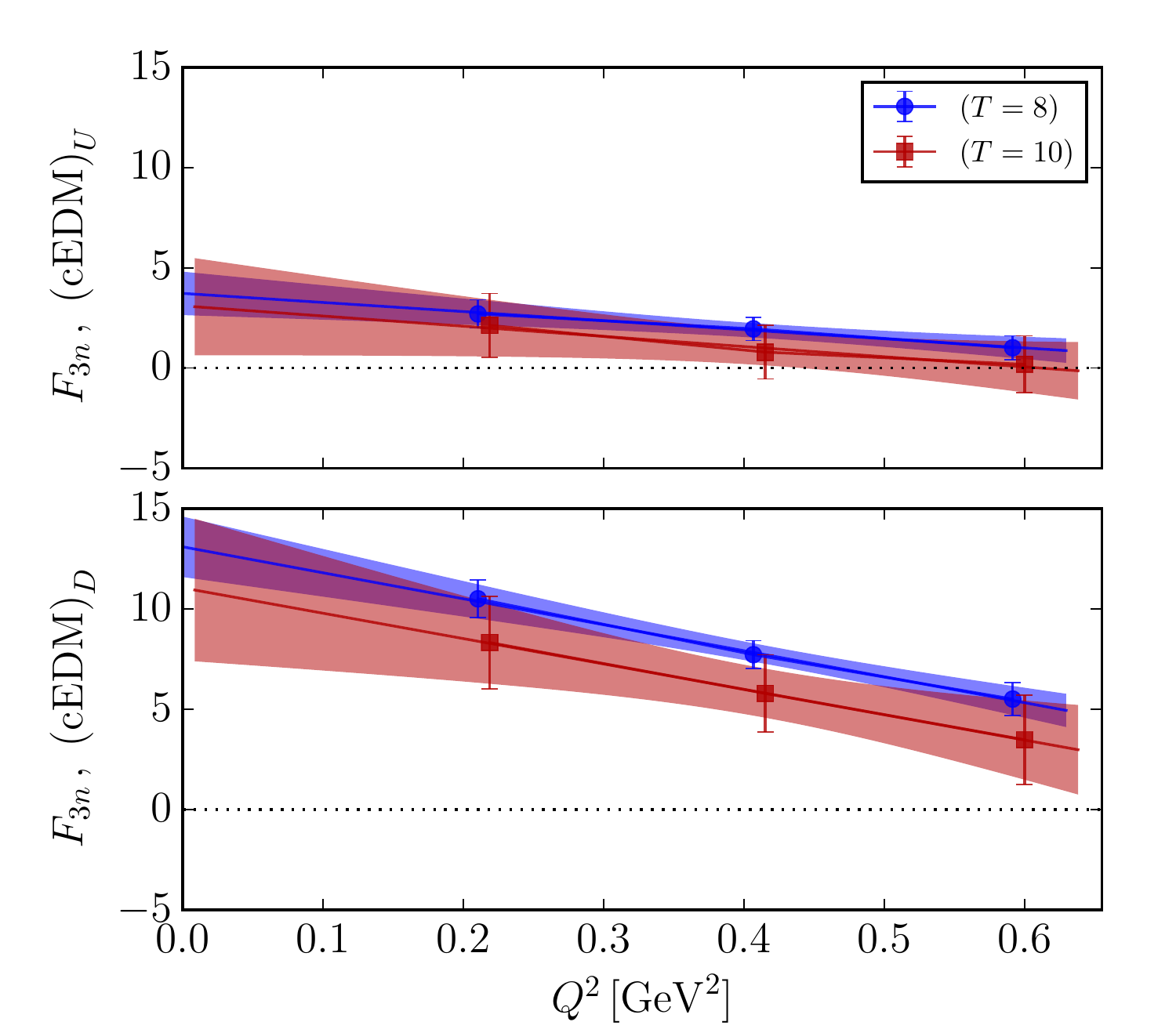}~
\includegraphics[width=0.49\textwidth]{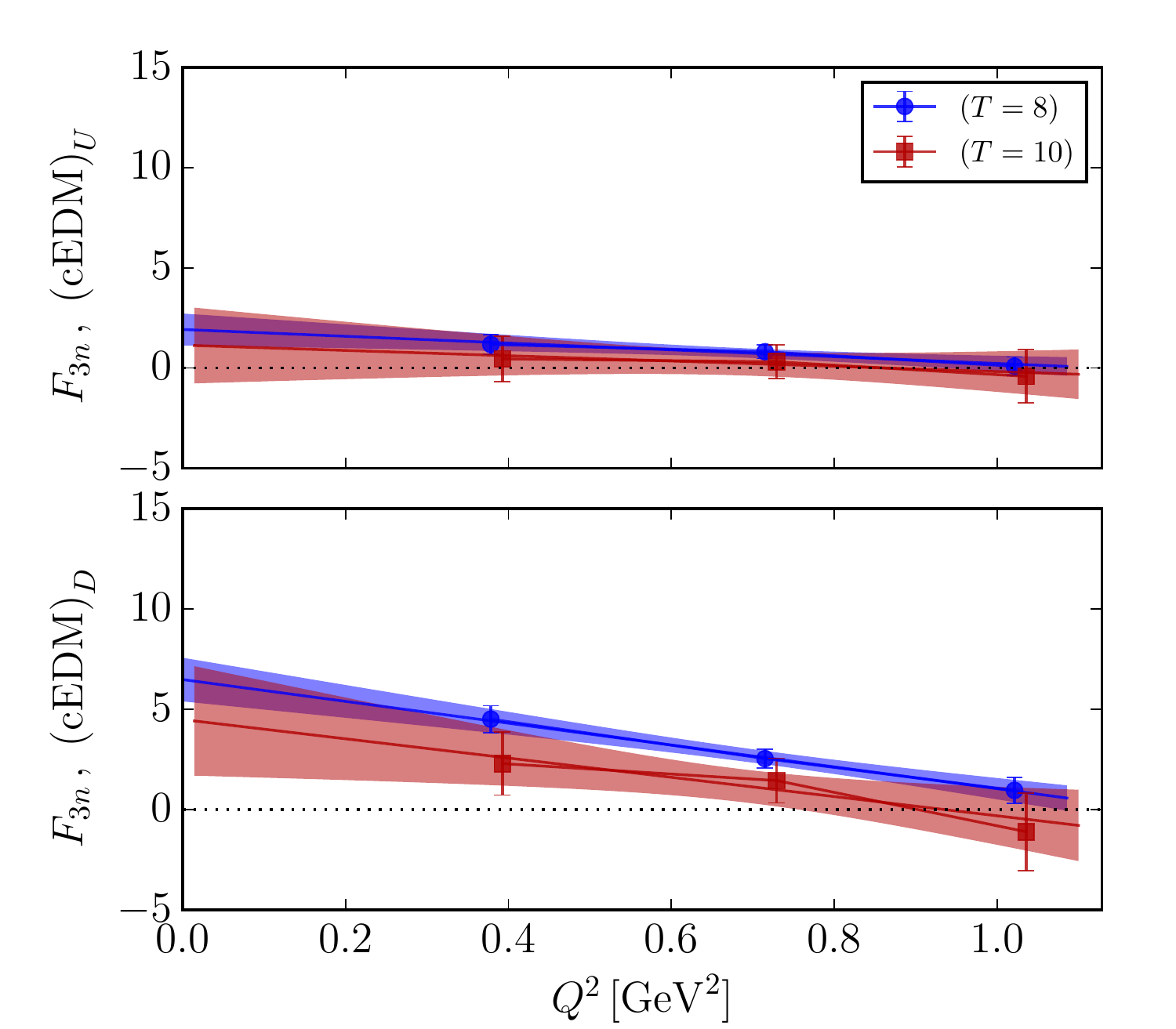}\\
\caption{Linear $Q^2$ fits to the neutron EDFF $F_3$ (same data as in
  Fig.~\ref{fig:nucl_cpviol_ff}) including only the three smallest $Q^2>0$ points and source-sink
  separations $T=8a,10a$.
  Results for the $24^3\times64$ (left) and $16^3\times32$ (right) lattices. 
  \label{fig:neut_cpviol_ff_linfits}}
\end{figure}

The electric dipole moment is determined by the value of the form factor $F_3(Q^2)$ at zero.
This value is not directly calculable, and one has to extrapolate the $Q^2>0$ data points to
$Q^2=0$. 
In Figure~\ref{fig:neut_cpviol_ff_linfits} we show linear extrapolation of these form factors
using the three smallest $Q^2>0$ points.
Other fit models are not warranted until the statistical precision is substantially improved.

\subsection{Neutron electric dipole moments from energy shifts}\label{sec:bgem}

Calculation of the dipole moment using uniform background field has an advantage that no form
factor extrapolation in momentum is required, because the energy shift depends on the forward
matrix element of the nucleon.
This calculation is easier for the neutron than for the proton, because in case of a charged 
particle there are additional complications due to its constant acceleration,
which makes analysis of its correlation functions more complicated~\cite{Detmold:2009dx}.

On the other hand, since the uniform background field is quantized on a lattice, these fields
cannot be made arbitrarily small.
In fact, the field quanta are very large and their magnitudes are comparable to the QCD scale, 
especially on the smaller $16^3\times32$ lattice.
Because of the fractional charges of quarks, there is additional factor of 3 in the minimal value 
of the electric field, which is quantized in multiples of $\mcE_0 = \frac{6\pi}{a^2 L_x L_t}$. 
The $\mcE_0$ values are shown in Tab.~\ref{tab:ens}, and for the smaller $16^3\times32$ 
lattice the minimal electric field is $\mcE_0=0.110\,\mathrm{GeV}^2 = 560\,\mathrm{MV}/\mathrm{fm}$.
Such electric field pulls the $u$ quark in the neutron with tension 
$\approx(270\,\mathrm{ MeV})^2$, or approximately 40\% of the QCD string tension,
and may deform the neutron too far away from the ground state.

We introduce the uniform electric field on a lattice as described in
Sec.~\ref{sec:edm_dirac_eshift_euc} along the $z$ direction.
Using modified QCD+$U(1)$ gauge links, we calculate the regular nucleon correlator 
$C_{N\bar N,\mcE}$, as well as the correlator with the insertion of 
$\CP$-odd interaction, in full analogy with Sec.~\ref{sec:evenodd_corr}, e.g.
\begin{align}
\delta^\CPbar C_{N\bar N,\mcE}(\vec p, t) 
  = \sum_{\vec y} e^{-i\vec p(\vec y - \vec x)} \langle N(\vec y, t) 
    \bar N(\vec x, 0)\cdot\mcO^\CPbar_{\psi G}\rangle_\mcE
\end{align}
The modified gauge links are used in both computing the propagators and construction of the
smeared sources and sinks.
In fact, since the individual quarks are charged, smearing their distributions with only 
the QCD gauge links breaks the covariance and makes the calculation dependent on the choice 
of the gauge of the electromagnetic potential.
The QCD links used in Gaussian smearing are first APE-smeared, and then the electromagnetic potential 
is applied to them.

\begin{figure}[ht!]
\centering
\includegraphics[width=0.49\textwidth]{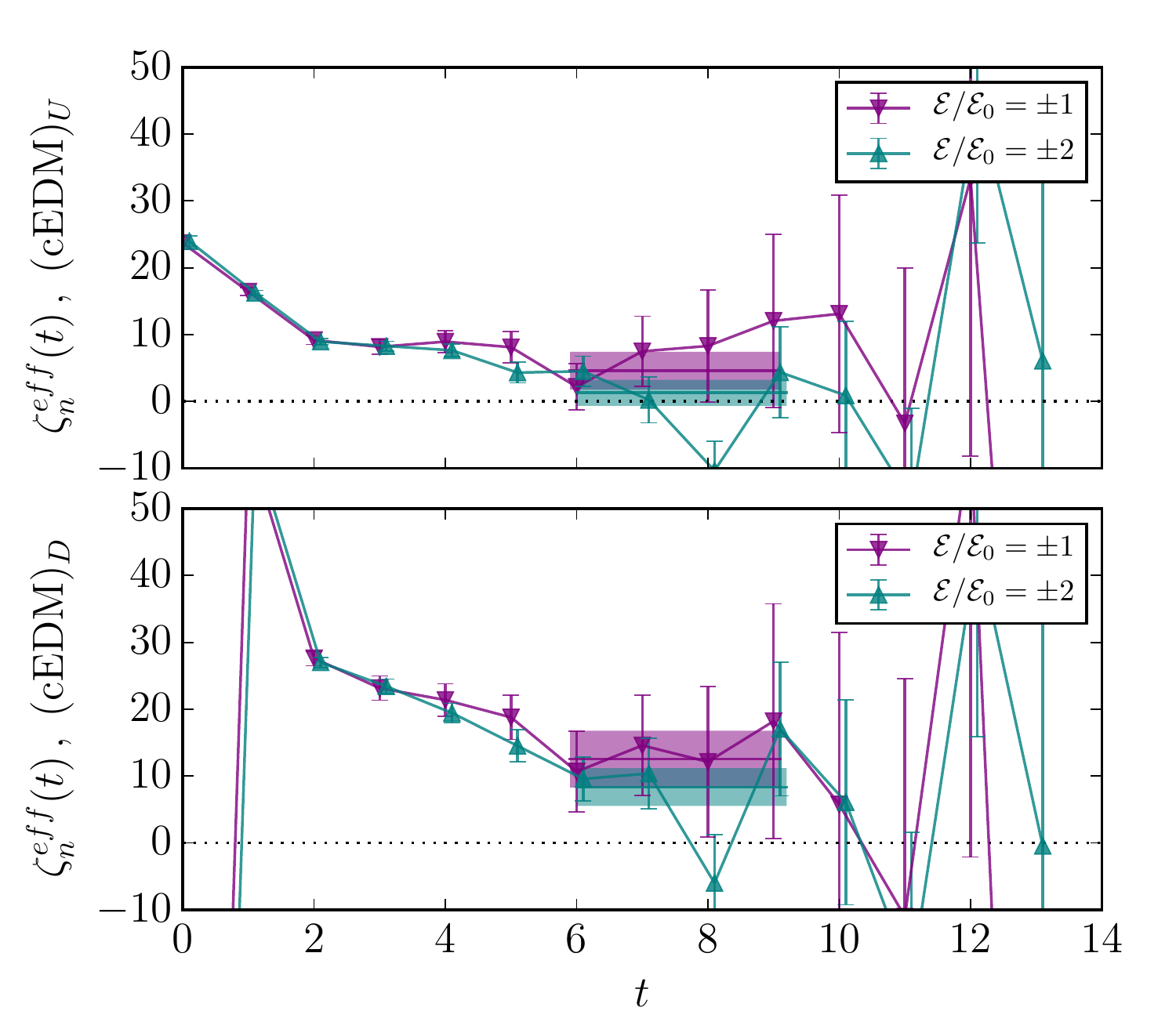}~
\includegraphics[width=0.49\textwidth]{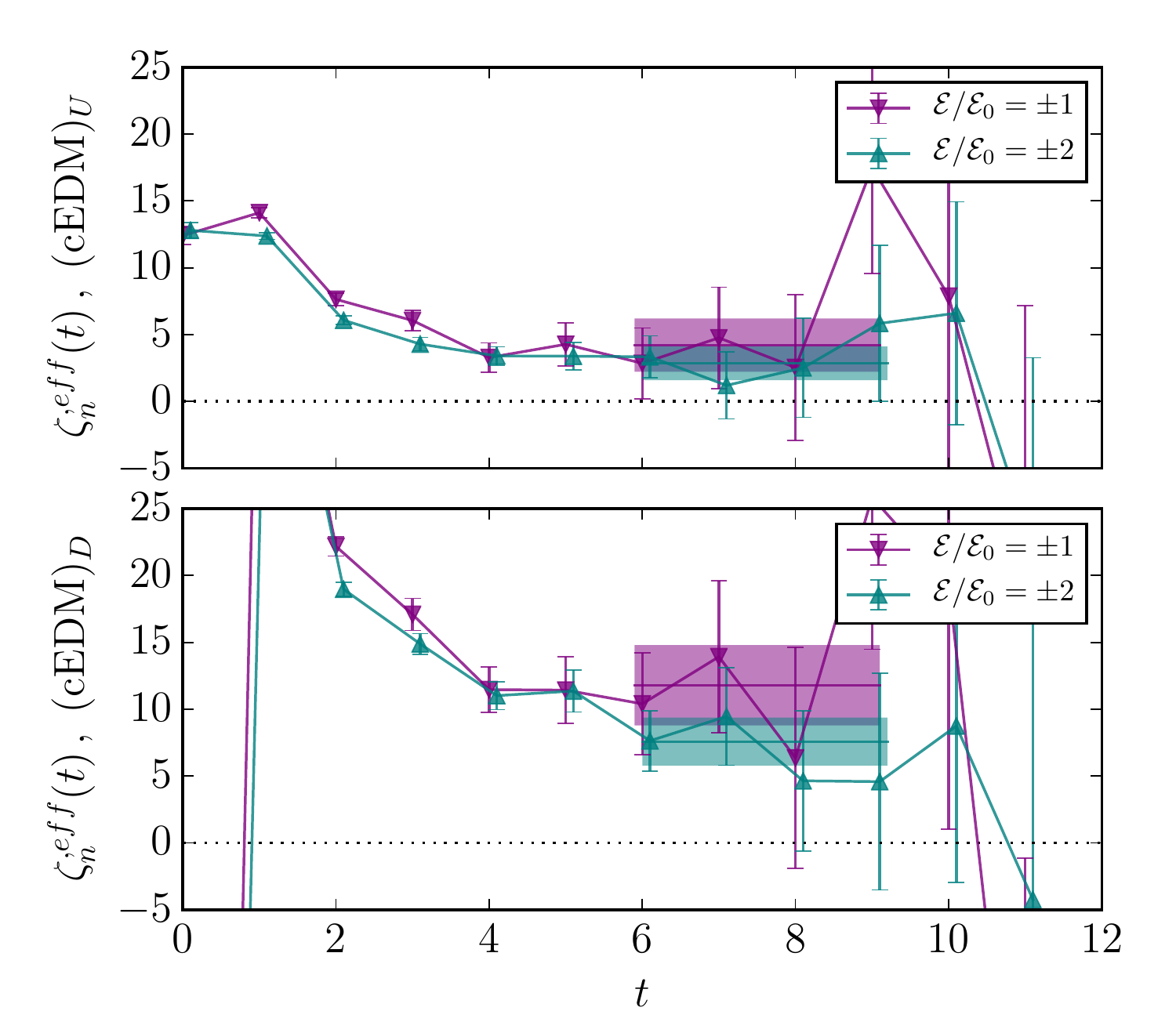}\\
\caption{The neutron EDM computed from energy shifts with two values of the electric field.
  The units are dimensionless and the scale is the same as for $F_3$.
  The values used in comparison are computed as the average from the $t=6\ldots9$ 
  conservative plateaus common for both ensembles and both cEDM flavors.
  Results for the $24^3\times64$ (left) and $16^3\times32$ (right) lattices. 
  \label{fig:neut_eshift_pltx}}
\end{figure}

From Eq.~(\ref{eqn:eshift_euc}), the energy of a particle on a lattice with the spin polarized 
along the electric field $\vec\mcE=\mcE\hat z$ is shifted by the imaginary value 
$\delta E = -(\zeta / 2m) i\mcE$.
The nucleon correlator at rest ($\vec p = 0$) thus must take the form
\begin{equation}
C^{\CPviol}_{N\bar N,\mcE}(\vec p=0, t) 
  = |Z_N|^2 e^{i\alfive\gamma_5} \frac{1+\gamma_4}2 \big[
      \frac{1+\Sigma_z}2 e^{-(m+\delta E)t}
     +\frac{1-\Sigma_z}2 e^{-(m-\delta E)t}
  \big]e^{i\alfive\gamma_5}
\end{equation}
As with the $\CP$-odd form factor $F_3$, expanding the correlator up to the first order in
$c_{\psi G}\sim\alfive\sim \delta E\sim \zeta$, we get
\begin{equation}
C_{N\bar N,\mcE} - i c_{\psi G} \,\delta^\CPbar C_{N\bar N,\mcE}
  \stackrel{t\to\infty}= |Z_N|^2 e^{-m_N t}\,
  \big[\frac{1 + \gamma_4}2 + i\alfive \gamma_5 
    - \Sigma_z\,\delta E\, t \big] 
\end{equation}
for the electric dipole moment we obtain the following estimator for the 
effective energy shift:
\begin{equation} 
\label{eqn:dneff} 
\zeta^\text{eff}(t) = 2 m_N d_N^\text{eff}(t) = -\frac{2m_N}{\mcE_z} \big[R_z(t+1) - R_z(t)\big]\,, 
\quad R_z(t) = \frac{\mathrm{Tr}\big[ T^+ \,\Sigma_z\,\delta^\CPbar C_{N\bar N,\mcE_z}(t)\big]} 
                    {\mathrm{Tr}\big[ T^+ \, C_{N\bar N,\mcE_z}(t)\big]}\,. 
\end{equation}
We have computed the neutron correlation functions with two values of the electric field 
$\mcE=\mcE_0$ and $2\mcE_0$.
The results for both ensembles are shown in Fig.~\ref{fig:neut_eshift_pltx}. 
We choose $t=6\ldots9$ as the common plateau to estimate the value of $\zeta$ on both ensembles and 
both flavors in the cEDM operator.
In the case of $d$-cEDM, we observe non-zero values of the energy shift.
Also the EDM values computed with $\mcE=\mcE_0$ and $2\mcE_0$ agree well with each other,
indicating that the energy shift is linear in $\mcE$ and our EDM result does not depend on the
polarizing effect of the electric field.


\subsection{Numerical comparison of the form factor and energy shift methods}\label{sec:ff_eshift_cmp}

\begin{figure}[ht!]
\centering
\includegraphics[width=0.49\textwidth]{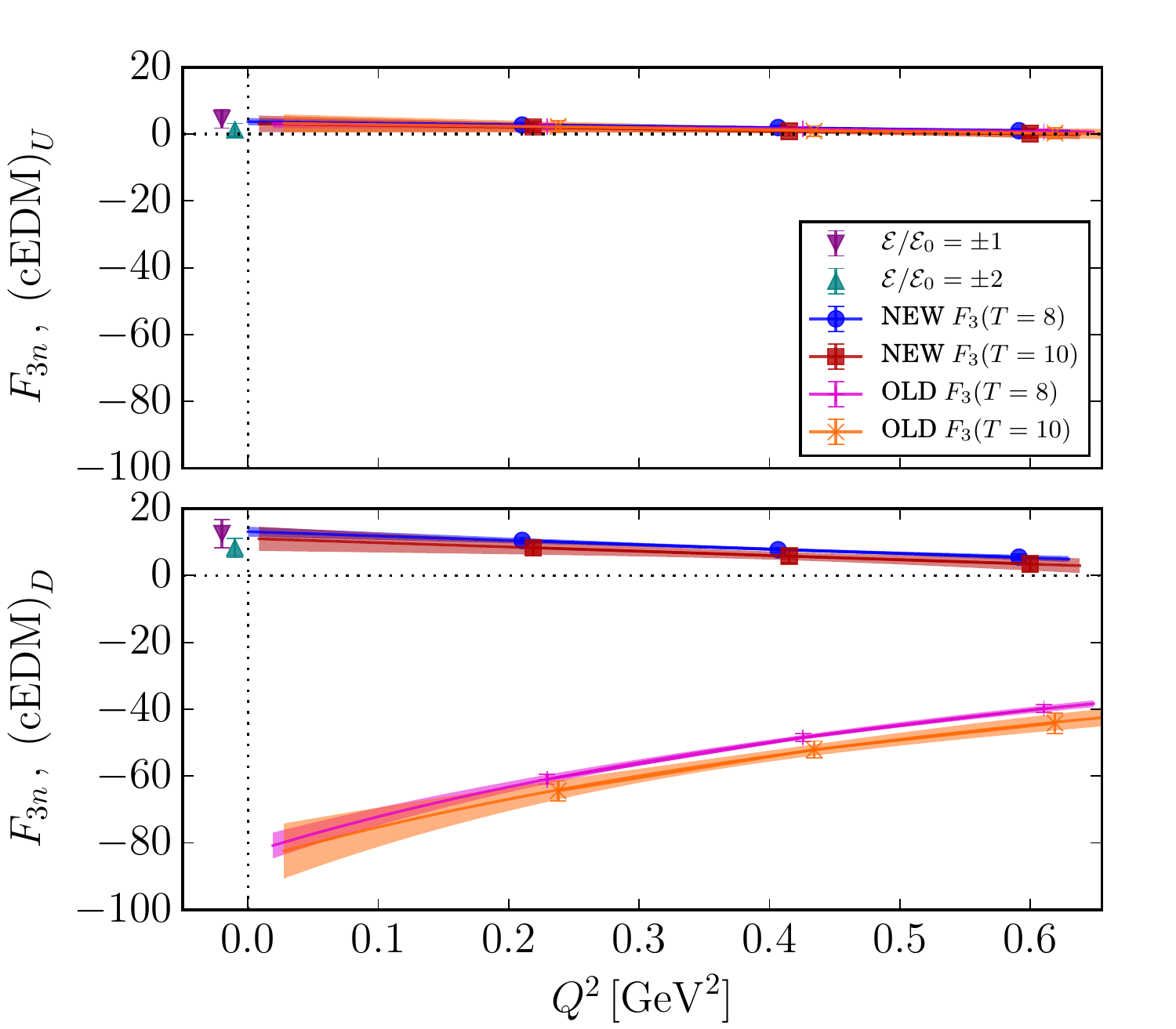}~
\includegraphics[width=0.49\textwidth]{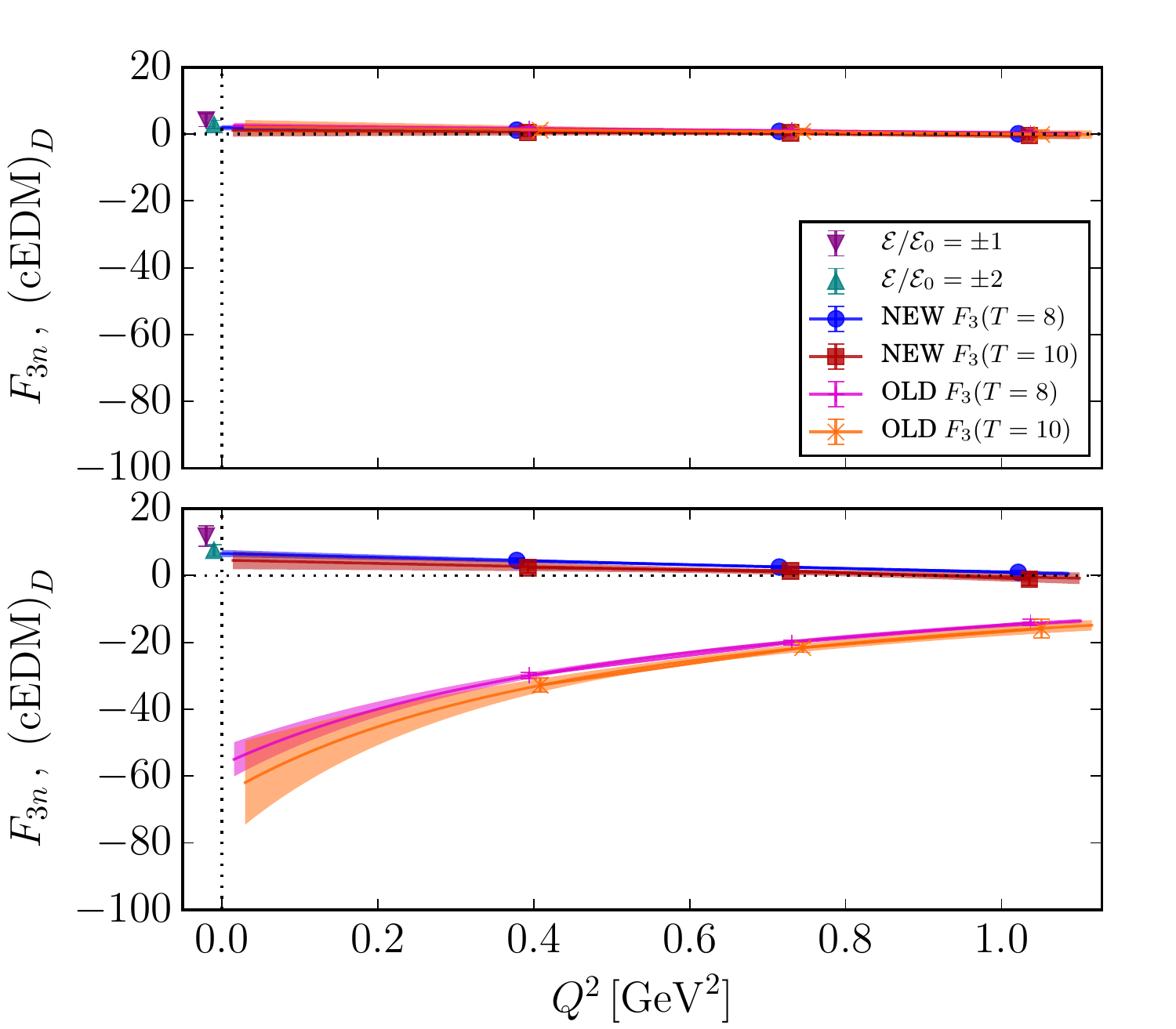}\\
\caption{Comparison of the neutron EDFF $F_{3n}(Q^2)$ computed with the 
  conventional (``OLD'')~\cite{Shintani:2005xg,Berruto:2005hg,Aoki:2008gv,Guo:2015tla,Shindler:2015aqa,
  Alexandrou:2015spa,Shintani:2015vsx} 
  and the ``NEW'' formula~(\ref{eqn:V4coeff_posSz})
  to the neutron EDM $\zeta$ computed from the energy shift (see 
  Fig.~\ref{fig:neut_eshift_pltx}).
  The ``OLD'' $F_{3n}(Q^3)$ data are extrapolated with the dipole fit, and the ``NEW'' with the
  linear fit.
  Data points are shifted horizontally for legibility.
  Results for the $24^3\times64$ (left) and $16^3\times32$ (right) lattices. 
  \label{fig:neut_cpviol_ff_cmp_eshift}}
\end{figure}

The normalization and the sign convention of the dimensionless EDM $\zeta$ in Sec.~\ref{sec:bgem}
are identical to those of $F_3(0)$ in Sec.~\ref{sec:lat_ff}, and we plot
them for comparison in Fig.~\ref{fig:neut_cpviol_ff_cmp_eshift}.
We observe satisfactory agreement between the values of $\zeta$ computed in the uniform background
method and the values obtained from the $Q^2\to0$ extrapolation of form factors $F_{3n}(Q^2)$.

In order to check how the spurious mixing affects the results, 
in Fig.~\ref{fig:neut_cpviol_ff_cmp_eshift} we also plot the values of form factors 
computed with the old formula used in Refs.~\cite{Shintani:2005xg,Berruto:2005hg,
Aoki:2008gv,Guo:2015tla,Shindler:2015aqa,Alexandrou:2015spa,Shintani:2015vsx}
\begin{equation}
\tilde F_3 = F_3 - 2\alfive F_2\,.
\end{equation}
This formula obviously gives a value for $\tilde{F}_3$ different from $F_3$ only if 
$\alfive$ is large.
In the case of $u$-cEDM, the value $\alfive$ for the neutron is small, and there is no observable
difference between $F_3$ and $\tilde F_3$.
However, in the case of $d$-cEDM, the difference is remarkable.
Neither of the three sources of uncertainty: excited state bias in the energy shift calculation, 
excited state bias in the form factor calculation, nor the $Q^2\to0$ extrapolation of the form
factors, can plausibly change the outcome of this comparison, due to the large value of $\alfive$.
The agreement between the new form factor extraction formula and the energy shift method is one of
the main results of this paper, and serves as a numerical cross-check of the analytic derivation.

We collect the values of $\alfive$, extrapolated $F_3(0)$, and $\zeta_n$ from the background field
method in Tab.~\ref{tab:neut_cpviol_ff_cmp_eshift}.

\begin{table}
\centering
\caption{
  Comparison of the neutron EDM $\zeta_n$ computed from the energy shift 
  to the neutron forward EDFF $F_3(0)$ computed with the new
  formula~(\ref{eqn:V4coeff_posSz}) and the old formula~\cite{Shintani:2005xg}.
  The parity mixing angle $\alfive$ is computed from the plateaus in Fig.~\ref{fig:alfive} (the
  flavors have been switched $u\leftrightarrow d$ to describe the neutron).
  \label{tab:neut_cpviol_ff_cmp_eshift}}
$24^3\times64$\\
\begin{tabular}{ll|ll}
\hline\hline
& & $(\mathrm{cEDM})_U$ & $(\mathrm{cEDM})_D$ \\
\hline
$\alfive$ & $t=8\ldots11$ &
  $-0.16(14)$ &  $-32.2(2)$ \\
\hline
$\zeta_n$ from $\Delta E(\mcE)$ & $\mcE/\mcE_0=1$ & 
  $\invplus4.6(2.8)$ &  $\invplus12.5(4.2)$ \\
& $\mcE/\mcE_0=2$ & 
  $\invplus1.3(1.9)$ &  $\invplus8.4(2.8)$ \\
NEW $F_{3n}(0)$ [(\ref{eqn:V4coeff_posSz})] & $T=8a$ & 
  $\invplus3.7(1.1)$ &  $\invplus13.1(1.5)$ \\
  & $T=10a$ & 
  $\invplus3.1(2.4)$ &  $\invplus11.0(3.5)$ \\
OLD $F_{3n}(0)$ \cite{Shintani:2005xg} & $T=8a$ & 
  $\invplus3.1(1.3)$ &  $-80.8(3.8)$ \\
  & $T=10a$ & 
  $\invplus3.2(2.7)$ &  $-82.4(8.2)$ \\
\hline\hline
\multicolumn{4}{c}{$16^3\times32$}\\
\hline\hline
$\alfive$ & $t=8\ldots11$ &
  $\invplus0.23(12)$ &  $-19.54(15)$  \\
\hline
$\zeta_n$ from $\Delta E(\mcE)$ & $\mcE/\mcE_0=1$ & 
  $\invplus4.2(2.0)$ &  $\invplus11.8(3.0)$ \\
& $\mcE/\mcE_0=2$ & 
  $\invplus2.8(1.3)$ &  $\invplus7.6(1.8)$ \\
NEW $F_3(0)$ [(\ref{eqn:V4coeff_posSz})] & $T=8a$ & 
  $\invplus1.9(8)$ &  $\invplus6.5(1.1)$ \\
  & $T=10a$ & 
  $\invplus1.1(1.9)$ &  $\invplus4.4(2.7)$ \\
OLD $F_3(0)$ \cite{Shintani:2005xg} & $T=8a$ & 
  $\invplus2.5(9)$ &  $-55.0(5.1)$ \\
  & $T=10a$ & 
  $\invplus2.1(2.1)$ & $-62.0(12.5)$ \\
\hline\hline
\end{tabular}
\end{table}

\section{Corrections to Existing $\theta$-induced nEDM Lattice Results}\label{sec:prev_works}

In Section~\ref{sec:edm_ff_mink} it has been shown that the commonly used formula for extracting the
form factor $F_3$ from $\CPviol$ nucleon matrix elements on a lattice is incorrect.
This formula has been used in all of the papers that compute QCD $\theta$-induced nucleon 
EDM~\cite{Shintani:2005xg,
Berruto:2005hg,Aoki:2008gv,Guo:2015tla,Shindler:2015aqa,Alexandrou:2015spa,Shintani:2015vsx}.
Fortunately, the correction has a very simple form~(\ref{eqn:F23_mixing}), in which
$\tilde{F}_{2,3}$ refer to the old results and $F_{2,3}$ refer to corrected results.
Unfortunately, Refs.~\cite{Shintani:2005xg,Berruto:2005hg,Aoki:2008gv,Guo:2015tla,
Shindler:2015aqa,Alexandrou:2015spa,Shintani:2015vsx} offer a broad spectrum of conventions 
for $\tilde{F}_3$ and $\alfive$ differing in sign and scale factors. 
However, by comparing expressions for \emph{polarized} $\CP$-odd 
matrix elements of the timelike component of the vector current $J_4$
we can deduce the appropriate correction 
using that reference's conventions.
For example, using Eq.(55) from Ref.~\cite{Alexandrou:2015spa},
\begin{equation}
\begin{aligned}
\Pi^0_{3pt,Q}(\Gamma_k=\frac i4(1+\gamma_0)\gamma_5\gamma_k)
  &\sim \frac{i Q_k}{2m_N} \big[\alpha^1 \big(F_1 + \frac{E_N + 3m_N}{2m_N} F_2\big) 
    + \frac{E_N + m_N}{2m_N} \tilde{F}_3\big]\\
  &= \frac{i Q_k}{2m_N} \big[ \alpha^1 G_E 
    + (1 + \tau)
\underbrace{
\big(\tilde{F}_3 + 2\alpha^1 F_2\big)
}_{F_3}
\big]\,,
\end{aligned}
\end{equation}
where $\tau=\frac{E_N-m_N}{2m_N}$ introduced in Eq.(\ref{eqn:tau_def}) and $G_E=F_1-\tau F_2$ is the
Sachs electric form factor.
Comparing the above equation to the expected form~(\ref{eqn:V4coeff_posSz}), 
for the corrected value of $F_3$  we obtain\footnote{
  Note that this correction is the opposite compared to Eq.~(\ref{eqn:F23_mixing}), which results
  from a difference in used conventions.}
\begin{equation}
\label{eqn:edm_refs_correction}
\begin{aligned}
F_3(Q^2) = \tilde{F}_3(Q^2) + 2\alpha^1 F_2(Q^2) \,, \\
\end{aligned}
\end{equation}
which should hold for any value of $Q^2$.

\begin{table}[t]
\centering
\caption{
  Corrections to the results reported in earlier calculations of $\bar\theta$-induced 
  nucleon EDM   for the nucleon ($n$) and the proton($p$). 
  Some of the used values are at non-zero momentum transfer $Q^2$, or at non-zero value of
  $\bar\theta$-angle.
  Both form factors $F_{2,3}$ are quoted as dimensionless (in ``magneton'' units $e/(2m_N)$).
  The errors for $F_3$ are taken equal to those of $\tilde{F}_3$
  except Ref.~\cite{Guo:2015tla}, in which the error are extracted from our interpolation 
  of the corrected $\bar{F}_3(\bar\theta)$ values (see Fig.~\ref{fig:cmpF3_Guo2015tla}).
  In the first row, the correction follows the original conventions~\cite{Alexandrou:2015spa} 
  exactly.
  In the following rows, the parity-mixing angles $\alpha$ have been transformed to $\alpha<0$ 
  to and the EDMs were corrected with $F_3 = \tilde{F}_3 + 2\alpha F_2$ using the assumption
  discussed in the text.
  \label{tab:end_refs_correctF3}}
\begin{tabular}{lc|cc|ll|ll}
\hline\hline
& & $m_\pi\,[\text{MeV}]$ &  $m_N\,[\text{GeV}]$ &  $F_2$ & $\alpha$ &
  $\tilde{F}_3$ &  $F_3$ \\
\hline
\cite{Alexandrou:2015spa} & 
$n$ &  $373$ &  $1.216(4)$ &
  $-1.50(16)$\footnotemark[2] &  $-0.217(18)$ &  
  $-0.555(74)$ &  $\invplus0.094(74)$ \\
\hline
\cite{Shintani:2005xg} &  
$n$ &  $530$ &  $1.334(8)$ &
  $-0.560(40)$ &  $-0.247(17)$\footnotemark[1] &  
  $-0.325(68)$ &  $-0.048(68)$ \\
& 
$p$ &  $530$ &  $1.334(8)$ &
  $\invplus0.399(37)$ &  $-0.247(17)$\footnotemark[1] & 
  $\invplus0.284(81)$ &  $\invplus0.087(81)$ \\
\cite{Berruto:2005hg} & 
$n$ &  $690$ &  $1.575(9)$ &
  $-1.715(46)$ &  $-0.070(20)$ &  
  $-1.39(1.52)$ &  $-1.15(1.52)$ \\
& 
$n$ &  $605$ &  $1.470(9) $ &
  $-1.698(68)$ &  $-0.160(20)$ &   
  $\invplus0.60(2.98)$ &  $\invplus1.14(2.98)$ \\
\cite{Guo:2015tla} & 
$n$ &  $465$ &  $1.246(7)$ &
  $-1.491(22)$\footnotemark[3] &  $-0.079(27)$\footnotemark[4] &  
  $-0.375(48)$ &  $-0.130(76)$\footnotemark[4] \\
& 
$n$ &  $360$ &  $1.138(13)$ &
  $-1.473(37)$\footnotemark[3] &  $-0.092(14)$\footnotemark[4] &
  $-0.248(29)$ &  $\invplus0.020(58)$\footnotemark[4] \\
\hline\hline
\end{tabular}
\footnotetext[1]{The value $f_{1n}$ was reported incorrectly in Ref.~\cite{Shintani:2005xg}
  with a factor of $\frac12$~\cite{Shintani_privcomm}.}
\footnotetext[2]{Estimated as $(-\frac12 F_2^v(0))$ from Ref.~\cite{Abdel-Rehim:2015jna} 
  assuming $F_2^s\approx00$.}
\footnotetext[3]{From Ref.~\cite{Shanahan:2014uka} where $F_2$ was computed with $\bar{\theta}=0$.}
\footnotetext[4]{Estimated from a linear+cubic fit to plotted $\bar\alpha(\bar\theta)$ 
  and $F_3^\theta$ data~\cite{Guo:2015tla}.}
\end{table}

\begin{figure}
\centering
\includegraphics[width=.49\textwidth]{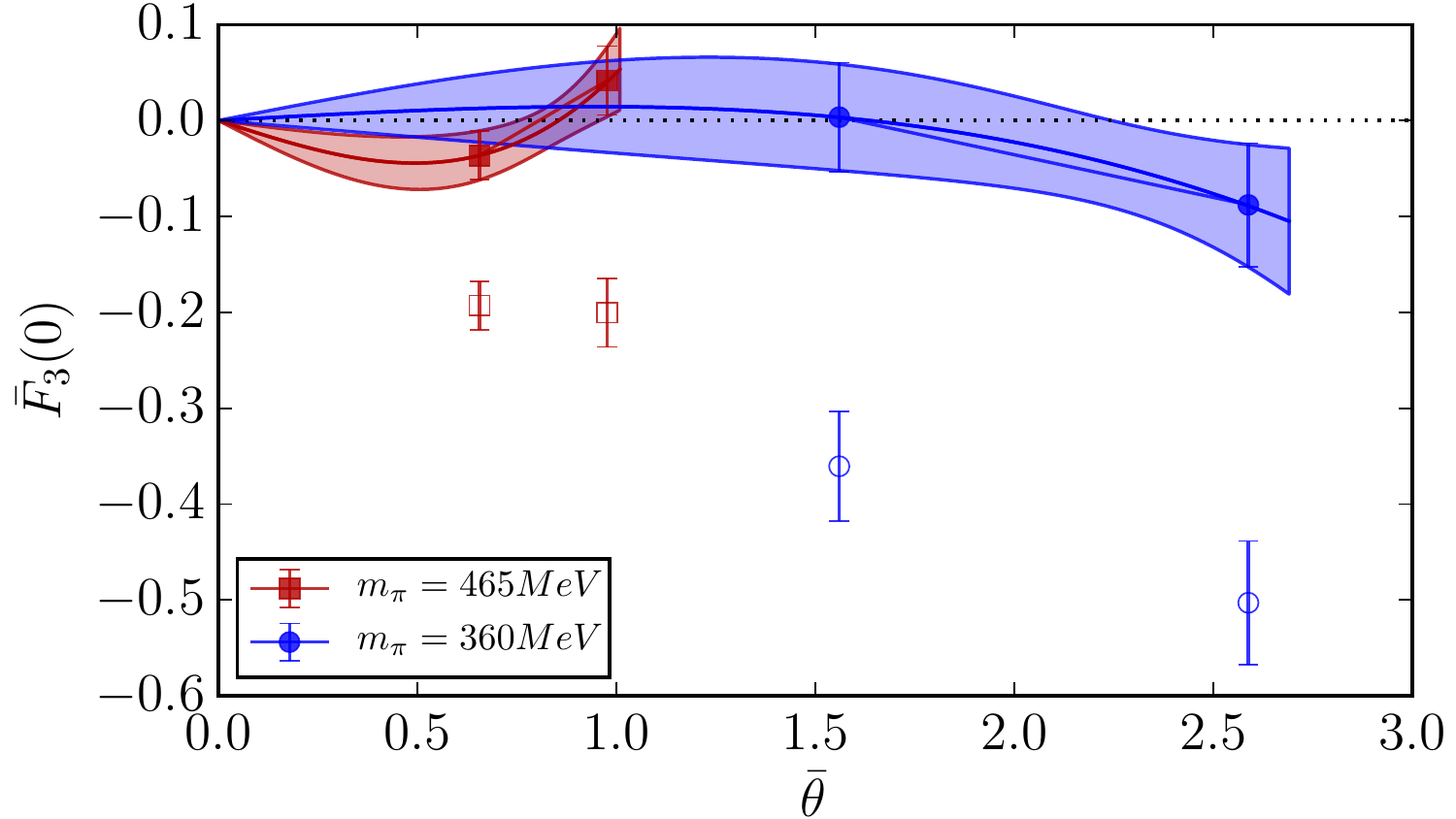}
\caption{Corrected (filled symbols) and original (open symbols) values for the neutron
  form factor $F_3$ at nonzero imaginary $\theta$-angle from Ref.~\cite{Guo:2015tla}.
  The linear parts in the limit $\theta\to0$ are shown in Tab.~\ref{tab:end_refs_correctF3}.
  \label{fig:cmpF3_Guo2015tla}}
\end{figure}

Although it is more suitable that the original authors of Refs.~\cite{Shintani:2005xg,
Berruto:2005hg,Aoki:2008gv,Guo:2015tla, Shindler:2015aqa,Alexandrou:2015spa,Shintani:2015vsx} 
reanalyze their data with these new formulas,
it is interesting to examine whether the presently available lattice calculations 
necessarily yield non-zero values $\bar\theta$-induced nucleon EDM 
after corrections similar to Eq.~(\ref{eqn:edm_refs_correction}) have been applied.
The most precise result for $F_{3n}(0)$ that also allows us to perform the correction
unambiguously is Ref.\cite{Alexandrou:2015spa}, 
which reports an $8\sigma$ non-zero value for $F_3(0)=-0.56(7)$ from calculations with dynamical 
twisted-mass fermions at $m_\pi=373\text{ MeV}$. 
However, when we apply the corresponding correction~(\ref{eqn:edm_refs_correction}), 
the value becomes $0.09(7)$ and essentially compatible with zero.

Calculations with finite imaginary $\theta$-angle~\cite{Aoki:2008gv,Guo:2015tla} yield the most precise 
values of the neutron EDM to date. 
However, they do not contain sufficient details to deduce the proper correction for $F_3$.
It must also be noted that it is not clear if the sign of the $\CP$-odd interaction $\sim\tilde{G}G$ 
is consistent in all of the Refs.~\cite{Shintani:2005xg,Berruto:2005hg,Aoki:2008gv,Guo:2015tla, 
Shindler:2015aqa,Alexandrou:2015spa,Shintani:2015vsx}.
On the other hand, all the reported non-zero results for the proton and neutron EDM agree in sign 
with $F_{3n}(0) < 0$ and $F_{3p}(0) > 0$, and it is \emph{reasonable to assume} that any 
differences in the conventions are compensated in each final reported EDM value.
Furthermore, because the $\theta$-angle is equivalent to a chiral rotation of quark fields, 
it is then \emph{reasonable to assume} that upon conversion to some common set of conventions,
e.g., those of Ref.~\cite{Alexandrou:2015spa}, the sign of the chiral rotation angle $\alpha$ 
agrees between different calculations.
Based on these plausible assumptions, we deduce that the results in \cite{Aoki:2008gv,Guo:2015tla} 
must be corrected as $F^\theta_3 = \tilde{F}^\theta_3 + 2\alpha(\theta) F_2$\footnote{
  Strictly speaking, for finite values of $\bar{\theta}$ and $\bar\alpha(\bar\theta)$, one has to 
  use the hyperbolic ``rotation'' formula $\cosh(2\alpha)F_3 = \tilde{F}_3 + \sinh(2\alpha) F_2$, 
  in which we neglect $O(\alpha^2)$ terms because $|\alpha|\lesssim0.15$, while the precision is 
  only $\approx10\%$.},
where $\alpha<0$, in analogy with Ref.~\cite{Alexandrou:2015spa}.
The data for $\bar\alpha^\theta$ and $\tilde F_3^\theta(0)$ at finite $\bar\theta$ values are 
extracted from figures in Ref.~\cite{Guo:2015tla}.
The original $\tilde F_3^\theta(0)$ and the corrected $F_3^\theta(0)$ values are shown in
Fig.~\ref{fig:cmpF3_Guo2015tla}.
Following Ref.~\cite{Guo:2015tla}, the corrected $F_3^\theta(0)$ values are interpolated to 
$\bar\theta\to0$ using a linear+cubic fit $F_3(0)\bar\theta + C\bar\theta^3$
and the resulting normalized values $F_3(0)=dF_3^\theta/d\bar\theta\big|_{\bar\theta=0}$ 
are given in Tab.~\ref{tab:end_refs_correctF3}.
We observe that the corrected values at both the finite and zero $\bar\theta$ values agree with
zero at $\lesssim2\sigma$ level.

Corrections to other results~\cite{Shintani:2005xg,Berruto:2005hg},
 may be done on the similar basis\footnote{
  Correction to results in Ref.~\cite{Aoki:2008gv} require the corresponding values for $F_2$,
  which we could not locate in published works.}.
The resulting values are also collected in Tab.~\ref{tab:end_refs_correctF3}, and in most cases
they are compatible with zero, deviating at most $2\sigma$. 
We emphasize that, apart from Ref.~\cite{Alexandrou:2015spa}, these corrections are made using the
sign assumptions discussed above.
If our assumptions are wrong, the corrected central values will be approximately twice as large
compared to the originally reported values.
Although we find our assumptions \emph{plausible}, and thus the corrected values in
Tab.~\ref{tab:end_refs_correctF3} most likely valid, it is up to the authors of
Refs.~\cite{Shintani:2005xg,Berruto:2005hg,Aoki:2008gv,Guo:2015tla,Shindler:2015aqa,Shintani:2015vsx}
to reanalyze their data and confirm or deny our findings.
It is possible that the difference between the lattice values of the neutron EDM and
phenomenological estimates $d_n\sim O(10^{-3}\ldots10^{-2})\,\bar\theta\,e\cdot\mathrm{fm}$~\cite{
Crewther:1979pi,Pospelov:1999ha,Guo:2012vf,Mereghetti:2010kp}, 
which has been ascribed to chiral symmetry breaking of lattice fermions and the 
heavy quark masses used in simulations, can disappear when the proper corrections are applied.

\section{Summary and Conclusions}

Among our most important findings in this paper is the new formula for analysis of nucleon-current
correlators computed in $\CPviol$ vacuum and extraction of the electric dipole form factor $F_3$.
We have demonstrated, both analytically and numerically, that the analysis of the 
$\bar\theta$-induced nucleon EDM in previous calculations~\cite{Shintani:2005xg,
Berruto:2005hg,Aoki:2008gv,Guo:2015tla,Shindler:2015aqa,Alexandrou:2015spa,Shintani:2015vsx} 
received contribution $(-2\alfive \kappa)$ from spurious mixing with the anomalous magnetic 
moment $\kappa$ of the nucleon.
Fortunately, the correction is very simple and requires only the values of the nucleon 
anomalous magnetic moments from calculations on the same lattice ensembles.
Applying this correction properly is somewhat complicated due to differences
in the conventions used in these works.
Under some \emph{plausible} assumptions we have demonstrated that, after the correction, 
even the most precise current lattice results for $\bar\theta$-nEDM may be compatible with zero.
If this finding is confirmed in detailed reanalysis of Refs.~\cite{Shintani:2005xg,
Berruto:2005hg,Aoki:2008gv,Guo:2015tla,Shindler:2015aqa,Alexandrou:2015spa,Shintani:2015vsx}, 
the precision of the current lattice QCD determination of $\bar\theta$-nEDM may be completely 
inadequate \emph{to constrain} the QCD $\bar\theta$ angle from experimental data.
The entire modern Physics program to search for fundamental symmetry violations as signatures
of new physics relies on our understanding of the effects of quark and gluon $\CPviol$ 
interactions on nucleon structure.
The importance and urgency of first-principles calculations of these effects
hardly needs more emphasis, and we have to conclude, that they will likely be 
even more difficult that thought before.

In this paper, we have performed calculations of nucleon electric dipole moments 
induced by $\CP$-odd quark-gluon interactions using two different methods. 
In the first method, we have successfully calculated the nucleon-current correlators 
with and without the $\CP$-odd interaction, evaluating up to four-point connected nucleon 
correlation functions.
We have demonstrated that this novel technique works well and we argue that it is both cheaper 
and has fewer uncertainties than the technique used in
~\cite{Bhattacharya:2016oqm,Bhattacharya:2016rrc} to compute the same observables with modified
Wilson action.
One of the obstacles to applying the technique of
Refs.~\cite{Bhattacharya:2016oqm,Bhattacharya:2016rrc} is that low-eigenmode deflation used to
accelerate calculations will be more expensive, because the eigenvectors have to be computed for 
every modification of the fermion action.
This may also be partially true for recently introduced multi-grid methods, in which
operator-dependent subspace null vectors have to be computed in the multi-grid setup phase, 
which has considerable cost.

In the second method, we computed the neutron EDM using its energy shift in uniform background
electric field and in the presence of the same $\CP$-odd interaction.
The energy-shift method to compute nucleon EDM has been used before~\cite{
Aoki:1989rx,Shintani:2006xr,Shintani:2008nt},
but our calculation is the first one that uses the uniform background electric
field that respects boundary conditions~\cite{Detmold:2009dx}.
We perform calculations with identical statistics in both methods and can directly compare
the central values and the uncertainties of the results.
We find that the EDM results agree if the new formula for extraction of the EDFF $F_3$ is used. 
Also, both methods yield comparable uncertainty, and the energy shift method may be preferable in
the future because it does not require forward-limit extrapolation and the excited states may be
easier to control~\cite{Bouchard:2016heu}.

Our calculations on a lattice are far from perfect and require improvement of the treatment of
excited states and forward-limit extrapolation of the form factors.
However, the associated systematic uncertainties are too small to cast doubt on the numerical
comparison of the energy shift and the form factor methods.
Although our calculations lack evaluation of the disconnected diagrams and 
renormalization and mixing subtractions of the quark chromo-EDM operator, 
these drawbacks apply equally to both methods, therefore do not affect said validation.

Future calculation of disconnected contributions to the $F_3$ form factors will 
be an extension to the present work, in which the quark-disconnected loops with 
insertions of the quark current, chromo-EDM, and both, will be evaluated and used 
together with the existing nucleon correlators.
The disconnected contractions do not require four-point correlators and are simpler 
to construct, although the stochastic noise will likely be a much bigger problem 
than for the connected contractions.
We expect that with advances in numerical evaluation of the disconnected
diagrams~\cite{Gambhir:2016jul}, this problem will be tractable.

\begin{acknowledgements}
T.B. is supported by US DOE grant DE-FG02-92ER40716.
T.I. is supported in part by US DOE Contract AC-02-98CH10886(BNL). 
T.I. is also supported in part by the Japanese Ministry of Education 
Grant-in-Aid, No. 26400261.
H.O. is supported by the RIKEN Special Postdoctoral Researcher program
S.N.S. was supported by the Nathan Isgur fellowship program at JLab
and by RIKEN BNL Research Center under its joint tenure track
fellowship with Stony Brook University.
This material is based upon work supported by the U.S. Department of Energy, Office of Science,
Office of Nuclear Physics under contract DE-AC05-06OR23177.
The U.S. Government retains a non-exclusive, paid-up, irrevocable, world-wide license to publish
or reproduce this manuscript for U.S. Government purposes.
S.A. and S.N.S are also grateful for the hospitality of Kavli Institute for Theoretical Physics 
(UC Santa Barbara) during the ``Nuclear16'' workshop.
This research was supported in part by the National Science Foundation 
under Grant No. NSF PHY11-25915. 
%
%
%
Gauge configurations with dynamical domain wall fermions used in this work
were generated by the RBC/UKQCD collaboration.
The computation was performed using the Hokusai supercomputer of the RIKEN ACCC facility
and Jlab cluster as part of the USQCD collaboration.
The calculations were performed with the ``Qlua'' software suite\cite{qlua-software}.
\end{acknowledgements}

\appendix
\section{Conventions}\label{sec:app_mink_euc}

In this appendix section, we collect conventions for $\gamma$-matrices implicitly or explicitly
used throughout the text.
In Table~\ref{tab:mink_euc_cmp}, we also provide notes the transformation between 
Minkowski ($\MTwoConv$) and Euclid ($\EucConv$) notations to avoid any ambiguities in matching
Minkowski and Euclidean form factor expressions for matrix elements and vertices.

In Minkowski space with metric $\{-1,-1,-1,+1\}$, we use the chiral $\gamma$-matrix basis
\begin{equation}
\label{eqn:gamma_chiral}
[\gamma^i]_\MTwoConv = \left(\begin{array}{cc} & \sigma^i \\ -\sigma^i & \end{array}\right)\,, \quad
[\gamma^4]_\MTwoConv = \left(\begin{array}{cc} & 1 \\ 1 & \end{array}\right)\,, \quad
\end{equation}
and with $\epsilon^{4123}=+1$ we define the chiral $\gamma_5$ matrix 
\begin{equation}
[\gamma_5]_\MTwoConv 
  = -\frac{i}{4!}[\epsilon^{\mu\nu\rho\sigma}
      \gamma_\mu\gamma_\nu\gamma_\rho\gamma_\sigma]_\MTwoConv
  = i[\gamma^4\gamma^1\gamma^2\gamma^3]_\MTwoConv
  = \left(\begin{array}{cc} -1 & \\ & 1\end{array}\right)\,,
\end{equation}
For the spin matrix $\sigma^{\mu\nu}=\frac{i}2[\gamma^\mu,\gamma^\nu]$ we 
will also need the relation 
\begin{equation}
\label{eqn:gamma5sigma}
[\sigma^{\mu\nu}\gamma_5]_\MTwoConv
  = \frac{i}2[\epsilon^{\mu\nu\rho\sigma}\sigma_{\rho\sigma}]_\MTwoConv
\end{equation}
In accordance with Tab.~\ref{tab:mink_euc_cmp}, the $\gamma$-matrices in Euclidean space are
\begin{equation}
\label{eqn:gamma_euc}
[\gamma^i]_\EucConv = \left(\begin{array}{cc} & -i\sigma^i \\ +i\sigma^i & \end{array}\right)\,, \quad
[\gamma^4]_\EucConv = \left(\begin{array}{cc} & 1 \\ 1 & \end{array}\right)\,, \quad
\end{equation}
in which $\gamma^{1,3}$ have the opposite sign compared to the deGrand-Rossi basis used in most 
of the lattice QCD software.
This difference is inconsequential because all results are manifestly covariant with respect 
to unitary basis transformations.
Finally, we use the $\gamma_5$ definition that agrees with the lattice software,
\begin{equation}
\label{eqn:gamma5_euc}
[\gamma_5]_\EucConv = [\gamma^1\gamma^2\gamma^3\gamma^4] = 
  \left(\begin{array}{cc} 1 & \\ & -1\end{array}\right)\,,
\end{equation}
and note that the kinematic coefficients for vector form factors derived in
Sec.~\ref{sec:app_kincoeff} depend on a particular $\gamma_5$ definition in terms of
$\gamma^\mu$, but the numerical lattice results are invariant as long as the same
$[\gamma_5]_\EucConv$ is used in both Eqs.~(\ref{eqn:qcedm_lat})
and~(\ref{eqn:nuc_cur_vertex_euc}).

\begin{table}
\caption{Correspondence between notations used in Minkowski $\MTwoConv$ 
  (metric $\{-,-,-,+\}$) and Euclidean $\EucConv$ space.
  Upon transition $\MTwoConv\leftrightarrow\EucConv$, the quantities in the corresponding 
  columns transform into each other.
  \label{tab:mink_euc_cmp}}
\begin{tabular}{l|r|r}
\hline\hline
What & $[*]_{\MTwoConv}$ & $[*]_{\EucConv}$\\
\hline
Coordinate
  & $(\vec x,t)=(x^i, x^4)$ &  $(x^i, -ix^4)$ \\
Momentum
  & $(\vec p,E)=(p^i, p^4)$ &  $(p^i, -ip^4)$ \\
Scalar product
  & $a^\mu b_\mu$ & $(-a^\mu b_\mu)$ \\
Plane wave 
  & $e^{-ipx}=e^{-iEt+i\vec p\vec x}$ & $e^{ipx}=e^{-Ex^4+i\vec p\vec x}$ \\
$\gamma$-matrices
  & $(\gamma^i, \gamma^4)$  & $(i\gamma^i, \gamma^4)$ \\
``Slashed'' vector
  & $\slashed{p}=p^\mu\gamma_\mu$
  & $(-i\slashed{p})=(-ip^\mu\gamma_\mu)$ \\ 
Dirac operator
  & $(\slashed{p} - m)$ 
  & $(i\slashed{p} + m)$ \\
Spin matrix
  & $( \sigma^{ij}, \sigma^{i4})$
  & $(-\sigma^{ij},i\sigma^{i4})$ \\
The Thing $\sigma^{\mu\nu}q_\nu$
  & $(\sigma^{i\nu}q_\nu, \sigma^{4\nu}q_\nu)$
  & $(\sigma^{i\nu}q_\nu,-i\sigma^{4\nu}q_\nu)$ \\
\hline\hline
\end{tabular}
\end{table}

\section{Electric and Magnetic Dipole Moments and Form Factors}\label{sec:edm_ff_rest}

In this Appendix section, we recall the connection between the form factors $F_{2,3}$ and
the magnetic and electric dipole moments of a spin-1/2 particle.
Although this is discussed in many textbooks, we find it useful to perform a rigorous derivation
expanding the matrix element~(\ref{eqn:ff_cpviol_mink}) in the momentum transfer $q=p^\prime-p$
and taking the limit $q\to0$.
For completeness and to avoid any ambiguities, in addition to the $\gamma$-matrices 
in Sec.~\ref{sec:app_mink_euc}, we collect all relevant conventions for E\&M fields, 4-spinors, 
and their interaction.
The discussion in this Section assumes Minkowski conventions $\MTwoConv$ with 
$g_{\mu\nu}=\mathrm{diag}\{-1,-1,-1,+1\}$.

The fermion-photon interaction is determined by the form of the ``long'' derivative,
\begin{equation}
D_\mu = \partial_\mu + ieA_\mu,
\quad 
\mcL = \bar{\psi}(iD_\mu\gamma^\mu - m) \psi 
  = \bar{\psi}(i\slashed{\partial} - m)\psi - e A_\mu J^\mu\,,
\end{equation}
which leads to the interaction Hamiltonian
\begin{equation}
\label{eqn:Hint_photon}
H_\text{int} = \int\,d^3x\,(-\mcL_\text{int})
  = e\int\,d^3x\, A_\mu J^\mu = e\,\int\,d^3x\, (\rho\phi - \vec J\cdot \vec A)
\end{equation}
where the EM potential $A^\mu=(\vec A, \phi)$, EM current $J^\mu=(\vec J, \rho)$,
and the electric coupling (charge) $e=|e|$. 

To evaluate the matrix element~(\ref{eqn:ff_cpviol_mink}) in the
interaction~(\ref{eqn:Hint_photon}), we use the chiral $\gamma$-matrix representation
summarized in Appendix~\ref{sec:app_mink_euc}.
The on-shell spinors satisfying the regular Dirac equation with a real-valued mass $m>0$ 
and energy $E^{(\prime)}=\sqrt{m^2+\vec p^{(\prime)2}}$
take the form
\begin{equation}
\label{eqn:spinor_expand}
\begin{aligned}
u_p 
  &= \left(\begin{array}{c}\sqrt{E-\vec p\vec\sigma}\xi \\ 
              \sqrt{E+\vec p\vec\sigma}\xi \end{array}\right) 
  = \sqrt{m}\big[1 + \frac{\vec p\vec\Sigma}{2m}\gamma_5 + O(\vec p^2)\big]
  \left(\begin{array}{c} \xi \\ \xi \end{array}\right)\,,
\\
\bar{u}_{p^\prime} 
  &= \left(\begin{array}{c}\sqrt{E^\prime-\vec p^\prime\vec\sigma}\xi^\prime \\ 
              \sqrt{E^\prime+\vec p^\prime\vec\sigma}\xi^\prime \end{array}\right)^\dag \gamma^4
  = \sqrt{m}\left(\begin{array}{c} \xi^\prime \\ \xi^\prime \end{array}\right)^\dag
     \big[1 - \frac{\vec p^\prime\vec\Sigma}{2m}\gamma_5 + O(\vec p^2)\big]
\end{aligned}
\end{equation}
where
$
\Sigma^k = \frac12\epsilon^{ijk}\sigma^{jk} 
  = \left(\begin{array}{cc} \sigma^k & \\ &\sigma^k\end{array}\right)\,.
$
We will use these spinors to evaluate matrix elements of the Hamiltonian~(\ref{eqn:Hint_photon}),
treating the E\&M field as classical background. 
Note that in order to treat these matrix elements as the interaction energy, 
they must be normalized as non-relativistic,
\begin{equation}
E_\text{int} 
  = \langle\vec p^\prime,\sigma^\prime|H_\text{int}|\vec p,\sigma\rangle_\text{NR}
  = eA_\mu \frac1{\sqrt{2E^\prime\cdot2E}} \bar{u}_{p^\prime}\Gamma^\mu u_p
  \doteq eA_\mu\llangle \Gamma^\mu\rrangle\,,
\end{equation}
where we introduced the notation 
$\llangle X\rrangle=\frac1{\sqrt{2E^\prime\cdot2E}}\bar{u}_{p^\prime} X u_p$ for convenience.
In the limit of small spatial momenta $|\vec p|,|\vec p^\prime|\to0$, only the spatial components
$\sigma^{ij}$ give non-vanishing contributions when contracted with
spinors~(\ref{eqn:spinor_expand}):
\begin{equation}
\llangle \sigma^{ij}\rrangle 
  = \frac1{\sqrt{2E\,2E^\prime}} \bar{u}_{p^\prime} \sigma^{ij} u_p 
  = \epsilon^{ijk}\,\xi^{\prime\dag} \sigma^k\xi + O(|\vec p|,|\vec p^\prime|)\,,
\quad
\llangle \sigma^{4k}\rrangle 
  = \bar{u}_{p^\prime} \sigma^{4k} u_p 
  = O(|\vec p|,|\vec p^\prime|)\,,
\end{equation}

Recalling the conventions~\cite{Landafshitz:v2} for the EM potential $A^\mu$,
\begin{align}
(\vec E)^i 
  &= -\frac{\partial}{\partial x^i} A^4 - \frac{\partial}{\partial t} (\vec A)^i\,,
\\
(\vec H)^i &= \big(\mathrm{curl}\vec A\big)^i
  = \epsilon^{ijk}\frac{\partial}{\partial x^j} (\vec A)^k\,,
\\
\end{align}
which result in the following field strength tensor $F_{\mu\nu}$ and its dual
$\tilde{F}_{\mu\nu}=\frac12\epsilon_{\mu\nu\rho\sigma} F^{\rho\sigma}$,
$\epsilon_{1234}=+1$
\begin{align}
\label{eqn:Fmunu}
F_{\mu\nu} &= \left(\begin{array}{r|rrrr}
  & 1 & 2 & 3 & 4\\
\hline
    1 &     0  & -H^3  &  H^2  & -E^1 \\
    2 &   H^3  &    0  & -H^1  & -E^2 \\
    3 &  -H^2  &  H^1  &    0  & -E^3 \\
    4 &   E^1  &  E^2  &  E^3  &    0 
\end{array}\right)\,,
\\
\label{eqn:Fmunu_dual}
\tilde{F}_{\mu\nu} &= \left(\begin{array}{r|rrrr}
  & 1 & 2 & 3 & 4\\
\hline
    1 &     0  &  E^3  & -E^2  & -H^1 \\
    2 &  -E^3  &    0  &  E^1  & -H^2 \\
    3 &   E^2  & -E^1  &    0  & -H^3 \\
    4 &   H^1  &  H^2  &  H^3  &    0 
\end{array}\right)\,,
\end{align}
where the rows and the columns are enumerated by $\mu$ and $\nu$, respectively.
With the following conventions for the fermion and photon fields with definite momenta
$p^{(\prime)}$ and $q$, respectively,
\begin{equation}
\label{eqn:field_momenta}
\psi_p(x)\sim e^{-ipx}\,,
\quad \bar{\psi}_{p^\prime}(x) \sim e^{ip^\prime x}\,,
\quad A_{q,\mu}(x)\sim e^{-i(p^\prime-p)x} = e^{-iqx}\,,
\end{equation}
the derivatives acting on these fields are translated into momentum factors,
\begin{align}
\label{eqn:psi_mom_deriv}
\slashed{\partial}\psi = \gamma^\mu\partial_\mu \psi 
  &\rightarrow \gamma^\mu (-ip_\mu) \psi = (-i)\slashed{p}\psi \,,
\\
\label{eqn:Fmunu_mom}
F_{\mu\nu}(x) = \partial_\mu A_\nu - \partial_\nu A_\mu 
  &\rightarrow (-i) ( q_\mu A_\nu - q_\nu A_\mu)\,.
\end{align}
Applying the Gordon identity to Eq.~(\ref{eqn:ff_cpviol_mink}) and omitting the $F_A$ form
factor, we get 
\begin{equation}
\langle p^\prime,\sigma^\prime |J^\mu|p,\sigma\rangle_{\CPviol}
  = \bar{u}_{p^\prime} \big[
    F_1 \frac{(p^\prime+p)^\mu}{2m} 
    + (G_M + i\gamma_5 F_3) \frac{i\sigma^{\mu\nu}q_\nu}{2m}
  \big] u_{p,\sigma}\,,
\end{equation}
where $G_M=F_1+F_2$ is the magnetic Sachs form factor determining the full magnetic moment 
$\mu=Q+\kappa=G_M(0)$.
The first term is independent of the spin and is equal to the electromagnetic interaction 
of a scalar particle, which we omit as irrelevant.
With the use of (\ref{eqn:gamma5sigma}) and (\ref{eqn:Fmunu_mom}), the spin-dependent 
interaction energy takes the form 
\begin{equation}
E_\text{int,spin} 
  = i q_\nu A_\mu \, \big[
        eG_M\,\frac{\llangle\sigma^{\mu\nu}\rrangle}{2m}
       - eF_3\,\frac12\epsilon^{\mu\nu\rho\sigma}\frac{\llangle\sigma_{\rho\sigma}\rrangle}{2m}
      \big]
  = \frac12 \big(\frac{e G_M}{2m} \, F_{\mu\nu} - \frac{e F_3}{2m} \, \tilde{F}_{\mu\nu}\big)\,
      \llangle\sigma^{\mu\nu}\rrangle\,.
\end{equation}
Neglecting all but the leading order in $O(|\vec p|,|\vec p^\prime|)$, 
we only have to keep the spatial components $\llangle\sigma^{ij}\rrangle$:
\begin{equation}
\label{eqn:Eint_EB}
E_\text{int,spin} 
  = -\frac{e G_M}{2m} \, \vec H \cdot \hat\Sigma  - \frac{e F_3}{2m} \vec E\cdot \hat\Sigma\,,
\end{equation}
where the unit spin vector $\hat\Sigma = \xi^{\prime\dag}\vec\sigma\xi$, $|\hat\Sigma|=1$.
The coupling coefficients to the magnetic and electric fields in the above equation have to be
identified with the magnetic and electric dipole moments, respectively,
\begin{equation}
\label{eqn:mdm_edm}
\mu_N = G_M(0)\,,
\quad d_N = F_3(0)\,.
\end{equation}
which both are expressed here in the particle magneton units $e/(2m)$.

Note that the above derivation could be repeated for the chirally-rotated spinors and the
nucleon-current vertex~(\ref{eqn:ff_cpviol_mink_wrong}). 
It can be easily shown that the only change compared to Eq.~(\ref{eqn:Eint_EB}) would be that the
magnetic and electric fields would couple to some orthogonal linear combinations of
$\tilde{F}_{2,3}$, and that these combinations would reproduce $F_2$ and $F_3$ exactly in
agreement with Eq.~(\ref{eqn:F23_mixing}).

Finally, we note that if one uses the chirally-rotated spinors to calculate the spatial 
matrix elements $\llangle\sigma^{ij}\rrangle$, they are reduced by a factor of $\cos(2\alfive)$
while the timelike matrix elements $\llangle\sigma^{4k}\rrangle$ become non-zero:
\begin{equation}
\begin{aligned}
e^{2i\alfive\gamma_5} \sigma^{ij} 
  &= \cos(2\alfive) \sigma^{ij} + \sin(2\alfive) \epsilon^{ijk} \sigma^{4k}\,,\\
e^{2i\alfive\gamma_5} \epsilon^{ijk}\sigma^{4k} 
  &= -\sin(2\alfive) \sigma^{ij} + \cos(2\alfive) \sigma^{4k}\,.
\end{aligned}
\end{equation}
As we noted above, $\bar{u}_{p^\prime}\sigma^{ij}u_p$ couples to the magnetic field, 
while $\bar{u}_{p^\prime}\sigma^{4k}u_p$ couples to the electric field.
This ``mixing'' of electric and magnetic fields compensates exactly the mixing in
Eq.~(\ref{eqn:F23_mixing}) induced by using the chirally-rotated spinors
$\bar{\tilde{u}},\,\tilde{u}_p$ instead of the regular spinors $\bar{u}_{p^\prime},\,u_p$.

\section{Kinematic coefficients}\label{sec:app_kincoeff}

In this section, we present expressions for the kinematic coefficients for form factors
$F_{1,2,3}$ on a Euclidean lattice.
We use two types of the polarization projectors, (1) spin-average $T^+$ and (2) polarized
$T^+_{S_z}$. 
Both projectors also select the upper (positive-parity) part of the nucleon spinors
\begin{equation}
\label{eqn:tpol_plus_sz}
T^+=\big[\frac{1+\gamma^4}2\big]_{\EucConv}\,,\quad
T^+_{S_z}=\big[\frac{1+\gamma^4}2(-i\gamma^1\gamma^2)\big]_\EucConv\,.
\end{equation}

Form factor $F_3$ can be extracted from $\mu=3,4$ components of the vector current matrix
elements between $S_z$-polarized nucleon states.
Using handy notations for the positive-parity nucleon spinor matrices,
\begin{equation}
\mcS = -i\slashed{p}_\EucConv + m
\,,\quad
\mcS^\prime = -i\slashed{p}_\EucConv^\prime + m\,,
\end{equation}
the form factor expression for the $\CPviol$ nucleon-current correlation function on a lattice 
$C^{\CPviol}_\text{3pt}=C^{\CPviol}_{NJ^\mu \bar N}$ can be written as
\begin{equation}
\label{eqn:vecff_CPviol}
\begin{aligned}
\mathrm{Tr}\big[T_\text{pol}\,C^{\CPviol}_{N J^\mu\bar N}(\vec p^\prime,t; \vec q,t_{op})]
&\quad=\frac{e^{-E^\prime (t-t_{op}) - Et_{op}}}{2E^\prime\cdot2E} 
  \Tr\big[ e^{i\alfive\gamma_5} T e^{i\alfive\gamma_5} \cdot
        \mcS^\prime \cdot\Gamma_\EucConv^\mu(p^\prime,p)\cdot \mcS\big]\\
&\quad=\frac{e^{-E^\prime (t-t_{op}) - Et_{op}}}{2E^\prime\cdot2E} 
  \Tr\big[ \big(T + i\alfive\{\gamma_5,T\} + O(\alfive^2)\big) \cdot
        \mcS^\prime \cdot\Gamma_\EucConv^\mu(p^\prime,p)\cdot \mcS \big]\,,
\end{aligned}
\end{equation}
where, assuming that the $\CP$-odd interaction is small, we have expanded in
the $\CP$-odd mixing angle $\alfive$.

Below we quote formulas for contributions to the last line of Eq.~(\ref{eqn:vecff_CPviol})
computed for zero sink momentum $\vec p^\prime=0$, 
\begin{align*}
\text{source }\vec p &= - \vec q\,, & E&=\sqrt{m^2+\vec q^2}\,,\\
\text{sink }\vec p^\prime &= \vec p + \vec q = 0\,, & E^\prime&=m\,,
\end{align*}
with the nucleon spin projectors $T^+$ and $T^+_{S_z}$.
The $\alfive$-independent contribution is 
\begin{align}
\Tr[T^+\mcS^\prime\Gamma_\EucConv^\mu\mcS] 
&= 4m^2\left(\begin{array}{rrrr}
  i q_1 / m & -i\tau q_1 / m  & 0 \\
  i q_2 / m & -i\tau q_2 / m  & 0 \\
  i q_3 / m & -i\tau q_3 / m  & 0 \\
  2(1+\tau) & -2\tau(1+\tau)  & 0
\end{array}\right)\,,
\\
\Tr[T^+_{S_z}\mcS^\prime\Gamma_\EucConv^\mu\mcS] 
&= 4m^2\left(\begin{array}{rrrr}
   -q_2 / m & -q_2 / m  & q_1 q_3 / (2m^2) \\
    q_1 / m &  q_1 / m  & q_2 q_3 / (2m^2) \\
          0 &         0 & q_3^2 / (2m^2) \\
          0 &         0 & -i(1+\tau) q_3 / m
\end{array}\right)\,,
\end{align}
where the rows correspond to the Lorentz components $\mu=1,2,3,4$ 
and the columns correspond to the form factors $F_{1,2,3}$.
We have also introduced the frequently used kinematic variable $\tau$
\begin{equation}
\label{eqn:tau_def}
\tau \doteq \frac{Q^2}{4m^2} 
\,\overset{\vec p^\prime=0}{\equiv}\, \frac{E-m}{2m}\,.
\end{equation}
The coefficients of the contributions $\sim\alfive$ are
\begin{align}
\Tr[\{\gamma_5,T^+\}\mcS^\prime\Gamma_\EucConv^\mu\mcS] 
&= 4m^2\left(\begin{array}{rrrr}
  0 & 0 & -\tau q_1 / m \\
  0 & 0 & -\tau q_2 / m \\
  0 & 0 & -\tau q_3 / m \\
  0 & 0 & 2i\tau(1+\tau) \\
\end{array}\right)\,,
\\
\Tr[\{\gamma_5,T^+_{S_z}\}\mcS^\prime\Gamma_\EucConv^\mu\mcS] 
&= 4m^2\left(\begin{array}{rrrr}
  0         & i q_1 q_3 / (2m^2)          & 0 \\
  0         & i q_2 q_3 / (2m^2)          & 0 \\
  -2i\tau   & -2i\tau + i q_3^2 / (2m^2)  & 0 \\
  -q_3 / m  & \tau q_3 / m                & 0  \\
\end{array}\right)\,.
\end{align}
Up to order $O(\alfive)$, the $\CPviol$ nucleon correlation functions are\footnote{
  Note that both $\alfive$ and $F_3$ are proportional to the $\CP$-odd perturbation, therefore we
  consider $F_3=O(\alfive)$ and drop terms $\alfive F_3$ and higher.
} 
\begin{align}
\label{eqn:V3coeff_pos}
\Tr[T^+ C^{\CPviol}_{N J^3 \bar N}]
  &= \mcK \Big[
      i\frac{q_3}{m} G_E
    + O(\alfive^2) \Big]\,,
\\
\label{eqn:V4coeff_pos}
\Tr[T^+ C^{\CPviol}_{N J^4 \bar N}]
  &= \mcK \Big[
      2(1+\tau)G_E
    + O(\alfive^2)\Big]\,,
\\
\label{eqn:V3coeff_posSz}
\Tr[T^+_{S_z} C^{\CPviol}_{N J^3 \bar N}]
  &= \mcK \Big[
      2\tau\alfive G_M
    - \alfive\frac{q_3^2}{2m^2}F_2
    + \frac{q_3^2}{2m^2} F_3 
    + O(\alfive^2)\Big]\,,
\\
\label{eqn:V4coeff_posSz}
\Tr[T^+_{S_z} C^{\CPviol}_{N J^4 \bar N}]
  &= \mcK \Big[
    - i\alfive\frac{q_3}{m} G_E
    - i(1+\tau)\frac{q_3}{m} F_3 
    + O(\alfive^2)\Big]\,,
\end{align}
where $G_E=F_1-\tau F_2$ is the electric and $G_M=F_1+F_2$ is the magnetic Sachs form
factor, and 
\begin{equation}
\label{eqn:kincoeff}
\mcK=\frac{m}{E}e^{-E^\prime (t_{sep}-t_{op}) - E t_{op}}
\end{equation}
is the time dependence combined with kinematic
factors.
In the analysis of the $C_{NJ\bar N}/C_{N\bar N}$ ratios~(\ref{eqn:c3_c2_ratio}), the exponential
time dependence is canceled, and the kinematic coefficients have to be modified 
to take into account the traces of the nucleon two-point functions:
\begin{equation}
\label{eqn:kincoeff_ratio}
\mcK_\mcR=\frac{m}{\sqrt{2E(m+E)}}\,.
\end{equation}

In addition, we evaluate the extra contributions to the kinematic coefficients
$\sim\alfive\{\gamma_5,\Gamma_\EucConv^\mu\}$ that comes from spurious mixing of $F_{2,3}$ 
\begin{align}
\Tr[T^+\mcS^\prime \{\gamma_5,\Gamma_\EucConv^\mu\} \mcS] 
&= 4m^2\left(\begin{array}{rrrr}
  0 & 0 &  2\tau q_1 / m \\
  0 & 0 &  2\tau q_2 / m \\
  0 & 0 &  2\tau q_3 / m \\
  0 & 0 & -4i\tau (1+\tau) \\
\end{array}\right)\,,
\\
\Tr[T^+_{S_z}\mcS^\prime \{\gamma_5,\Gamma_\EucConv^\mu\} \mcS] 
&= 4m^2\left(\begin{array}{rrrr}
  0 & -i q_1 q_3 / m^2    & -2 i q_2 / m \\
  0 & -i q_2 q_3 / m^2    &  2 i q_1 / m \\
  0 & -i q_3^2 / m^2      &  0 \\
  0 & -2 (1+\tau) q_3 / m &  0 \\
\end{array}\right)\,,
\end{align}
which in Refs.~\cite{Shintani:2005xg,
Berruto:2005hg,Aoki:2008gv,Guo:2015tla,Shindler:2015aqa,Alexandrou:2015spa,Shintani:2015vsx}
contributes to the polarized nucleon-current correlators as
\begin{align}
\label{eqn:wrongcorrV3coeff_posSz}
\delta \Tr[T^+_{S_z} C^{\CPviol}_{N J^3 \bar N}]
  &\overset?= \mcK \Big[
      \alfive\frac{q_3^2}{m^2}F_2
    + O(\alfive^2)\Big]\,,
\\
\label{eqn:wrongcorrV4coeff_posSz}
\delta\Tr[T^+_{S_z} C^{\CPviol}_{N J^4 \bar N}]
  &\overset?= \mcK \Big[
    - 2i\alfive(1+\tau)\frac{q_3}{m} F_2
    + O(\alfive^2)\Big]\,,
\end{align}
If the terms~(\ref{eqn:wrongcorrV3coeff_posSz},\ref{eqn:wrongcorrV4coeff_posSz}) 
are erroneously added to the kinematic 
coefficients~(\ref{eqn:V3coeff_posSz},\ref{eqn:V4coeff_posSz}), analysis of the same 
lattice correlation functions will result in incorrect values of EDFF 
$\tilde{F}_3 = F_3 - 2\alfive F_2$, in full agreement with Eq.~(\ref{eqn:F23_mixing}).

\bibliographystyle{aip}
\bibliography{bib}

\end{document}